\documentclass{pasj01}
\usepackage{url}

\def\uncatcodespecials{\def\do##1{\catcode`##1=12}\dospecials}%
{\catcode`\`=\active\gdef`{\relax\lq}}
\def\setupcode 
    {\tt %
     \spaceskip=0pt \xspaceskip=0pt 
     \catcode`\`=\active
     \uncatcodespecials \obeyspaces
     \catcode`\{=1\catcode`\}=2
    }%
\def\SETUPCODE 
    {\tt %
     \spaceskip=0pt \xspaceskip=0pt 
     \catcode`\`=\active
     \uncatcodespecials \obeyspaces
    }%
\def\docode#1{#1\endgroup}%
\def\code{\begingroup\setupcode\docode}%

\begin{document}
\SetRunningHead{Aihara et al.}{HSC-SSP PDR1}

\title{First Data Release of the Hyper Suprime-Cam\\Subaru Strategic Program}

\author{
Hiroaki Aihara\altaffilmark{1},
Robert Armstrong\altaffilmark{2},
Steven Bickerton\altaffilmark{3},
James Bosch\altaffilmark{2},
Jean Coupon\altaffilmark{4},
Hisanori Furusawa\altaffilmark{5},
Yusuke Hayashi\altaffilmark{5},
Hiroyuki Ikeda\altaffilmark{5},
Yukiko Kamata\altaffilmark{5},
Hiroshi Karoji\altaffilmark{6,2}
% ...
}
\altaffiltext{1}{Department of Physics, University of Tokyo, Tokyo 113-0033, Japan}
\altaffiltext{2}{Department of Astrophysical Sciences, Princeton University, 4 Ivy Lane, Princeton, NJ 08544}
\altaffiltext{3}{ 	Orbital Insight, 100 W. Evelyn Ave. Mountain View, CA 94041}
\altaffiltext{4}{	Department of Astronomy, University of Geneva, ch. d’\'Ecogia 16, 1290 Versoix, Switzerland}
\altaffiltext{5}{	National Astronomical Observatory of Japan, 2-21-1 Osawa, Mitaka, Tokyo 181-8588, Japan}
\altaffiltext{6}{	National Institutes of Natural Sciences, 4-3-13 Toranomon, Minato-ku, Tokyo, JAPAN}
% ...
\email{masayuki.tanaka@nao.ac.jp}


\maketitle
\definecolor{gray}{rgb}{0.6, 0.6, 0.6}
\newcommand{\commentblue}[1]{\textcolor{blue} {\textbf{#1}}}
\newcommand{\commentred}[1]{\textcolor{red} {\textbf{#1}}}
\newcommand{\commentgray}[1]{\textcolor{gray} {\textbf{#1}}}


\begin{abstract}
  The Hyper Suprime-Cam Subaru Strategic Program (HSC-SSP) is a three-layered imaging survey aimed
  at addressing some of the most outstanding questions in astronomy today, including the nature of
  dark matter and dark energy.  The survey has been awarded 300 nights of observing time at the Subaru
  Telescope and it started in March 2014.  This paper presents
  the first public data release of HSC-SSP.  This release includes data taken in the first 1.7 years
  of observations (61.5 nights) and each of the Wide, Deep, and UltraDeep layers covers about 108, 26, and
  4 square degrees down to depths of $i\sim26.4$, $\sim26.5$, and $\sim27.0$~mag, respectively
  ($5\sigma$ for point sources). 
  All the layers are observed in five broad bands ($grizy$), and the Deep and UltraDeep layers are
  observed in narrow bands as well.
  We achieve an impressive image quality of 0.6 arcsec in the $i$-band in the Wide layer.
  We show that we achieve $1-2$ per cent PSF photometry (rms) both internally and externally (against Pan-STARRS1),
  and $\sim10$~mas and $40$~mas internal and external astrometric accuracy, respectively.   Both the calibrated
  images and catalogs are made available to the community through dedicated user interfaces and
  database servers.
  In addition to the pipeline products, we also provide
  value-added products such as photometric redshifts and a collection of public spectroscopic
  redshifts.  Detailed descriptions of all the data can be found online.  The data release website is
  \url{https://hsc-release.mtk.nao.ac.jp}.
\end{abstract}

\section{Introduction}

Hyper Suprime-Cam (HSC; \cite{2012SPIE.8446E..0ZM,miyazaki17}) is an optical imaging camera installed at the prime focus of
the Subaru Telescope.  It offers the widest field of view, 1.5 degree diameter, on existing 8-10m class telescopes;
combined with the telescope aperture, HSC is currently the most efficient survey instrument in terms of \'{e}tendue,
which is defined as the product of telescope aperture squared and the area of the field of view.
The focal plane is filled with 116 full-depletion Hamamatsu CCDs, of which 104 are used for science
and the remaining 12 are for guiding and focusing.  Each CCD has 2048$\times$4096 pixels and each pixel
subtends 0.168 arcsec on the sky  (15$\mu m$ physical).  These CCDs are 200$\mu m$ thick and are very sensitive even at $\sim1\mu m$,
making deep imaging at such long wavelengths possible.  An overview of the camera and results
from engineering runs can be found in \citet{miyazaki17}.  Details of the dewar system
are given in \citet{komiyama17}.  HSC has five broad band filters, $grizy$, as well as
a number of narrow band filters designed to study emission line objects at high redshifts.
\citet{kawanomoto17} describe detailed measurements of the system response functions.

A large imaging survey with HSC is being conducted as a Subaru Strategic Program (SSP).
The survey is led by an international collaboration of the Japanese community, Taiwan,
and Princeton University.  We have been awarded 300 nights over 5-6 years, which is
the largest program ever approved at the Subaru Telescope.  The goal of the survey is
to address outstanding astrophysical questions such as the nature of
dark matter and dark energy, the cosmic reionization, and galaxy evolution over cosmic
time.  Due to the legacy value of the survey data, it also allows us to tackle other important
scientific questions in many areas of astrophysics.  The HSC-SSP website\footnote{\url{http://hsc.mtk.nao.ac.jp/ssp/}}
gives details of our science goals.

As described in the survey description paper \citep{aihara17} and also on the HSC-SSP website,
the survey consists of three layers: Wide, Deep and UltraDeep.
The Wide survey aims to cover 1,400 square degrees of the sky in all five
broad band filters ($grizy$). The survey fields are mostly located around the equator;
two long stripes around the spring and autumn equator with an additional stripe
around the Hectomap region \citep{2011AJ....142..133G}.  The integration times are 10 min in $gr$ and 20 min in $izy$
broken into 4 ($gr$) and 6 ($izy$) dithers,
going down to $i\sim26$ at $5\sigma$ for
point sources (see the next section for details about the depths).
We also take a 30 second exposure to increase the dynamic range at the bright end.
We apply a large dither ($\sim1/3$ of field of view) between exposures to ensure uniform coverage.
Priority is given to $i$-band observations when the seeing is good ($\lesssim0.75$ arcsec).
in order to carry out precise shape measurements for weak-lensing science.

The Deep survey has four separate fields; XMM-LSS, Extended-COSMOS (E-COSMOS), ELAIS-N1
and DEEP2-F3.  These fields are widely separated in R.A. and
at least one of them is observable in any observing runs.  Each field consists of
4 HSC pointings, except for XMM-LSS which is 3 pointings, amounting to about 27 square
degrees in total for the 4 fields. 
Our goal is to expose for a few hours in the broad bands as well as in 3 narrow band
filters, NB387, NB816, and NB921 (the numbers indicate the central wavelength in nm).
We aim to reach $i\sim27$ and
the target exposure times are given in the next section.
We apply a small 5-point dither, 150 arcsec in R.A. and 75 arcsec in Dec., on top of
a larger random dither with $r<450$ arcsec from the fiducial center.
The larger dither is needed to have the same objects on different CCDs for better calibration.
A 30 second exposure is taken in each field as is done for the Wide layer.

The UltraDeep layer has two fields, COSMOS and SXDS (or equivalently UDS), with one pointing each.
We aim to obtain very deep images of these fields both in the broad and narrow band
filters (NB816, N921, NB101),
reaching down to $i\sim28$.
We repeatedly visit these fields in order to enable transient sciences.
The dither pattern is the same as in Deep.  When the seeing is very bad ($>1.3$ arcsec), we tend to
observe the UltraDeep fields so the data can be useful for transient science.

This paper presents the first public data release of the HSC-SSP, which includes 
data from the first 1.7 years of observation.  
We first give an overview of the release in Section 2.  We then move on to describe 
the data processing and resultant data products in Sections 3 and 4, respectively.
Section 5 presents a number of quality assurance tests as well as a list of known
problems in our data.  Section 6 briefly describes our catalog and image archive servers,
followed by our plans for the future data releases in Section 7.
Finally, the paper is summarized in Section 8.

\section{Overview of the Release}

\subsection{The release}
\label{sec:the_release}

\begin{figure*}
 \begin{center}
  \includegraphics[width=18cm]{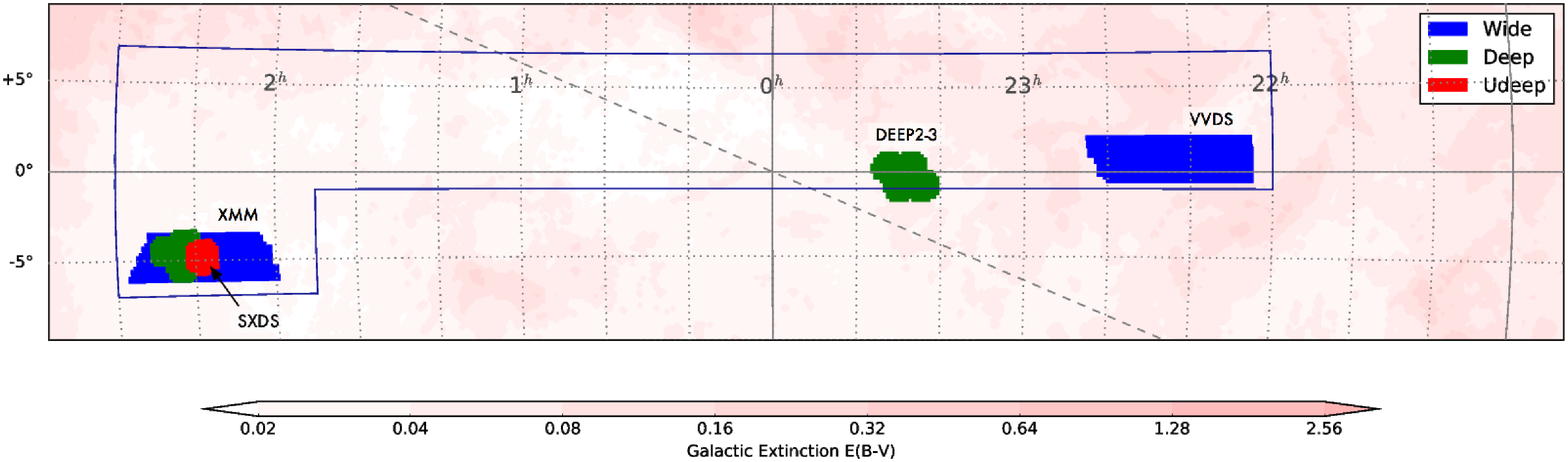}\vspace{1cm}
  \includegraphics[width=18cm]{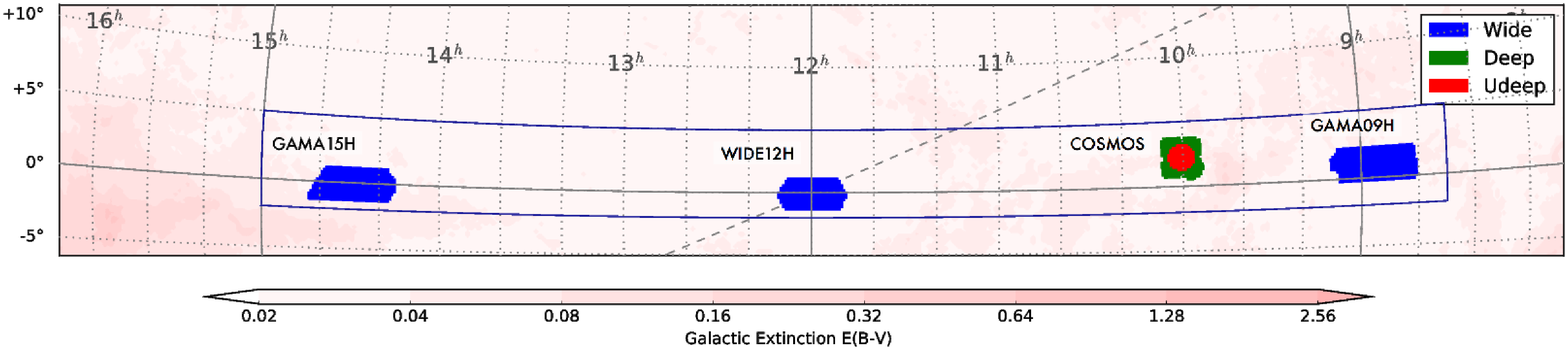}\vspace{1cm}
  \includegraphics[width=18cm]{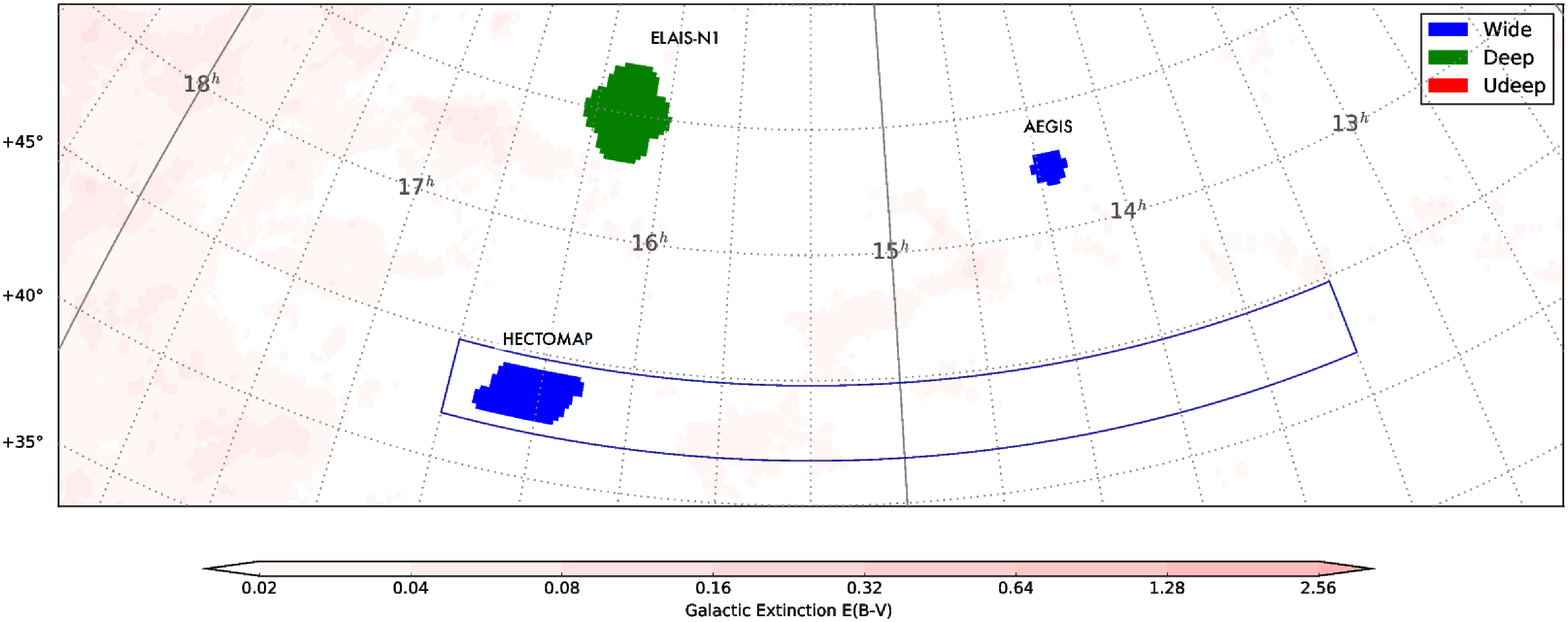} 
 \end{center}
 \caption{
   The area covered in this release shown in equatorial coordinates.
   The blue, green and red areas show the Wide,
   Deep, and UltraDeep layers, respectively included in the data release.
   The boxes indicate the approximate boundaries of the three disjoint regions that will make up the final Wide survey.
   Note that AEGIS is a calibration field observed
   at the Wide depth and is not formally a part of the Wide survey.
   The Galactic extinction map from \citet{1998ApJ...500..525S} is also shown as a grayscale.
 }
 \label{fig:sky_coverage}
\end{figure*}

This data release includes HSC data taken between March 2014 and November 2015 over
a total of 61.5 nights.
The data are processed with \texttt{hscPipe} \citep{bosch17}, a version of the LSST stack
\citep{2008arXiv0805.2366I,2010SPIE.7740E..15A,2015arXiv151207914J}, and both image and
catalog products are made available to the community through dedicated database servers
and user interfaces.  The image products include photometrically and astrometrically
calibrated CCD images, warped images, and coadds (we define the terminology in Section \ref{sec:terminology}).
The catalog products include both forced and unforced measurements of object positions and
object fluxes measured in various ways, together with measurement flags to indicate
the reliability of the measurements.
Precise galaxy shape measurements are withheld in this release because they are still being validated,
but they will be released in our future incremental release (see Section 7).
The catalog fits files that contain the shape measurements will also be made available in the same release.
In addition to the pipeline products, value-added products such as photometric redshifts
and a collection of public spectroscopic redshifts are available to the community.

The sky covered in this release is shown in Fig. \ref{fig:sky_coverage}.  For convenience,
we give names to each of the observed fields, as summarized in Table \ref{tab:field_names}.
A figure of our survey geometry with footprints of some of the extant large
imaging and spectroscopic surveys overlaid is available at the data release site.
The Wide data cover about 108 square degrees of the sky, mostly around the equator
in the 5 bands at the nominal survey depth.  The Deep and UltraDeep data are
shallower than the target depths but include the 5 broad bands over the full area,
plus partial coverage in two narrow bands, NB816 and NB921.
Table \ref{tab:exptime} summarizes the approximate exposure time, seeing, $5\sigma$ depth and saturation magnitudes for point sources
measured from the data as well as target exposures and depths.
As we will discuss in Section \ref{sec:data_quality}, we use flux uncertainties from the coadds to estimate
the depths and the depth quoted here may be somewhat optimistic.
As we discuss in Section \ref{sec:survey_depth}, the $5\sigma$ limits roughly correspond
to $50\%$ completeness limits.
Note as well that the seeing is derived from Gaussian-weighted moments \citep{2002AJ....123..583B}.
In total, this release includes $\sim7\times10^7$ objects and the data volume exceeds 70 Tbyes
as summarized in Table \ref{tab:data_releases}.
The data release site\footnote{\url{https://hsc-release.mtk.nao.ac.jp/}} describes in more
detail the available data products as well as how to use our online/offline tools.  In addition,
the site maintains a up-to-date list of known problems (see Section \ref{sec:known_problems} and FAQs.  Note that only the processed
data are available at the data release site.  The raw data are made available through
SMOKA\footnote{\url{http://smoka.nao.ac.jp/}}.

\begin{table}[htbp]
  \begin{center}
    \begin{tabular}{cc}
      \hline
      Layer      &  Field Name\\
      \hline
      UltraDeep  &  COSMOS\\
      UltraDeep  &  SXDS\\
      Deep       &  XMM-LSS\\
      Deep       &  E(xtended)-COSMOS\\
      Deep       &  ELAIS-N1\\
      Deep       &  DEEP2-3\\
      Wide       &  XMM-LSS\\
      Wide       &  GAMA09H\\
      Wide       &  WIDE12H\\
      Wide       &  GAMA15H\\
      Wide       &  HECTOMAP\\
      Wide       &  VVDS\\
      ---        &  AEGIS\\
      \hline 
    \end{tabular}
  \end{center}
  \caption{
    List of the observed fields.  AEGIS is observed as a photometric redshift calibration field at
    the Wide depth.
  }
  \label{tab:field_names}
\end{table}

\subsection{Survey Progress}
\label{sec:survey_progress}

\begin{figure}
  \begin{center}
  \includegraphics[width=9cm]{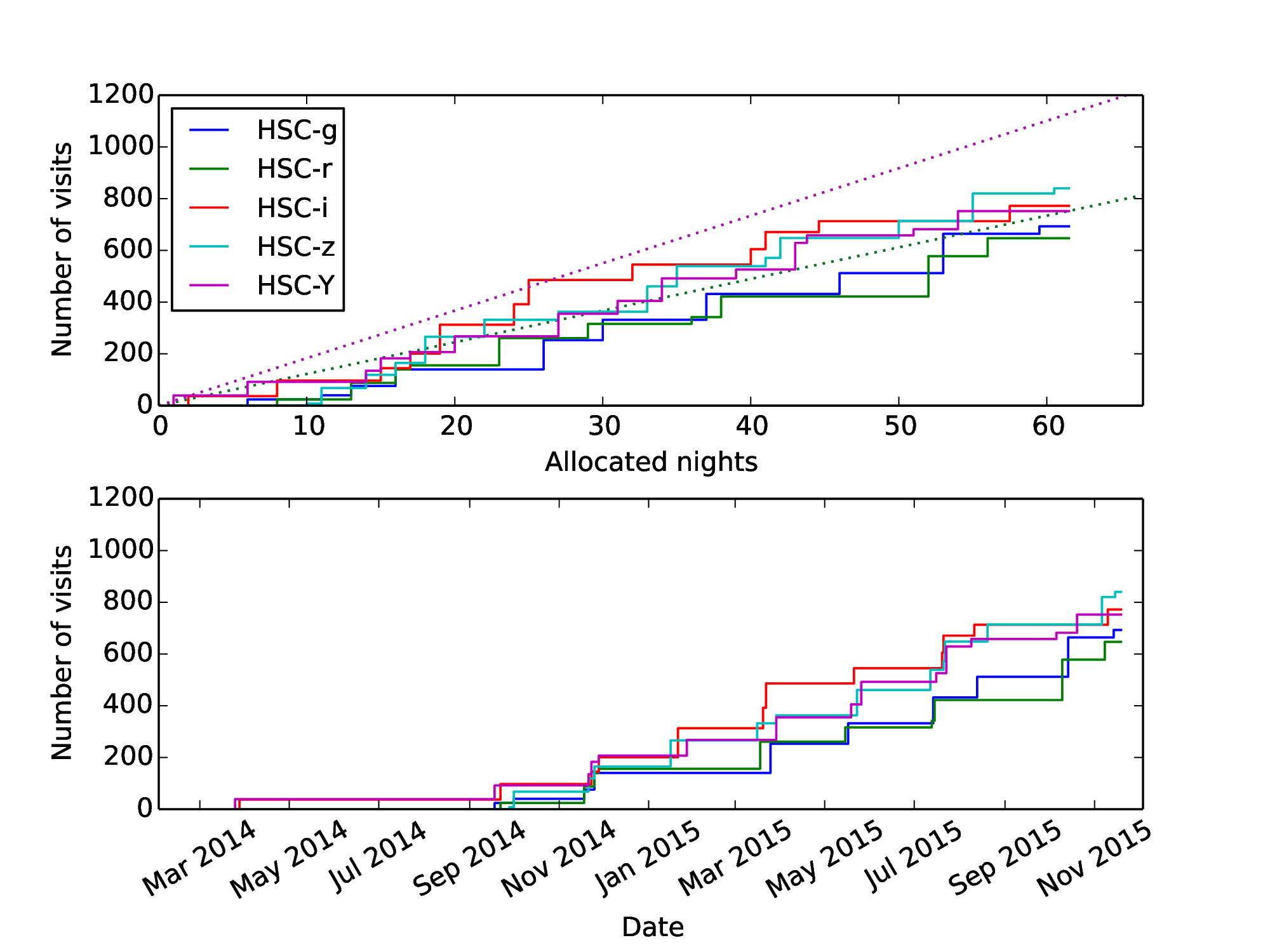} 
  \end{center}
 \caption{
   Allocated number of nights and number of visits acquired.
   The top panel shows the cumulative number of visits for the Wide layer obtained as
   a function of the number of observing nights.  The dashed lines indicate the average
   numbers of visits required to complete the survey in 300 nights in the $gr$ (bottom line) and
   $izy$ filters (top line), respectively.  The bottom panel shows the cumulative number of visits as a function of time.
 }
 \label{fig:survey_progress}
\end{figure}

Let us briefly discuss our survey progress so far.
Fig. \ref{fig:survey_progress} shows the growth of the number of visits (exposures;
we will define our terminology in the next section) as a function of observing nights
for the Wide layer.  About 2/3 of the total observing time is for the Wide survey
and thus the Wide layer most directly shows our survey progress.  Our progress is
somewhat slower than expected.  On average, we are about 10 nights behind the planned schedule.
This is primarily due to the rather poor observing conditions in the early observing runs,
and we hope the weather cooperates in our future runs.  We note that, in the early
phase of the survey, fewer nights were allocated for SSP than originally planned
because of operational reasons.  The time allocation is increasing to catch up with
the original plan.

\begin{figure}
  \begin{center}
  \includegraphics[width=8cm]{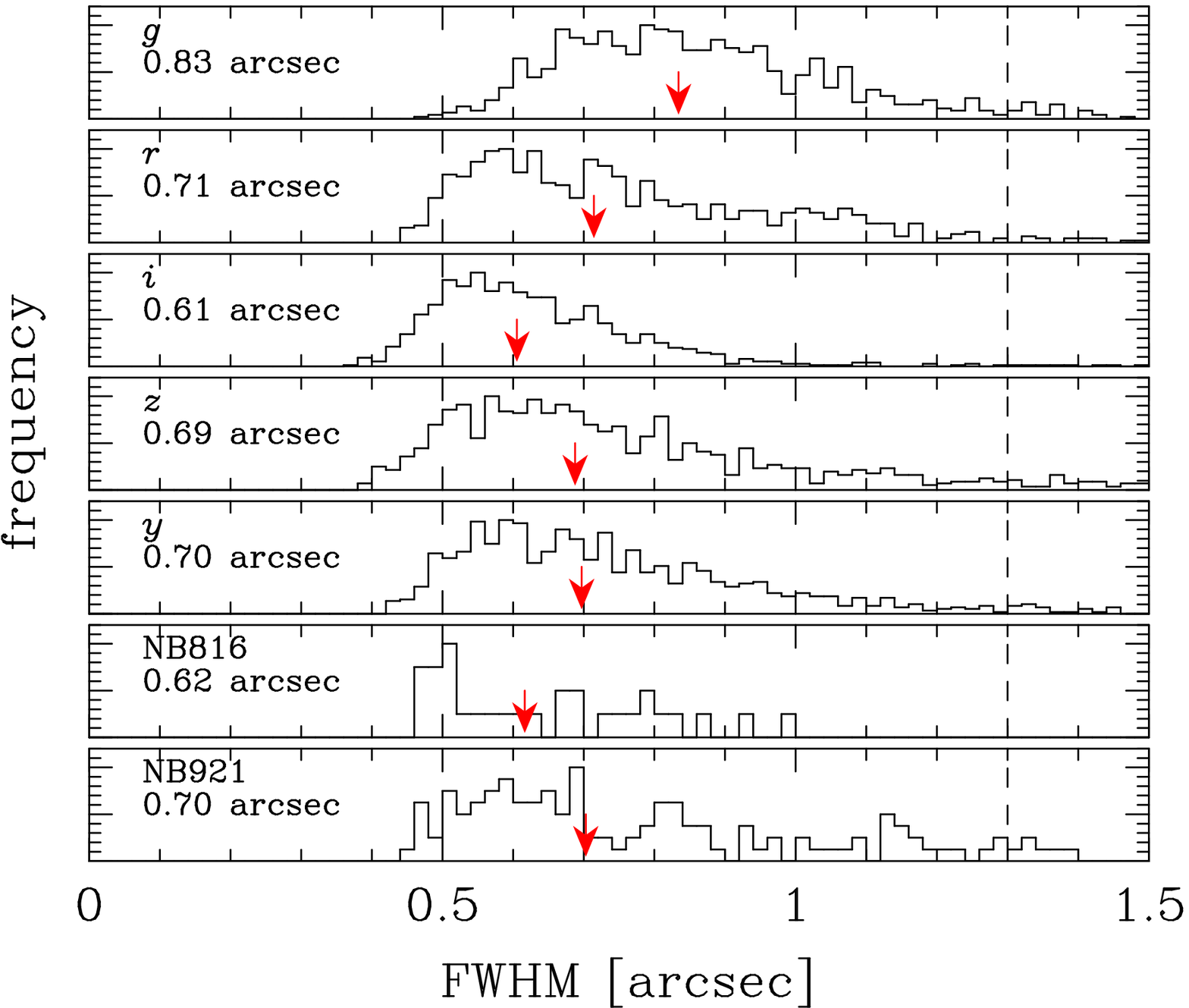} 
  \end{center}
 \caption{
   Seeing distribution of individual visits for each filter.  The seeing measured by the on-site system is used
   and only visits with sky transparency greater than 0.3 are plotted here (note that
   only data with transparency $>0.3$ are used in the main processing; see Section \ref{sec:data_processing}).
   The numbers and arrows show the median seeing.  The vertical dashed lines indicate
   the seeing threshold (1.3 arcsec) below which visits are used in the processing.  Note that seeing
   shown is as measured and is not corrected for airmass.
 }
 \label{fig:seeing_distrib}
\end{figure}

Despite the small delay, the data we have collected thus far have exquisite quality.
Fig. \ref{fig:seeing_distrib} shows the seeing distribution of the acquired data.
The seeing measured by the on-site reduction system (Furusawa et al. 2017) is used here.
A significant fraction of our data are taken under seeing conditions better than 0.7 arcsec.
The $g$-band is worse than the other bands but its median seeing is still 0.8 arcsec.
As described in the survey description paper \citep{aihara17}, we give priority to the $i$-band observations
when the seeing is good and the median seeing in the $i$-band is 0.6 arcsec.
This is superior to the typical seeing achieved in the Dark Energy Survey ($\sim$0.9 arcsec;
\cite{2016MNRAS.460.1270D}), making more precise shape measurements as well as deeper imaging possible.
The Kilo Degree Survey achieves similar seeing but is much shallower \citep{2017arXiv170302991D}.
The combination of excellent seeing and depth
is one of the strengths of our survey.
We will elaborate on the depth of our data in Section \ref{sec:survey_depth}.

\begin{figure}
  \begin{center}
  \includegraphics[width=8cm]{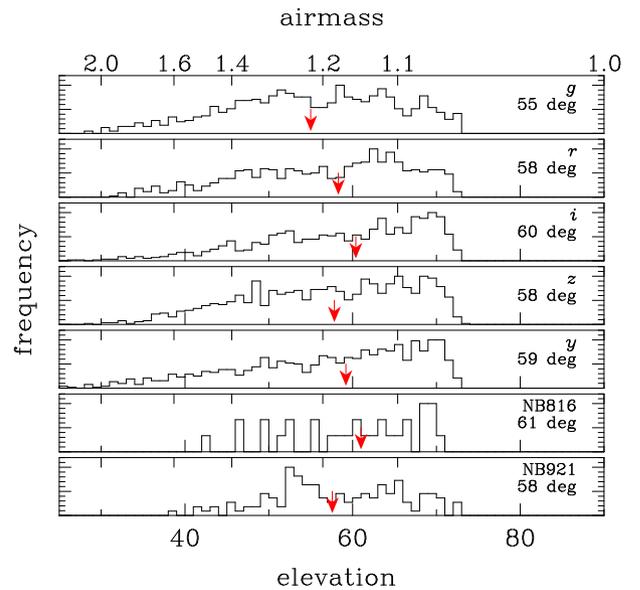} 
  \end{center}
 \caption{
   Distribution of elevation at which visits are taken for each filter.
   The top ticks show the corresponding airmass.
   The numbers and arrows show the median elevation in each band.
 }
 \label{fig:airmass_distrib}
\end{figure}

Fig. \ref{fig:airmass_distrib} shows the airmass distribution of our observations.  Most of
the visits are taken around 60 degrees (airmass$\sim1.2$), but there is a tail towards lower elevation.
Most of our fields are located around the celestial equator and they do not
go above $\sim70$ degrees elevation.

\begin{table*}[htbp]
  \begin{center}
    \begin{tabular}{l|ccccccccc}
      \hline\hline
      {\bf Wide}            &  $g$  &  $r$  &  $i$  &  $z$   &  $y$  &          &          &           & \\
      exposure (min)        &  10   &  10   &   20  &   20   &   20  &          &          &           & \\
      seeing (arcsec)       &  0.72 &  0.67 &  0.56 &  0.63  &  0.64 &          &          &           & \\
      depth (mag)           &  26.8 &  26.4 &  26.4 &  25.5  &  24.7 &          &          &           & \\
      {saturation (mag)}   & { 17.8} & { 17.8} & { 18.4} & { 17.4} & { 17.1} &          &          &           & \\ 
      \hline
      target exposure (min) &   10  &  10   &   20  &   20   &  20   &          &          &           & \\
      target depth (mag)    & 26.8  &  26.4 &  26.2 &  25.4  &  24.7 &          &          &           & \\
      \hline \hline
      {\bf Deep}            &  $g$  &  $r$  &  $i$  &  $z$   &  $y$  &  $NB387$ & $NB816$  &  $NB921$  & \\
      exposure (min)        &  20   &  15   &   30  &   35   &   20  &  ---     &  45      &  60       & \\
      seeing (arcsec)       & 0.83  &  0.68 &  0.55 &  0.69  &  0.59 &  ---     &  0.53    &  0.65     & \\
      depth (mag)           & 26.8  &  26.6 &  26.5 &  25.6  &  24.8 &  ---     &  25.9    &  25.6     & \\
      { saturation (mag)}   & { 17.9} & { 18.2} &  { 18.8} & { 17.6} & { 17.3} & ---   &  { 17.2}  & { 17.0} & \\ 
      \hline
      target exposure (min) &   84  &  84   &  126  &  210   & 126   &  84      & 168      &  252      & \\
      target depth (mag)    & 27.8  &  27.4 &  27.1 &  26.6  &  25.6 &  24.8    & 26.1     &  25.9     & \\
      \hline \hline
     {\bf  UltraDeep}       &  $g$  &  $r$  &  $i$  &  $z$   &  $y$  &          & $NB816$  &  $NB921$  & $NB101$\\
      exposure (min)        &  70   &  70   &  130  &  130   &  210  &          & 200      &  270      & ---\\
      seeing (arcsec)       & 0.74  &  0.62 &  0.64 &  0.59  &  0.74 &          & 0.60     &  0.76     & ---\\
      depth (mag)           & { 27.4}  & { 27.3} & { 27.0} & { 26.4} & { 25.6} &            & 26.3       & { 25.8}    & ---\\
      { saturation (mag)}   & { 18.3}  & { 19.0} & { 18.7} & { 18.2} & { 17.3} &            & { 17.2} & { 16.6}    & ---\\ 
      \hline
      target exposure (min) &  420  &  420  &  840  &  1134  & 1134  &          & 630      &  840      & 1050\\
      target depth (mag)    & 28.4  &  28.0 &  27.7 &  27.1  &  26.6 &          & 26.8     &  26.5     & 25.1\\
       \hline \hline
    \end{tabular}
  \end{center}
  \caption{
    Approximate exposure time, seeing, $5\sigma$ depth for point sources,
    and saturation magnitudes (also for point sources)
    for each filter
    and survey layer. For the Deep and UltraDeep layers, the numbers are for the data
    collected thus far and we expect to reach much deeper later in the survey.  The target
    exposure times and expected depths are also shown for reference.  Note that the expected
    depths are for point sources and are in reasonable agreement with
    the measured depths in the Wide layer.  The $5\sigma$ limiting mags within 2 arcsec diameter
    apertures, which may be more relevant for extended sources, are shallower by 0.3~mags than
    the point source limits.   The seeing measurements are derived from
    Gaussian-weighted moments of stars, transformed to FWHM by assuming a Gaussian profile.
    Note that there is a significant spatial
    variation of all the values listed here over the survey area.
  }
  \label{tab:exptime}
\end{table*}

\subsection{Previous internal releases}

We have made several internal data releases so far.  As the data from these internal releases
are used in our science papers, we briefly summarize them in Table \ref{tab:data_releases}.
We started with a test release (S14A0) including data from the first observing run,
followed by larger data releases for science twice a year.  Major updates in the processing
pipeline were made in each release and the data quality steadily improved.
Five internal data releases have been made to date, and the S15B release forms the basis of this public release.
We present the exact definition of the data included in this public release in Section \ref{sec:the_release_data}.

\begin{table*}[htbp]
  \begin{center}
    \begin{tabular}{l|ccccrrc}
      \hline
      Release               & Date       & Layer     & N      & \multicolumn{1}{c}{Area} & \multicolumn{1}{c}{Files}    & \multicolumn{1}{c}{N}             & Version \\
                            &            &           & filter & \multicolumn{1}{c}{(deg$^2$)}   & \multicolumn{1}{c}{(TBytes)} & \multicolumn{1}{c}{object} & hscPipe \\
      \hline \hline
      Public Data Release 1 & 2017-02-28 & UltraDeep & 7 &   4       &   8.6 &   3,225,285 & 4.0.1   \\
                            &            & Deep      & 7 &  26       &  16.6 &  15,959,257 & 4.0.1   \\
                            &            & Wide      & 5 & 108 (100) &  57.1 &  52,658,163 & 4.0.1   \\
      \hline \hline
      S14A0                 & 2014-09-04 & UltraDeep & 5 &   2       &   2.2 &     880,792 & 2.12.4a \\
                            &            & Wide      & 2 &  24       &   2.6 &  10,548,142 & 2.12.4a \\
      \hline
      S14A0b                & 2015-02-10 & UltraDeep & 5 &   4       &   6.4 &   2,183,707 & 2.12.4d \\
                            &            & Wide      & 5 &  94  (23) &  18.6 &  63,954,672 & 3.4.1   \\
      \hline
      S15A                  & 2015-09-01 & UltraDeep & 6 &   4       &   7.2 &   2,973,579 & 3.8.5   \\
                            &            & Deep      & 6 &  24       &  17.7 &  14,747,568 & 3.8.5   \\
                            &            & Wide      & 5 & 203  (82) &  40.7 &  64,073,662 & 3.8.5   \\
      \hline
      S15B                  & 2016-01-29 & UltraDeep & 7 &   4       &   8.6 &   3,225,285 & 4.0.1   \\
                            &            & Deep      & 7 &  26       &  16.6 &  15,959,257 & 4.0.1   \\
                            &            & Wide      & 5 & 413 (111) & 145.2 & 157,423,778 & 4.0.1   \\
      \hline
      S16A                  & 2016-08-04 & UltraDeep & 7 &   4       &   7.5 &   3,208,918 & 4.0.2   \\
                            &            & Deep      & 7 &  28       &   8.0 &  16,269,129 & 4.0.2   \\
                            &            & Wide      & 5 & 456 (178) & 245.0 & 183,391,488 & 4.0.2   \\
      \hline 
      \hline   
    \end{tabular}
  \end{center}
  \caption{
    Summary of this public release and previous internal data releases. 
    The area is estimated by using the HEALPix index system ($N_{side}=2^{11}$) and
    mosaicking information from the pipeline processing.
    The 5th column gives the survey area in square degrees.
    The full-color full-depth area in the Wide survey is shown in parenthee.
    Only the full-color full-depth Wide area is included in this release, but the area
    in the brackets in the top row is smaller than the total area.  This is primarily because
    the release area is determined on a patch by patch basis, but a fraction of the area
    in the patches on the field borders actually do not reach the full depth.
    The 7th column shows the number of objects.  Since the deblender became functional in
    the S15A release, the numbers for the subsequent releases are for primary objects (\texttt{detect\_is\_primary}=True;
    see Section \ref{sec:selecting_objects_with_clean_photometry}).
  }
  \label{tab:data_releases} 
\end{table*}

\section{Data Processing}
\label{sec:data_processing}

\subsection{Terminology}
\label{sec:terminology}

In order to describe our data, we introduce a series of HSC/LSST-specific terms.
They are also defined in Bosch et al. (2017) in detail.
These terms are defined based on the terminology in the Subaru Telescope's
data archive system (Subaru Telescope ARchive System; STARS\footnote{\url{https://stars.naoj.org/}}) as well as data
handling needs in our data analysis.

\texttt{visit} is an integer number uniquely assigned to each exposure, i.e.,
a single shot of an image with HSC. A \texttt{visit} comprises 112 CCD images and
is always an even number incremented by 2 every exposure.
\texttt{ccd} is also an integer number between 0 and 111 to refer to the individual
CCDs of a \texttt{visit}, which is equivalent to the FITS header keyword \texttt{DET-ID}.
The pair of numbers, \texttt{visit} and \texttt{ccd}, is used to identify the CCD data of a given visit.
In STARS, each raw CCD data is stored as a separate FITS image file, each of
which is assigned a unique id called \texttt{FRAMEID}.
CCD data can be also identified by \texttt{FRAMEID} and there is a one-to-one mapping
between \texttt{FRAMEID} and a pair of \texttt{visit} and \texttt{ccd}.

The data are processed in several separate stages -- single-visit
processing followed by several multi-visit processing stages.  The single-visit processing is done  
for each visit separately.  The multi-visit processing which follows is
performed on a group of multiple visits, generating combined \texttt{coadd}
products as well as source catalogs measured from the coadds.
In the latter stage, the data sets are processed separately in 
equi-area rectangular regions on the sky.  The regions, called
\texttt{tracts}, are pre-defined as an iso-latitude tessellation, where 
each \texttt{tract} covers approximately $1.7\times 1.7$ square degrees of the sky.
Neighboring tracts have a small overlap, $\sim1$ arcmin around the equatorial fields.
A \texttt{tract} is further divided into $9\times 9$ sub-areas, each of which is
4200 pixels on a side (approximately 12 arcmin) and is called a \texttt{patch}.
Adjacent \texttt{patches} have an overlap of 100 pixels on their edges.
These \texttt{tracts} and \texttt{patches} are the two major areal units
introduced to parallelize the processing.

In the final steps of the coadd analysis, we detect sources and
measure their properties on the coadd image. We first measure sources in each band separately,
and then combine these measurements to perform consistent photometry across
the bands.  We refer to the first measurement as \texttt{unforced} measurement and
the latter as \texttt{forced}.

We will describe each procedure in detail in the next sections, but we use this
terminology defined here throughout the paper as well as many of our science papers.

\subsection{Data screening}

Data sets for the processing are selected on the basis of the results from
the data evaluation by the on-site quality assurance (QA) system (Furusawa et al. 2017),
which is located at the Hilo facility of the Subaru Telescope.  The onsite QA system
records data quality information such as seeing and sky transparency, as well as
observers' notes in a dedicated database.
The first step in the data screening is to select visits taken with exposure time
of 30.0 sec or longer.  The visit list is further screened to include data with
decent quality by applying the following conditions:
(1) background count $\leq 45000$ ADUs (a constraint on sky brightness, but few visits are
removed by this cut),
(2) seeing FWHM $\leq 1.3$ arcsec, and
(3) sky transparency $\geq 0.3$.
We further filter the visit list by
carefully reviewing the observers' notes to generate the final visit list for processing.
About 90\% of the visits pass all the screening.
This does not mean that the weather is good 90\% of the time, but it simply
means that we could obtain good data for 90\% of the time we could observe on the sky.
Data collected through November 2015 are included in the processing.
During the validation phase, we discovered a PSF modeling problem in VVDS
and we have further removed a small fraction of visits from that field (Section \ref{sec:VVDS}).

\subsection{Single-visit processing -- Detrending and Calibration}
\label{sec:single_visit_processing}

In the following subsections, we briefly describe how we process the data.
We refer the reader to the pipeline paper (Bosch et al. 2017) for algorithmic details.
Fig. \ref{fig:processing_flow} summarizes the flow of the processing as well as
data products generated in each processing step.

The single-visit processing is a procedure to correct cosmetics of the
CCD data and homogenize and linearize counts, and to determine photometric
zeropoints and astrometric solution per CCD.  Each processed CCD image is
stored in a separate file.

The single visit processing starts with detrending -- overscan subtraction, two-dimensional bias and dark
subtraction, and flatfielding.  We use the dome flat for flatfielding as
it is a stable flat source.  It does not give a uniform illumination across
the field of view, but it will be corrected for in the joint calibration
process described below.  Fringes are subtracted in the $y$ and NB921
bands because they are evident only in these bands.  Variance and mask images
are generated from a science image and are processed as with the science image.
Dedicated mask values are used to indicate the known bad pixels, detected
cosmic rays, crosstalk, saturated pixels, and etc, are all defined in the FITS header. 
After the bias subtraction, the linearity is corrected for using a set of
predefined linear coefficients for each CCD based on the laboratory measurements.
In addition, the brighter-fatter effect, whereby brighter stars have a broader PSF
due to detector physics \citep{2014JInst...9C3048A}, is also corrected (see \cite{bosch17} for details).
We measure the sky in grids of 128 pixels on a side and fit a 2D Chebyshev polynomial to model
the sky background taking into account the inverse variance in each grid.
The sky model is then subtracted from the original image.

We characterize the point spread function (PSF) by using a customized
version of PSFEx \citep{2011ASPC..442..435B}.  For each
CCD, we fit a pixelized image as a function of position to selected stellar candidates,
in order to reproduce the PSF at any given position. 
On average, we select $\sim$70 
candidate stars per CCD with a typical $S/N\sim100-200$, which roughly corresponds to
20-22 magnitudes (exact numbers depend on filter and observing conditions).
Based on this PSF information, cosmic rays are detected and interpolated by
the surrounding pixels.  
Sources are detected by applying the maximum likelihood technique and
their pixel coordinates and fluxes are measured on each CCD.  Aperture
corrections, which are required to account for fluxes outside of
the \texttt{sinc} aperture used in the zero-point determination (see below),
as a function of coordinates are also estimated in this step.
The typical aperture correction is a few per cent.

Photometric zero-points are determined on a CCD-by-CCD basis by comparing
\texttt{sinc} fluxes (12 pixel radius aperture; \cite{2013MNRAS.431.1275B}) of bright point sources and
their fluxes from the Pan-STARRS 1 (PS1) 3$\pi$ catalog (see Section~\ref{sec:refcatalog}). 
We apply color-terms to translate the zero-points from PS1 into
the native HSC system (Kawanomoto et al. 2017).
Some of our data are taken under non-photometric conditions, but effects
of clouds are largely removed by calibrating against the external catalog.
Astrometry is calibrated against PS1 as well and the WCS (TAN-SIP) is
fitted across the entire focal plane (i.e., 104 CCDs) with 9th-order
non-linear terms.
We do not warp the images in the single-visit processing.
The WCS includes the correction for the optical distortion.

\subsection{Multi-visit processing -- mosaic, joint calibration, and coadding}
\label{sec:multi_visit_processing}

The multi-visit processing stage coadds the detrended CCD images from
multiple visits into a deeper stack to achieve a higher S/N.
The first step, mosaicking, solves for relative positions and flux
scales of each CCD. This is done on a tract-by-tract basis.
A matched list of reference sources (bright stars) from each CCD in a given
tract is first generated and we solve for a set of spatially-varying 
terms and per-CCD scaling to minimize the difference in the coordinates and fluxes of overlapping sources
on different CCDs/visits by the least-squares method.  This procedure is similar to
\"{u}ber-calibration in SDSS \citep{2008ApJ...674.1217P}.
%
%
This process corrects for the systematic flux error
introduced by the dome flats as well as zero-point and astrometric errors
in the individual CCDs.
The resultant internal photometry and astrometry show a smaller scatter by $\sim10$\%.

Utilizing the improved photometry and astrometry, the individual CCD images are warped
onto patches.  For each patch, a coadded image is created weighted by the inverse of the mean
variance for each input image (i.e., all the pixels in an input image have the same weight).
We do not apply a global sigma clipping algorithm as that will adversely impact the shape
of the PSF on the coadd.  Instead, we identify regions in individual visits that are significantly
different from other visits.  The pixels from these regions are then clipped from the coadd.
Refer to Section 3.3.2 of \citet{bosch17} for more details.  While the coadd PSF
for objects from these regions is not correct, such objects are flagged and can be
ignored for scientific analyses.  This process will also reduce the occurrence of transient objects such as
satellite trails, ghosts, and cosmic rays.



\subsection{Multi-band measurements}

\label{sec:multi_band_measurement}


We then move on to detect and measure sources on the coadds.
In order to measure photometry consistently across the bands, we follow
the following steps.  Firstly, we detect sources on the coadds for each band
separately using the same algorithm as SDSS (i.e., maximum likelihood detection).
In short, we convolve the science image with the PSF and search for above-threshold
pixels ($5\sigma$).  Our detection is thus optimized for point sources.
See Section 4.7 of \citep{bosch17} for further details.
The lists of the detected objects in a given patch are merged
into a master detection catalog.  This catalog contains positions and
pixel coverages (\texttt{footprints}) of all the detected sources for detailed
measurements.

We then perform \texttt{unforced} measurements of coordinates, fluxes,
and shapes of each of the sources listed in the merged detection catalog.
Objects are deblended to child objects when needed and measurements are also performed
on the children.
In this step, the centroids and object shapes are allowed to vary from
band to band.  This is why we call this unforced.
As the centroids and shapes are different from band to band, the measurement
does not give good colors of objects, but unforced CModel\footnote{
  Composite model photometry.  It fits a linear combination of
  exponential profile and de Vaucouleurs profile convolved with PSF to objects \citep{SdssPhoto,2004AJ....128..502A,bosch17}.
}
and Kron
fluxes are likely a better proxy for total fluxes in each band than the forced measurements
described below.
From the unforced measurements, we choose one
reference band for each object.  We refer to Bosch et al. (2017) for the detailed algorithm to choose
the reference filter, but the $i$-band is the reference
band for most objects.


Finally, we perform forced measurements.  In this last step, objects'
centroids and shapes from the reference band are applied to all the other bands.
Thus, we perform photometry consistently across the bands.  However, we apply
no smoothing to equalize the PSF across the bands in our processing and fixed aperture photometry does not deliver
consistent colors of objects.  Measurements that explicitly incorporate the PSF
in each band such as PSF flux and CModel flux should be used
for colors.  The multi-band catalogs from the forced measurements
should be the most useful catalogs for a wide range of scientific applications.


\subsection{Afterburner}
\label{sec:afterburner}

The above procedures are the main processing steps, but we have performed additional
processing described in this and the following subsections.

The deblender tends to fail in very crowded areas such as cores of galaxy clusters.
The failure results in poor photometry, causing cluster finders
to miss clusters or misidentify the brightest cluster galaxies \citep{2017arXiv170100818O}.  To mitigate the problem,
we apply Gaussian smoothing to a set of three target FWHMs, 0.6, 0.85, and 1.1 arcsec,
to perform PSF-matched aperture photometry for each FWHM
(this is only an approximate PSF equalization because the true PSF is not Gaussian).
When the native seeing is worse than the target seeing,
we do not make an attempt to deconvolve.  We instead give a flag to indicate
a measurement failure.  The photometry is done on both the parent and child images at the positions of parent and all children,
but the measurements on the parent are most useful to mitigate the deblender problem.
As shown in Section \ref{sec:deblending_failure_in_crowded_areas}, the PSF-matched photometry delivers
better colors than CModel
in crowded fields.
This seems to be the case for isolated objects as well because photometric redshifts
using the PSF-matched fluxes are better than those using the CModel fluxes \citep{tanaka17}.
But, the PSF-matched photometry does not necessarily give better total fluxes.

In addition to the PSF-matched photometry, the junk suppression algorithm, which was
mistakenly disabled in the main processing as described in Section \ref{sec:disabled_junk_suppression},
is turned on and objects that should have been eliminated are flagged.

These processes were run as an afterburner and the photometry and flags
are stored in a separate database table termed \texttt{afterburner}, which can be joined
with other tables by object IDs.  Note that the PSF-matched photometry will be
a part of the main processing in our future runs and will not be stored in a separate table.

\subsection{VVDS reprocessing}
\label{sec:VVDS}

Some of the visits in VVDS have excellent seeing -- better than 0.4 arcsec.
However, the exquisite seeing unfortunately caused problems in the PSF modeling
possibly because the PSF is undersampled (recall that the pixel scale is 0.168 arcsec).
We have not fully understood yet the root cause of the problem (see \cite{mandelbaum17} for more discussion),
but as a temporary solution, we have excluded these visits (20 in total)
with too good seeing and reprocessed the VVDS field.  A significant area (roughly 5 square degrees) in VVDS is affected
by this problem in the $i$-band, but it is also seen in the $z$-band over a much smaller area (about 20 patches).
As the affected $z$-band area is not large, we chose to reprocess only the $i$-band.
The affected $z$-band photometry should not be used and we provide a database
table to identify the problematic patches (see Section \ref{sec:catalog_and_data_archives}).
Note that only the reprocessed data are available in the database.  We anticipate that further improvements in the pipeline
will allow us to use the currently-excluded data in the future.

\subsection{COSMOS Wide-depth stacks}

There is a wealth of deep multi-wavelength data in the COSMOS field \citep{2007ApJS..172....1S}
and the field can be used for various tests and calibrations.
In order to perform photo-$z$ calibrations at the depth of the Wide survey, we have stacked
a subsample of the COSMOS data to a depth approximately similar to the Wide layer\footnote{
  The exposure is not exactly the same because the individual exposure
  times are different between the Wide and UltraDeep layers.
}.
The large number of visits taken under various observing conditions allows us
to generate Wide-depth stacks for a range of seeing sizes.  We have generated best, median,
and worst seeing stacks with FWHMs of roughly 0.5, 0.7, and 1.0 arcsec in all the bands.
Two to four visits are included in the processing depending on the band.
The multi-band processing is then run to generate photometric catalogs.
These Wide-depth stacks are stored in separate database tables from the main tables.

\subsection{The release data}
\label{sec:the_release_data}

As our survey is still in progress, our current data are far from uniform
in terms of both depth and the number of filters observed (i.e., not all the area is
covered in all the five filters), especially in the Wide layer.
We choose to release the full-color full-depth Wide area to the community to ensure the data uniformity.
We define the full-color full-depth area,
using \texttt{countInputs}, which is a number of visits contributed to a patch,
and require the mean \texttt{countInputs} in a patch to be larger than 5/6 of
the nominal number of visits
in each of the 5 filters
(4,4,6,6,6 visits in $grizy$)\footnote{
  A slightly different
  definition of the full-color full-depth area may be found in our science papers,
  but it is driven by scientific needs and its definition is explicitly spelled out
  in each paper.
}.  As mentioned earlier, many of
our first science papers are based on the full-color full-depth data and thus
this is also important from the point of view of reproducibility of our science results.
For Deep and UltraDeep, the full area has already been observed.  The current depths
are much shallower than those we expect to reach at the end of the survey, but
the data are already very useful for scientific exploration.  For this reason,
we release all of the Deep and UltraDeep data to
the community (exactly the same data as the S15B internal data release).

\section{Data Products}
\label{sec:data_products}



\begin{figure}
 \begin{center}
  \includegraphics[width=8cm]{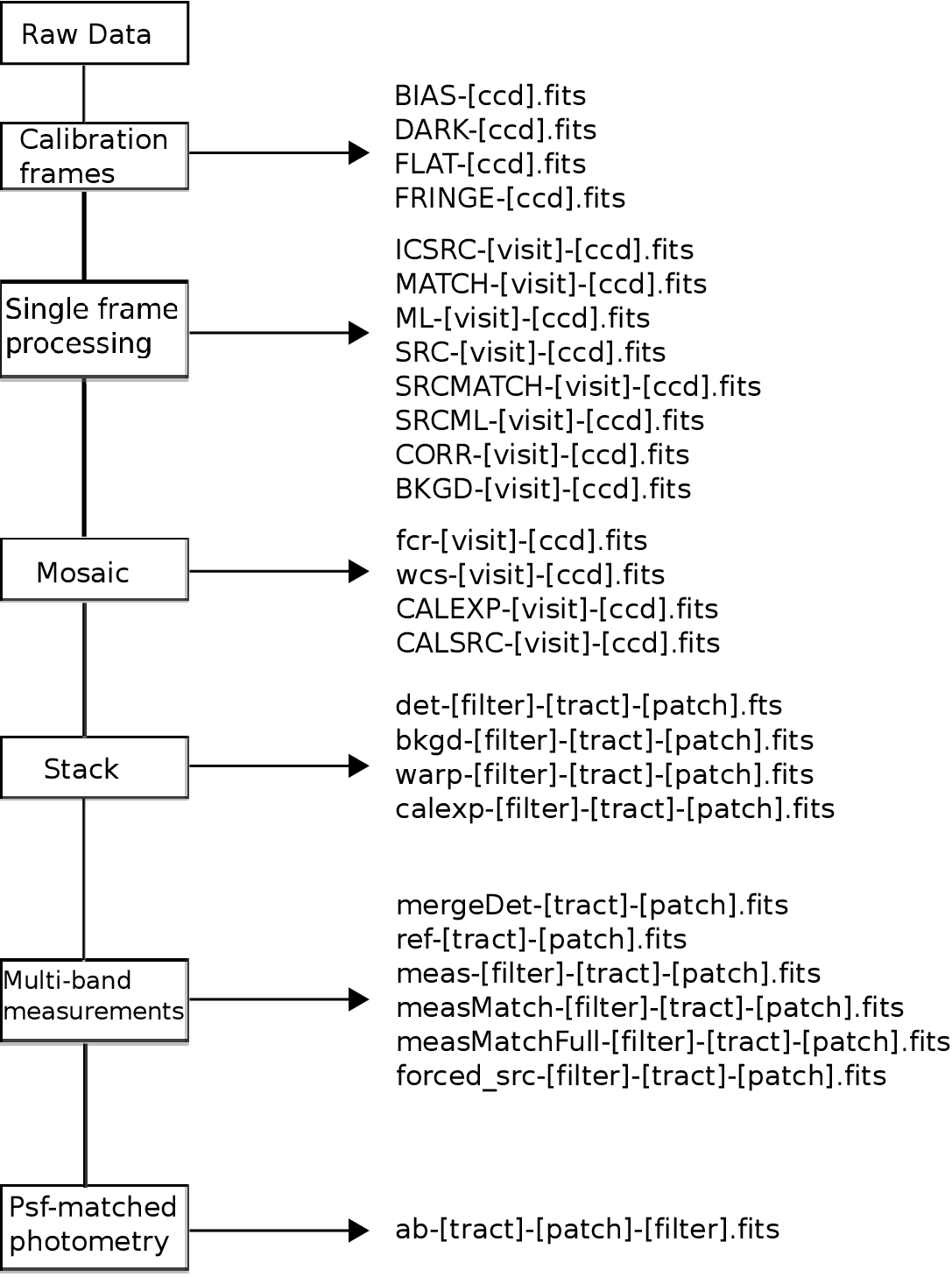} 
 \end{center}
 \caption{
   Schematic view of the flow of the processing and data products generated   at each stage.
 }
 \label{fig:processing_flow}
\end{figure}

This section describes the data products generated in the processing 
detailed in the previous section.  Understanding our data products requires knowing about
algorithmic details in the processing and we once again refer the reader
to Bosch et al. (2017).  We first focus on image products and then turn our
attention to catalog products, as images and catalogs can be retrieved from our data release
site in different ways (i.e.,  image file access vs. database query).
Fig. \ref{fig:processing_flow} summarizes the data products generated at each processing
stage.  The figure gives a nearly complete list of the products, but not all of them
are important for scientific use (e.g., some are used for data validation), 
and we focus on the most important ones here.  The data release site describes all the products.


\subsection{Image data}


The processed images, both individual CCD images and coadds, are stored in the standard FITS format with three image
layers and multiple binary tables. A basic FITS header representing the 
characteristics of data and the processing record is placed in the primary HDU. 
The science image, mask image, and variance image are 
stored in the next three HDUs. The definition of the mask bits
can be found in the header.  Photometric zero-points are also given in the header,
although they are not fully meaningful without aperture corrections.
The binary tables contain information about, e.g., PSF models and aperture corrections,
and users normally do not need to read these tables directly --- they can be most
easily read using the pipeline functions.

Detrended individual CCD images with photometric and astrometric calibrations applied are called
\texttt{CORR} images.  The photometry and astrometry are updated in
the joint calibration step (\texttt{mosaic} in Fig. \ref{fig:processing_flow}) and
stored as \texttt{CALEXP} images, which should be used
for analysis on individual CCDs requiring decent accuracy of flux and/or coordinates of objects, 
such as identifying moving objects, assessing light curves of variables, and so on.
These CCD images are then warped to patches (\texttt{warp}) and coadded (\texttt{calexp}).
The latter is often referred to as a patch image.  The coadd images
have a homogenized photometric zero-point of 27.0 mag/ADU.
Objects are detected and the unforced and forced measurements are both performed on
the \texttt{calexp} images.  \texttt{calexp} has an overlap of 100 pix ($\sim17$~arcsec)
on each side with adjacent patches.



\subsection{Catalog data}

A number of catalogs are also generated during the processing.
All the catalogs are FITS binary tables and the column names are in many cases
self-explanatory.  In addition, the FITS header gives brief explanations of the tabulated quantities.
More detailed descriptions may be found in the online documentation at the data release site.
Although some of the FITS catalog tables are withheld from this release,
we describe them below for the sake of completeness.  We will make them available
along with the shape catalog in our future incremental release.

Source catalogs (\texttt{SRC}) have detailed measurements of detected objects in
each CCD, and \texttt{meas} and \texttt{forced\_src} have unforced and forced
measurements on the coadds, respectively.  The latter two catalogs as well as
closely related catalogs are loaded to the database, which offers an easy
way to retrieve the measurement results.  Measurements include extensive flag bits,
which indicate the reliability of the measurements.  Measurement flags should be
applied for scientific use of our data.
We summarize some of the most frequently used flags in Table \ref{tab:important_flags}.

\subsection{Selecting objects with clean photometry}
\label{sec:selecting_objects_with_clean_photometry}

As noted above, patches and tracts overlap each other, and objects in the overlapping
regions are detected and measured multiple times.  They are all in the database.
In order to eliminate duplicates, users should apply the flags \texttt{detect\_is\_patch\_inner}
and \texttt{detect\_is\_tract\_inner}, which select unique objects.  Also, measurements are performed for both
parent (i.e., before deblending) and children (i.e., after deblending).  In order to
select objects after deblending, one needs to impose \texttt{deblend\_nchild}=0.
In practice, one can use \texttt{detect\_is\_primary}, which does all the above; it selects
objects in the inner tract and patch and without any children.  Also, one useful
parameter for deblending is \texttt{blendedness\_abs\_flux}, which shows the ratio of the
flux of the child objects to the total flux, indicating how strongly
objects are blended (see Murata et al. in prep. for details).

In order to have a set of objects with clean photometry, one is advised to apply
further flags.  The pixel flags in Table \ref{tab:important_flags} are among the most important
ones.  The \texttt{saturated} and \texttt{interpolated} flags come in two variants;
\texttt{any}, meaning that at least one pixel in the object footprint is saturated
or interpolated over (typically because of a cosmic ray or bad pixel column),
and \texttt{center}, meaning that the pixel in question lies within the central
0.5 arcsec (3 pixels) of the center.
The latter can
be used in most cases because the interpolation outside of the central region
should be reasonable, and if an object is saturated in the outer parts, it should
be saturated in the center as well.
There are other pixel flags such as \texttt{pixel\_bad} and \texttt{pixel\_cr\_center},
but due to improper flag propagation (see Section \ref{sec:known_problems}) in the coadds,
they are not very effective.  Most of the objects that should have these flags set can
be identified with the interpolation flag.
In addition, many of the measurement algorithms
require object centroids in the first place and it is a good practice to ensure
good centroids with \texttt{centroid\_sdss\_flags}.
Finally, flux measurement
flags such as \texttt{flux\_psf\_flags} should also be applied to ensure clean
photometry.  Note that each photometry technique has its own flags and the flags
are given for each band separately.  Flags should be applied to all
the filters of interest.

\begin{table*}[htbp]
  \begin{center}
    \begin{tabular}{l|l}
      \hline
      flag / parameter                            &  description \\
      \hline \hline
      \texttt{centroid\_sdss\_flags}              & Object centroiding failed\\
      \texttt{flags\_pixel\_interpolated\_center} & Any of the central 3x3 pixels of an object is interpolated\\
      \texttt{flags\_pixel\_interpolated\_any}    & Any of the pixels in an object's footprint is interpolated\\
      \texttt{flags\_pixel\_saturated\_center}    & Any of the central 3x3 pixels of an object is saturated\\
      \texttt{flags\_pixel\_saturated\_any}       & Any of the pixels in an object's footprint is saturated\\
      \texttt{detect\_is\_patch\_inner}           & Object is in an inner region of a patch\\
      \texttt{detect\_is\_tract\_inner}           & Object is in an inner region of a tract\\
      \texttt{detect\_is\_primary}                & Object is a primary object, meaning that it does not have any children and is\\
                                                  & in inner tract and patch\\
      \texttt{deblend\_nchild}                    & Number of children.  0 if object is not deblended.\\
      \texttt{blendedness\_abs\_flux}             & Measure of how strongly object is blended defined as $1-flux(child)/flux(total)$.\\
      \texttt{flux\_psf\_flags}                   & PSF flux measurement failed\\
      \texttt{flux\_kron\_flags}                  & Kron flux measurement failed\\
      \texttt{flux\_cmodel\_flags}                & CModel flux measurement failed\\
      \hline
    \end{tabular}
  \end{center}
  \caption{
    Some of the most important flags and parameters stored in the database.  The flags for fluxes
    are given for each filter.  There are also filter-independent flags such as
    \texttt{detect\_is\_primary}.
  }
  \label{tab:important_flags}
\end{table*}











\subsection{Value-added Catalogs}

In addition to the pipeline products described above, value-added
products such as photometric redshifts (photo-$z$'s) are available in separate
database tables.  We briefly describe them here.

\subsubsection{Photometric Redshifts}

The HSC photo-$z$ working group has computed photo-$z$'s for this public release.
Catalog products such as photo-$z$ point estimates and confidence intervals are
available from the database and full probability distributions are available in
the FITS format for download from the data release site.  Due to a technical
problem during the photo-z production phase, we are unable to release
photo-$z$'s for the Wide area in the Data Release 1, but
they are included in the first incremental data release happened in June 2017.
Refer to \citet{tanaka17} for details.

\subsubsection{Public spectroscopic redshifts}

Partly for the purpose of the photo-$z$ calibrations,  we have collected public
spectroscopic redshifts from the literature:
zCOSMOS DR3 \citep{2009ApJS..184..218L}, UDSz \citep{2013MNRAS.433..194B,2013MNRAS.428.1088M},
3D-HST \citep{2014ApJS..214...24S,2016ApJS..225...27M}, FMOS-COSMOS \citep{2015ApJS..220...12S},
VVDS \citep{2013A&A...559A..14L}, VIPERS PDR1 \citep{2014A&A...562A..23G},
SDSS DR12 \citep{2015ApJS..219...12A}, GAMA DR2 \citep{2015MNRAS.452.2087L},
WiggleZ DR1 \citep{2010MNRAS.401.1429D}, DEEP2 DR4 \citep{2003SPIE.4834..161D,2013ApJS..208....5N},
and PRIMUS DR1 \citep{2011ApJ...741....8C,2013ApJ...767..118C}.
These redshifts as well as confidence flags are stored in a database table and matched
with the HSC objects by position.  Each survey has its own flagging scheme to
indicate the redshift confidence and we have a homogenized flag for each object for
easy selection of objects with reliable redshifts.
The online documentation 
gives the details.  {\it It is important to emphasize that users should acknowledge the original
data source(s) when using this table.}
It is straightforward to identify which survey observed a given object; we have a set of
database flags to indicate that.





\section{Data Quality}
\label{sec:data_quality}

\begin{figure*}
 \begin{center}
  \includegraphics[width=18cm]{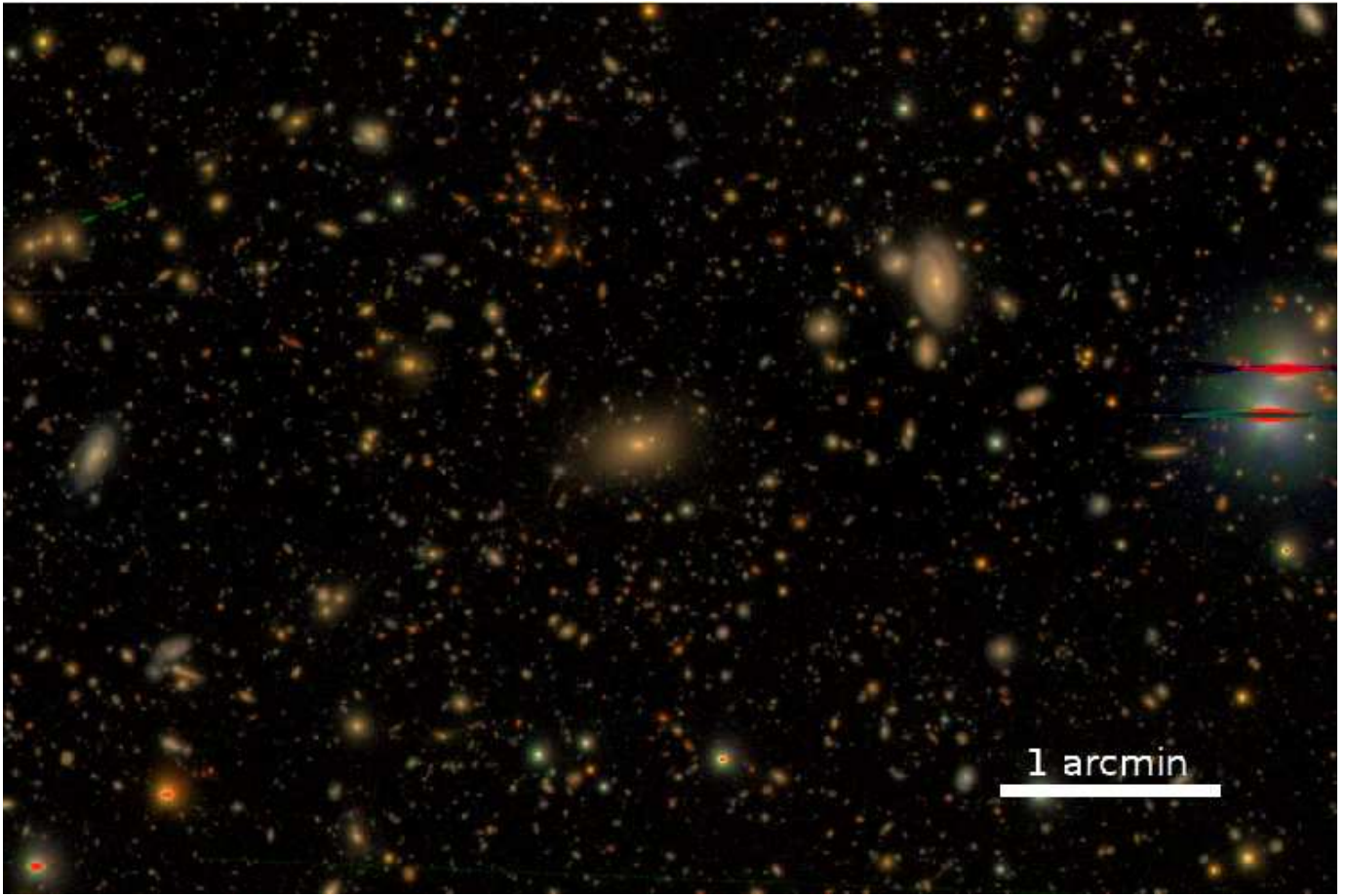} 
 \end{center}
 \caption{
   Blow-up of the COSMOS UltraDeep area in $riz$ using the color scheme of \citet{2004PASP..116..133L}
   centered at R.A.=$10^h01^m48^s.0$ and Dec.=$+02^\circ03'04''$.
   This is a 6 arcmin $\times$ 4 arcmin area, which is roughly 0.3\% of the total area of COSMOS.
   North is up and East is to the left.  The image reaches $i\sim27.5$ at $5\sigma$ for point sources.
 }
 \label{fig:cosmos_ud}
\end{figure*}

We now discuss the quality of our data.  We have performed a number of validation
tests and here we present some of the key results to illustrate our data quality.
For the sake of simplicity, we show only a few plots for each test, but more plots can be
found online.  
First of all, we demonstrate the quality of our data with the UltraDeep COSMOS image
in Fig. \ref{fig:cosmos_ud}.  As shown later in the section, this image reaches
$\sim26-27.5$~mag with seeing FWHM between 0.6 and 0.9 arcsec in the five broad bands.
The image shows a tiny fraction of the whole COSMOS field and we detect
as many as $\sim1.7\times10^6$ objects over the entire COSMOS area, allowing us
to peer deep into the distant universe.  This is a powerful dataset when combined with
a wealth of ancillary data available in this field.  We note that 
deeper COSMOS data with the combined HSC-SSP and University of Hawaii data
are made available in our first
incremental release (see \cite{2017arXiv170600566T} and Section \ref{sec:future_releaes}).

\subsection{Reference catalog}
\label{sec:refcatalog}

The HSC astrometry and photometry are calibrated relative to the Pan-STARRS 1 (PS1) 3$\pi$ catalog \citep{2013ApJS..205...20M}.
We chose this catalog because it covers all of our survey regions to a reasonable depth (allowing for a few magnitudes of overlap, from saturation of HSC to the detection limit of PS1) with a similar set of bandpasses ($grizy$; in particular, the availability of the $y$-band greatly simplifies the photometric calibration; \cite{2012ApJ...750...99T}).
This data release is calibrated against PS1 Processing Version 2 (PV2) data, which were made available to the PS1 Science Consortium members in December 2014.
Internal comparison \citep{2012ApJ...756..158S} and comparison with the Sloan Digital Sky Survey \citep{2016ApJ...822...66F} has shown that the PS1 photometric calibration is accurate to approximately 1\% in all bands.
The PS1 PV2 astrometry is itself referenced to 2MASS, and the failure to correct for proper motions has left zonal errors \citep{2013Icar..223..625T} up to 100~mas; our HSC calibration will inherit this.
Note that the recent public release of PS1 data \citep{2016arXiv161205560C} is from Processing Version 3 (PV3), which has had different astrometric \citep{2016AJ....152...53B} and photometric calibrations applied, and has generally superior quality; we plan to adopt PV3 for future data releases.

In the following sections, we make several internal and external comparisons.  We use stars brighter than $20$th~mag when comparing against external catalogs and those brighter than $21.5$~mag for internal comparisons.

\subsection{Astrometry}

Astrometric calibration is performed against PS1~PV2 (Section \ref{sec:refcatalog}) in two stages.
First, we derive an approximate astrometric solution for each individual CCD (section \ref{sec:single_visit_processing}).
This allows us to match sources between visits, which we use to derive a consistent astrometric solution for multiple overlapping exposures (section \ref{sec:multi_visit_processing}).
This solution is typically accurate to $<20$~mas; this is our internal accuracy.

\begin{figure*}
 \begin{center}
   \includegraphics[width=12cm]{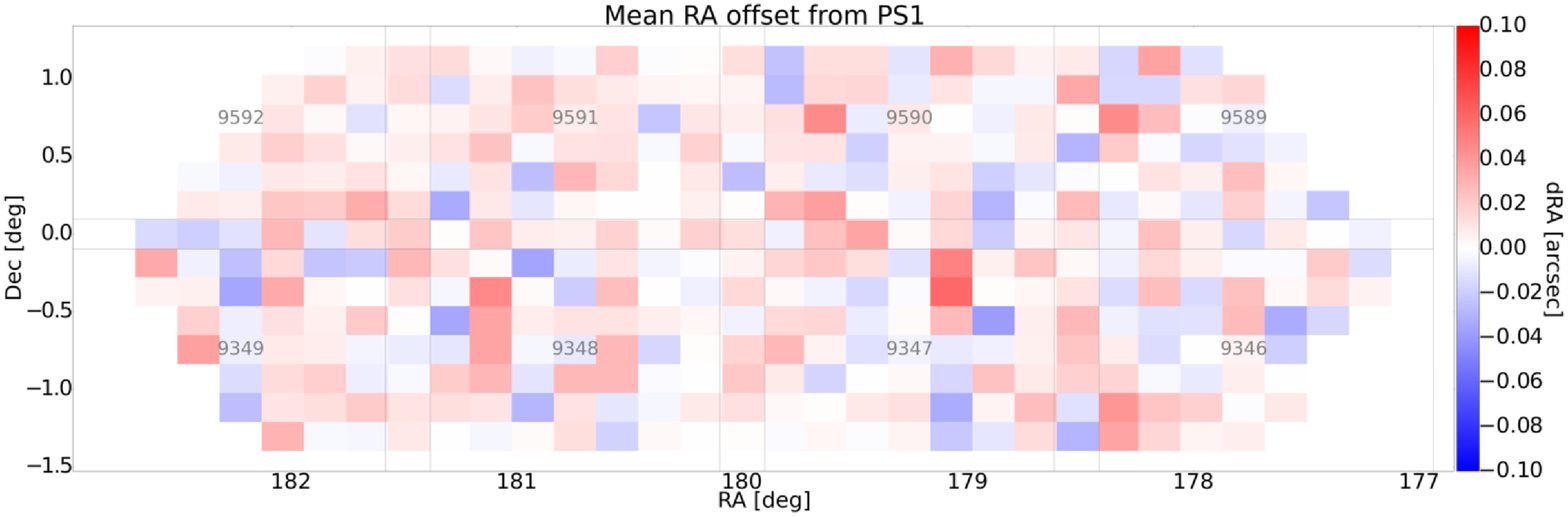}\\\vspace{0.5cm}
   \includegraphics[width=12cm]{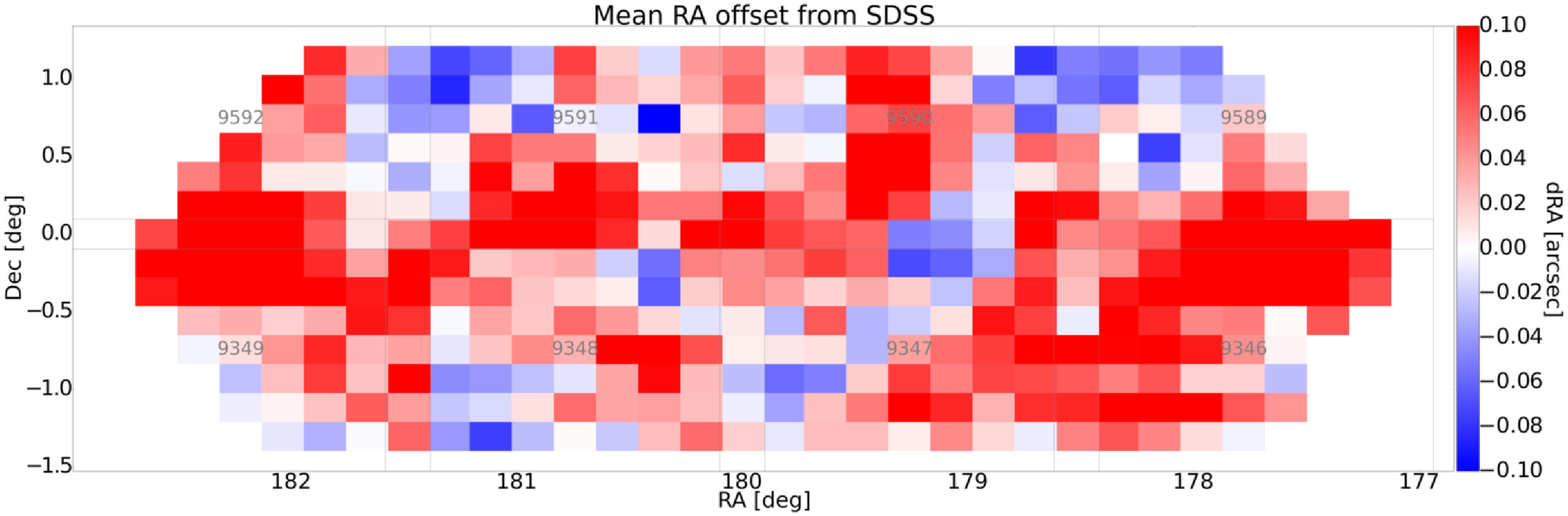}\\\vspace{0.5cm}
   \includegraphics[width=12cm]{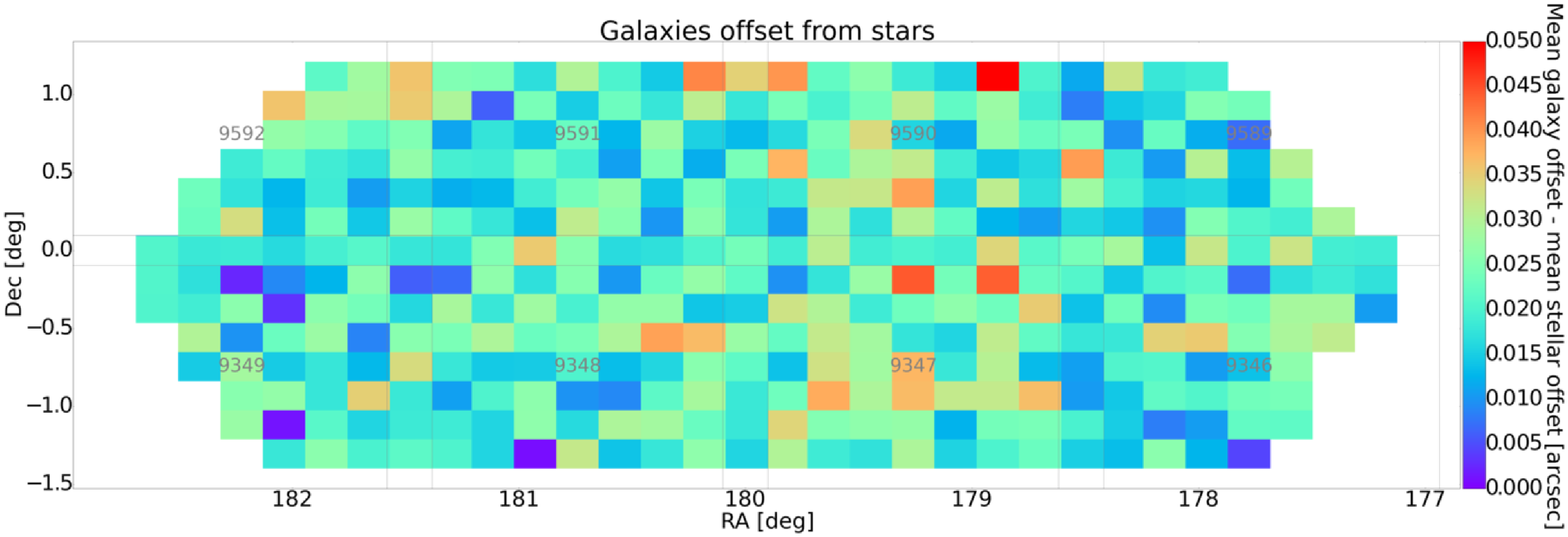}
 \end{center}
 \caption{
   Astrometric quality measures plotted for an example survey component, the WIDE12H region in the $i$-band.
   The first and second plots show the mean RA offset per patch against the PS1 and SDSS-DR9 reference catalogs.
   The third plot shows the astrometric offset between stars and galaxies.
   Each rectangle corresponds to a patch.  The tract IDs and tract borders are shown in gray.
 }
 \label{fig:astrometry}
\end{figure*}

Table~\ref{tab:astrometry} presents detailed measurements of our astrometric performance by survey region and filter.
HSC stellar positions measured on the coadds have RMS residuals (in R.A. and Dec. separately) against PS1 of $\sim 40$~mas and against SDSS~DR9 \citep{2012ApJS..203...21A} of $\sim 90$~mas.
Figure~\ref{fig:astrometry} shows an example field.
The RA (and Dec; not shown) offset against PS1 (first panel) does not show any systematic trends, but 
the residuals against SDSS show small-scale ($\sim 1\degree$) systematic trends (Figure~\ref{fig:astrometry}, second panel).
Similar systematic trends are visible when comparing the PS1 PV2 catalog with SDSS.
It is not clear which catalog has the problem, but it is beyond the scope of this paper to further investigate it.

We can test the astrometry for compact and extended sources separately. 
CModel photometry asymptotically approaches PSF photometry for compact sources
and they have very small magnitude differences. The parameter \texttt{classification\_extendedness} is
based on this difference and is a simple but useful star/galaxy classifier.  The classification is done
in each band separately, but the $i$-band is generally the best band for its superb seeing.
When we divide our sources into stars and galaxies using this parameter and compare their positions with the PS1 catalog, we find a differential offset between the stars and galaxies.
This offset varies in position angle on the sky from field to field, with an amplitude $\sim 30$~mas.
In the example field shown in Fig.~\ref{fig:astrometry} (third panel), 
the offset between stars and galaxies is relatively small (20~mas), but varies as a function of position.
We currently do not understand the origin of this effect, but it is not sufficiently large to prevent most scientific uses of our survey data.
It may be another effect of ignoring proper motions in our reference catalog.

\begin{longtable}{llrrrrr}
 \caption{
   Measurements of the quality of astrometric measurements on HSC coadds by survey region and filter.
   The first four statistical columns are the RMS of residuals of the stated quantity against the stated reference catalog (PS1: \protect{\cite{2016arXiv161205560C}}; SDSS: \protect{\cite{2012ApJS..203...21A}}) for stars.
   We use stars brighter than 20th mag for comparisons against SDSS and PS1, otherwise we use stars brighter than 21.5mag.
   No corrections for proper motion have been made to the reference catalog positions.
   The final statistical column is the mean of the residual offset against PS1 between stars (identified as sources with \code{classification.extendedness = 0}) and galaxies.
   In calculating statistics, we use all suitable sources in the stated region observed in the stated filter, clip at $3\sigma$ (where $\sigma$ is estimated from the inter-quartile range assuming a Gaussian distribution) and then calculate the mean or RMS as appropriate.
 }
 \label{tab:astrometry}
 \hline
 Region & Filter & \centering RA vs PS1 & Dec vs PS1 & RA vs SDSS & Dec vs SDSS & Star-Galaxy offset \\
 & & \multicolumn{1}{c}{(mas)} & \multicolumn{1}{c}{(mas)} & \multicolumn{1}{c}{(mas)} & \multicolumn{1}{c}{(mas)} & \multicolumn{1}{c}{(mas)} \\
 \endfirsthead
 \endhead
 \hline
 \endfoot
 \hline
 \endlastfoot
 \hline
             & $g$ & $32$ & $35$ & $105$ & $100$ & $22$\\
            & $r$ & $34$ & $35$ & $118$ & $102$ & $21$\\
      AEGIS & $i$ & $35$ & $36$ & $122$ & $113$ & $15$\\
            & $z$ & $37$ & $38$ & $125$ & $113$ & $22$\\
            & $y$ & $40$ & $38$ & $125$ & $113$ & $25$\\
\hline
            & $g$ & $38$ & $34$ & $93$ & $88$ & $21$\\
            & $r$ & $36$ & $32$ & $96$ & $89$ & $27$\\
   W-XMMLSS & $i$ & $36$ & $32$ & $96$ & $89$ & $18$\\
            & $z$ & $37$ & $33$ & $96$ & $89$ & $26$\\
            & $y$ & $40$ & $35$ & $98$ & $90$ & $28$\\
\hline
            & $g$ & $24$ & $24$ & $73$ & $74$ & $ 7$\\
            & $r$ & $23$ & $24$ & $79$ & $77$ & $ 5$\\
  W-GAMA09H & $i$ & $24$ & $24$ & $82$ & $80$ & $ 4$\\
            & $z$ & $27$ & $27$ & $84$ & $81$ & $ 2$\\
            & $y$ & $26$ & $26$ & $84$ & $82$ & $11$\\
\hline
            & $g$ & $34$ & $31$ & $101$ & $88$ & $19$\\
            & $r$ & $34$ & $29$ & $110$ & $88$ & $19$\\
  W-WIDE12H & $i$ & $37$ & $31$ & $114$ & $92$ & $20$\\
            & $z$ & $40$ & $34$ & $118$ & $96$ & $29$\\
            & $y$ & $41$ & $35$ & $118$ & $96$ & $23$\\
\hline
            & $g$ & $32$ & $30$ & $110$ & $100$ & $14$\\
            & $r$ & $31$ & $28$ & $116$ & $104$ & $11$\\
  W-GAMA15H & $i$ & $30$ & $27$ & $118$ & $105$ & $14$\\
            & $z$ & $36$ & $32$ & $123$ & $109$ & $21$\\
            & $y$ & $33$ & $30$ & $121$ & $108$ & $22$\\
\hline
            & $g$ & $25$ & $30$ & $83$ & $98$ & $15$\\
            & $r$ & $23$ & $30$ & $85$ & $102$ & $ 8$\\
 W-HECTOMAP & $i$ & $25$ & $31$ & $88$ & $104$ & $ 9$\\
            & $z$ & $27$ & $33$ & $91$ & $109$ & $20$\\
            & $y$ & $27$ & $34$ & $90$ & $108$ & $15$\\
\hline
            & $g$ & $30$ & $27$ & $78$ & $77$ & $ 9$\\
            & $r$ & $29$ & $26$ & $80$ & $77$ & $10$\\
     W-VVDS & $i$ & $31$ & $27$ & $84$ & $78$ & $ 7$\\
            & $z$ & $33$ & $30$ & $85$ & $79$ & $10$\\
            & $y$ & $34$ & $29$ & $85$ & $79$ & $17$\\
\hline
\hline
            & $g$ & $35$ & $34$ & $86$ & $83$ & $11$\\
            & $r$ & $36$ & $34$ & $90$ & $84$ & $24$\\
            & $i$ & $31$ & $29$ & $89$ & $82$ & $16$\\
   D-XMMLSS & $z$ & $36$ & $33$ & $92$ & $86$ & $32$\\
            & $y$ & $39$ & $36$ & $93$ & $87$ & $33$\\
            & NB816 & ---   &  ---   &  ---   &  ---   & ---\\
            & NB921 & ---   &  ---   &  ---   &  ---   & ---\\
\hline
            & $g$ & $34$ & $33$ & $102$ & $102$ & $16$\\
            & $r$ & $33$ & $31$ & $113$ & $100$ & $11$\\
            & $i$ & $33$ & $31$ & $115$ & $103$ & $14$\\
   D-COSMOS & $z$ & $35$ & $34$ & $113$ & $104$ & $15$\\
            & $y$ & $36$ & $33$ & $116$ & $103$ & $32$\\
            & NB816 & ---   &  ---   &  ---   &  ---   & ---\\
            & NB921 & $35$ & $32$ & $117$ & $107$ & $20$\\
\hline
            & $g$ & $27$ & $34$ & $82$ & $95$ & $ 6$\\
            & $r$ & $26$ & $32$ & $88$ & $101$ & $ 9$\\
            & $i$ & $27$ & $33$ & $89$ & $104$ & $ 7$\\
 D-ELAIS-N1 & $z$ & $29$ & $35$ & $92$ & $109$ & $15$\\
            & $y$ & $30$ & $36$ & $91$ & $105$ & $20$\\
            & NB816 & ---   &  ---   &  ---   &  ---   & ---\\
            & NB921 & $30$ & $37$ & $93$ & $108$ & $11$\\
\hline
            & $g$ & $34$ & $29$ & $104$ & $101$ & $10$\\
            & $r$ & $34$ & $28$ & $108$ & $99$ & $15$\\
            & $i$ & $35$ & $28$ & $109$ & $97$ & $11$\\
  D-DEEP2-3 & $z$ & $36$ & $29$ & $112$ & $99$ & $16$\\
            & $y$ & $40$ & $32$ & $112$ & $100$ & $23$\\
            & NB816 & $38$ & $30$ & $118$ & $102$ & $17$\\
            & NB921 & $40$ & $31$ & $118$ & $103$ & $22$\\
\hline
\hline
            & $g$ & $33$ & $31$ & $99$ & $101$ & $ 4$\\
            & $r$ & $32$ & $29$ & $111$ & $104$ & $ 7$\\
            & $i$ & $36$ & $31$ & $118$ & $106$ & $12$\\
  UD-COSMOS & $z$ & $39$ & $35$ & $118$ & $111$ & $12$\\
            & $y$ & $36$ & $33$ & $116$ & $108$ & $24$\\
            & NB816 & ---   &  ---   &  ---   &  ---   & ---\\
            & NB921 & $37$ & $34$ & $121$ & $113$ & $15$\\
\hline
            & $g$ & $42$ & $34$ & $87$ & $79$ & $ 6$\\
            & $r$ & $42$ & $33$ & $90$ & $76$ & $19$\\
            & $i$ & $38$ & $31$ & $92$ & $80$ & $14$\\
    UD-SXDS & $z$ & $50$ & $36$ & $101$ & $83$ & $21$\\
            & $y$ & $43$ & $35$ & $95$ & $81$ & $25$\\
            & NB816 & $41$ & $34$ & $95$ & $80$ & $22$\\
            & NB921 & $46$ & $39$ & $99$ & $85$ & $25$\\
\hline

 \hline
\end{longtable}

\subsection{Photometry}

\subsubsection{Internal and external comparisons}

Like the astrometric calibration, the photometric calibration is performed against PS1~PV2 (Section \ref{sec:refcatalog}) in two stages.
Individual CCDs are first calibrated with a single zero-point against the reference catalog; this zero-point is used for the processing of individual CCDs.
Then, in the mosaic stage, we use the multiple observations of sources in dithered exposures to fit a polynomial correction over the focal plane,
while accounting for individual CCD offsets.
This corrects for imperfections in the dome flats (e.g., imperfect illumination, scattered light and optical scale changes), resulting in point sources having consistent corrected magnitudes in the dithered exposures.
This correction typically has an RMS of $\sim 10$~mmag.

Table \ref{tab:photometry} summarizes our photometric performance.  When we compare to the external data, we apply color terms for fair comparisons.
HSC PSF fluxes for stars measured on the coadds have RMS residuals against PS1 of $\sim 20$~mmag and against SDSS~DR9 (with fluxes corrected using PS1 photometry; \cite{2016ApJ...822...66F}) of $\sim 25$~mmag (in $gri$) at bright magnitudes.
The scatter is larger in the $z$ and $y$ bands for SDSS, but this is because we extrapolate the SDSS photometry to the HSC $z$ and $y$ bands.
Since these catalogs individually are believed to be accurate to $\sim 10$~mmag, assuming that the errors between our catalogs are uncorrelated, this suggests our coadd photometry is accurate to $\sim 17$~mmag.
This value slightly exceeds our goal of 1\% photometry, but 
the scatter is also at least partly due to the spatial variation of the filter transmission; the $r$ and $i$-band
transmission curves change slightly as a function of radius (see \cite{kawanomoto17} for details).
These filters have been replaced with new ones with much smaller spatial variation and have been used in our observations since 2016 (but not used in this release).
With an updated PS1 reference catalog, more careful calibration, as well as new $r$ and $i$-band filters, we expect that we will be able to
surpass our goal in future reductions.

One way of checking the internal precision and consistency of our catalog is through comparing measurements made with different flux measurement algorithms.
The principal flux measurement algorithms we use are PSF, Kron aperture \citep{1980ApJS...43..305K} and CModel.
Each of these measurements are aperture corrected, and hence different flux measurements should measure the same values for compact sources, i.e., stars.
The width of the distribution of the magnitude difference of two flux measurements of stars is therefore a measure of the quality of each of those flux measurement algorithms, and therefore is a check on the internal photometric precision (it of course says nothing about calibration).
We use stars brighter than 21.5~mag here.
Our results are summarized in the last two columns in Table \ref{tab:photometry}.
The distribution of PSF-CModel is always quite tight (3~mmag), except for UD-SXDS NB816 which suffers from a few problematic patches.  This small scatter reflects the fact that the CModel collapses to a PSF measurement for stars, but PSF-Kron gives us an opportunity to evaluate the quality of the PSF modeling because it is being compared to an aperture measurement.
Our PSF-Kron widths are $\sim 10$~mmag, which indicates that systematic errors in the PSF photometry is about that order.



\begin{figure*}
 \begin{center}
  \includegraphics[width=12cm]{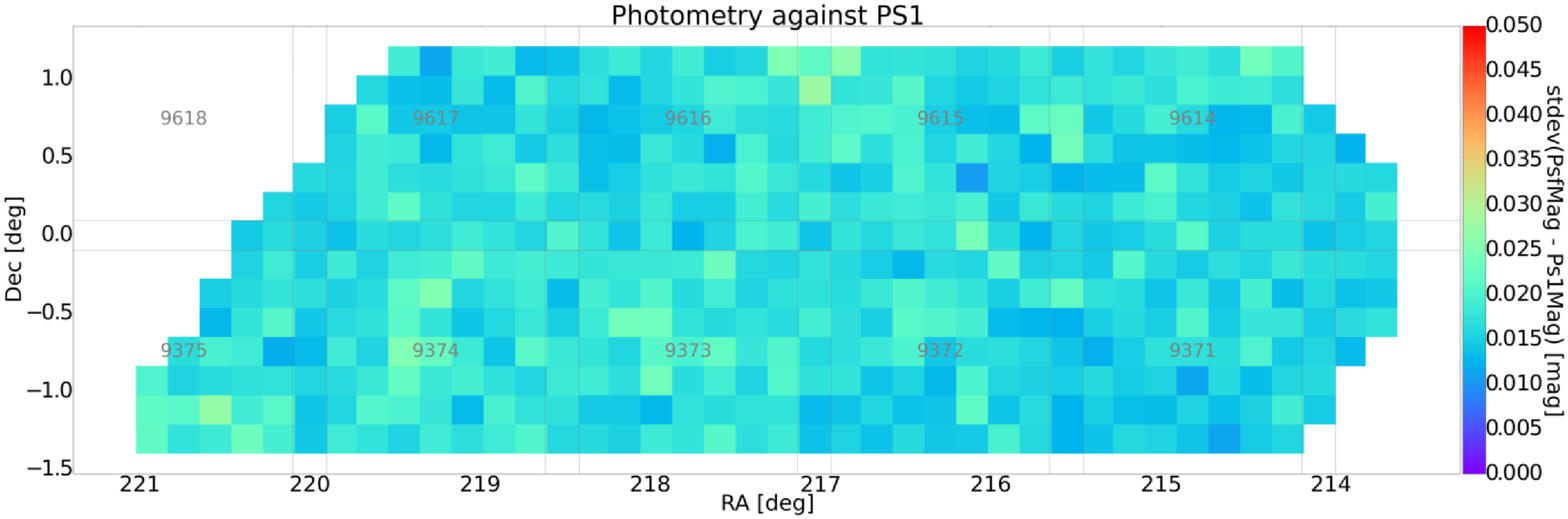}\\\vspace{0.5cm}
  \includegraphics[width=12cm]{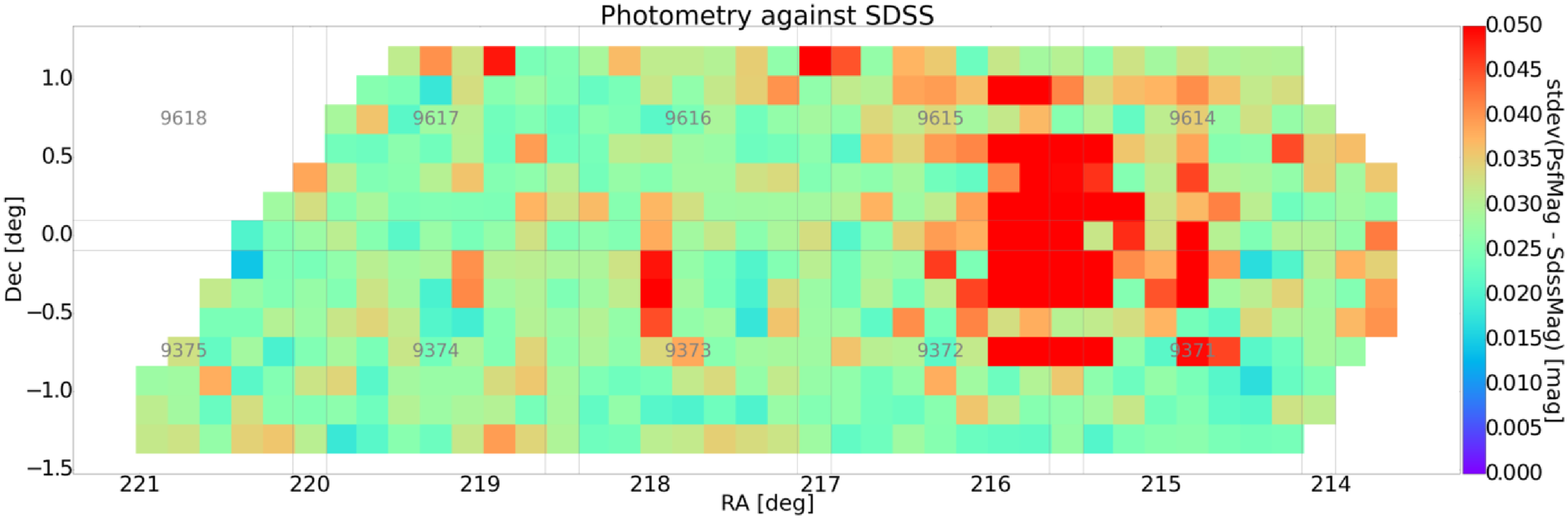}\\\vspace{0.5cm}
  \includegraphics[width=12cm]{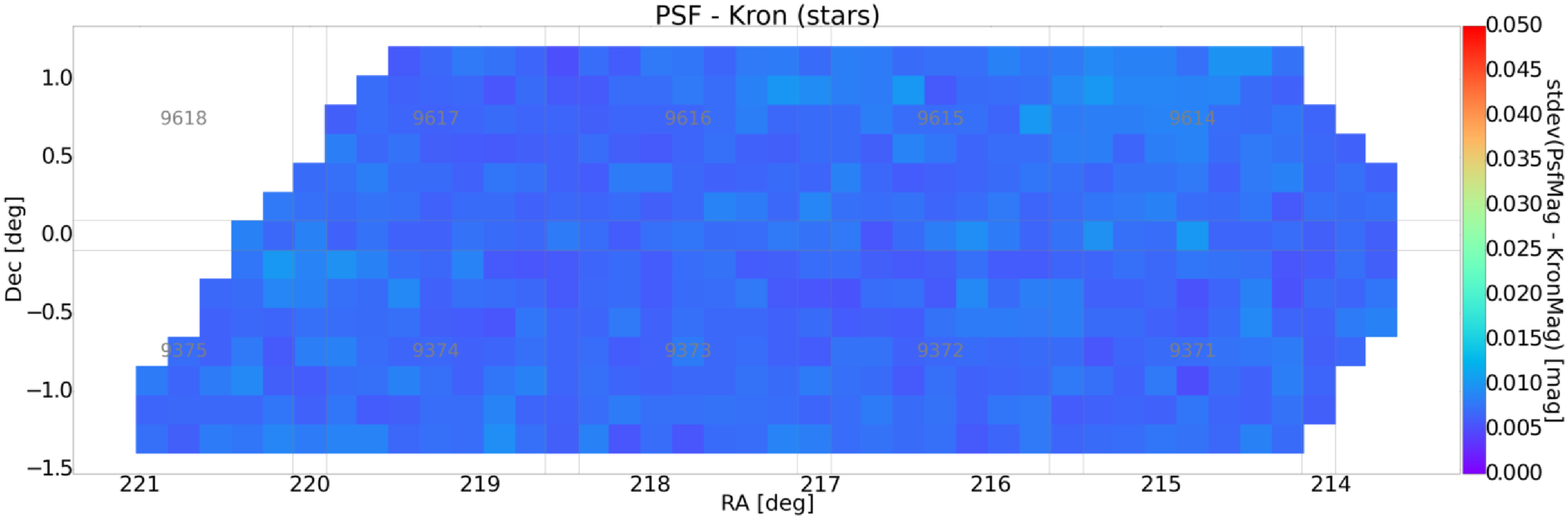}
 \end{center}
 \caption{
   Photometric quality measures plotted for an example survey component, the GAMA15H region in the $i$-band.
   The first and second plots show the width of the distribution of the difference in stellar PSF magnitudes from that in the PS1 and SDSS-DR9 reference catalogs done\
 separately on each patch.
   The third plot shows the width of the distribution of the difference between the PSF and Kron magnitudes.
   Each rectangle represents a patch.
 }
 \label{fig:photometry}
\end{figure*}

See \ref{fig:photometry} for the statistics in an example field, the GAMA15H region in the $i$-band.
Our photometry is fairly uniform across the field compared to PS1 with a scatter of 17~mmag.
However, there is a zonal offset over a degree scale compared to SDSS.  We observe a similar feature
in some of other fields.  The internal consistency between the PSF and Kron photometry is better than
10~mmag.  Overall, this field is calibrated well and this is the typical photometric quality of
our survey.  The only exception is the VVDS field, in which a small number of patches suffer from the PSF modeling
problem as mentioned earlier.

\begin{longtable}{llrrrr}
 \caption{
   Measurements of photometric quality by survey region and filter.
 The first two statistical columns are the RMS of residuals of the calibrated PSF magnitude against the stated reference catalog (PS1: \protect{\cite{2016arXiv161205560C}}; SDSS: \protect{\cite{2012ApJS..203...21A}}) for stars.
 The last two statistical columns are the RMS of the difference between the two stated magnitudes measured by the pipeline for stars.
 In all cases, stars are identified as sources with \code{classification.extendedness = 0}.
 We use stars brighter than 20th mag for comparisons against SDSS and PS1, otherwise we use stars brighter than 21.5mag.
 In calculating statistics, we use all suitable sources in the stated region observed in the stated filter, clip at $3\sigma$ (where $\sigma$ is estimated from the inter-quartile range assuming a Gaussian distribution) and then calculate the mean or RMS as appropriate.
  }
  \label{tab:photometry}
  \hline
 Region & Filter & \centering Psf vs PS1 & Psf vs SDSS & Psf - Kron & Psf - CModel \\
 & & \multicolumn{1}{c}{(mmag)} & \multicolumn{1}{c}{(mmag)} & \multicolumn{1}{c}{(mmag)} & \multicolumn{1}{c}{(mmag)} \\
 \endfirsthead
 \endhead
  \hline
 \endfoot
  \hline
 \endlastfoot
  \hline
            & $g$ & $27.3$ & $22.0$ & $10.2$ & $3.6$\\
            & $r$ & $20.1$ & $25.0$ & $8.6$ & $2.5$\\
      AEGIS & $i$ & $21.6$ & $34.5$ & $9.6$ & $3.2$\\
            & $z$ & $15.5$ & $50.8$ & $9.4$ & $2.3$\\
            & $y$ & $31.7$ & $67.4$ & $13.6$ & $3.2$\\
\hline
            & $g$ & $24.1$ & $24.3$ & $10.4$ & $3.4$\\
            & $r$ & $21.4$ & $27.3$ & $14.2$ & $3.0$\\
   W-XMMLSS & $i$ & $17.7$ & $31.0$ & $9.3$ & $2.7$\\
            & $z$ & $16.2$ & $55.5$ & $11.9$ & $2.5$\\
            & $y$ & $31.7$ & $69.8$ & $16.1$ & $3.0$\\
\hline
            & $g$ & $19.9$ & $16.7$ & $8.2$ & $2.2$\\
            & $r$ & $17.4$ & $21.0$ & $9.3$ & $2.0$\\
  W-GAMA09H & $i$ & $21.6$ & $28.5$ & $10.4$ & $2.2$\\
            & $z$ & $16.6$ & $55.7$ & $10.0$ & $2.1$\\
            & $y$ & $30.5$ & $71.3$ & $12.3$ & $2.3$\\
\hline
            & $g$ & $21.3$ & $19.7$ & $9.3$ & $2.9$\\
            & $r$ & $19.7$ & $24.1$ & $7.7$ & $2.9$\\
  W-WIDE12H & $i$ & $25.0$ & $28.9$ & $8.3$ & $2.5$\\
            & $z$ & $14.8$ & $50.1$ & $8.2$ & $2.3$\\
            & $y$ & $29.7$ & $69.6$ & $13.6$ & $3.0$\\
\hline
            & $g$ & $19.4$ & $19.3$ & $8.9$ & $2.8$\\
            & $r$ & $18.5$ & $24.1$ & $9.7$ & $2.5$\\
  W-GAMA15H & $i$ & $17.2$ & $31.4$ & $7.5$ & $2.4$\\
            & $z$ & $16.1$ & $59.9$ & $9.5$ & $2.2$\\
            & $y$ & $29.1$ & $79.8$ & $13.5$ & $2.7$\\
\hline
            & $g$ & $18.8$ & $18.3$ & $10.6$ & $2.7$\\
            & $r$ & $15.7$ & $23.4$ & $7.6$ & $2.2$\\
 W-HECTOMAP & $i$ & $18.3$ & $31.5$ & $8.0$ & $2.9$\\
            & $z$ & $13.3$ & $52.7$ & $11.7$ & $2.2$\\
            & $y$ & $28.2$ & $73.0$ & $14.8$ & $3.0$\\
\hline
            & $g$ & $21.2$ & $18.4$ & $12.1$ & $2.8$\\
            & $r$ & $16.7$ & $22.7$ & $10.7$ & $2.6$\\
     W-VVDS & $i$ & $16.3$ & $27.3$ & $8.5$ & $2.5$\\
            & $z$ & $19.1$ & $57.4$ & $16.2$ & $2.6$\\
            & $y$ & $35.2$ & $71.1$ & $19.2$ & $2.8$\\
\hline
\hline
            & $g$ & $26.5$ & $29.4$ & $12.3$ & $3.6$\\
            & $r$ & $28.5$ & $33.6$ & $14.9$ & $4.0$\\
            & $i$ & $19.2$ & $36.4$ & $12.3$ & $2.6$\\
   D-XMMLSS & $z$ & $14.9$ & $55.8$ & $14.3$ & $2.4$\\
            & $y$ & $34.1$ & $78.6$ & $18.1$ & $2.7$\\
            & NB816 & ---   &  ---   &  ---   &  ---\\
            & NB921 & ---   &  ---   &  ---   &  ---\\
\hline
            & $g$ & $20.3$ & $24.1$ & $11.8$ & $3.0$\\
            & $r$ & $21.0$ & $28.4$ & $8.9$ & $3.1$\\
            & $i$ & $28.1$ & $41.0$ & $19.0$ & $3.0$\\
   D-COSMOS & $z$ & $19.4$ & $69.4$ & $9.6$ & $2.3$\\
            & $y$ & $35.2$ & $79.7$ & $14.0$ & $3.0$\\
            & NB816 & ---   &  ---   &  ---   &  ---\\
            & NB921 & $23.9$ & $61.8$ & $9.5$ & $2.2$\\
\hline
            & $g$ & $24.2$ & $35.6$ & $15.5$ & $3.6$\\
            & $r$ & $16.3$ & $29.0$ & $9.6$ & $2.4$\\
            & $i$ & $19.2$ & $37.5$ & $9.0$ & $2.7$\\
 D-ELAIS-N1 & $z$ & $15.9$ & $59.2$ & $13.1$ & $2.0$\\
            & $y$ & $33.8$ & $83.3$ & $12.4$ & $3.2$\\
            & NB816 & ---   &  ---   &  ---   &  ---\\
            & NB921 & $23.0$ & $62.2$ & $10.2$ & $2.6$\\
\hline
            & $g$ & $21.7$ & $21.4$ & $11.9$ & $2.8$\\
            & $r$ & $19.1$ & $24.9$ & $9.3$ & $2.7$\\
            & $i$ & $21.4$ & $31.2$ & $12.3$ & $2.8$\\
  D-DEEP2-3 & $z$ & $18.1$ & $54.8$ & $12.7$ & $2.0$\\
            & $y$ & $34.3$ & $75.7$ & $10.7$ & $3.3$\\
            & NB816 & $24.3$ & $35.2$ & $11.2$ & $3.4$\\
            & NB921 & $23.9$ & $56.2$ & $11.5$ & $3.1$\\
\hline
\hline
            & $g$ & $19.6$ & $24.7$ & $11.5$ & $2.9$\\
            & $r$ & $22.8$ & $36.1$ & $10.9$ & $2.9$\\
            & $i$ & $25.7$ & $40.3$ & $11.6$ & $2.4$\\
  UD-COSMOS & $z$ & $20.5$ & $75.7$ & $13.4$ & $2.4$\\
            & $y$ & $34.5$ & $85.7$ & $15.1$ & $2.1$\\
            & NB816 & ---   &  ---   &  ---   &  ---\\
            & NB921 & $23.4$ & $67.7$ & $10.3$ & $1.9$\\
\hline
            & $g$ & $28.9$ & $33.4$ & $11.2$ & $3.5$\\
            & $r$ & $27.4$ & $34.2$ & $17.0$ & $3.0$\\
            & $i$ & $21.4$ & $41.7$ & $9.5$ & $2.4$\\
    UD-SXDS & $z$ & $26.8$ & $86.5$ & $24.6$ & $2.8$\\
            & $y$ & $39.9$ & $95.0$ & $12.7$ & $2.6$\\
            & NB816 & $26.2$ & $47.2$ & $29.9$ & $58.6$\\
            & NB921 & $25.1$ & $65.1$ & $14.1$ & $2.0$\\
\hline

 \hline
\end{longtable}

We perform further tests using SynPipe, an HSC synthetic galaxy pipeline
\citep{huang17}. This is a  Python-based module that interfaces with hscPipe
and which can inject realistic synthetic stars and galaxies at desired locations in
CCD images. We use SynPipe to examine the photometric
performance of hscPipe.  Details are given in \citet{huang17}, but in brief,
we find that the typical uncertainties of HSC forced PSF photometry for stars range
from around 0.01~mag at $i{\sim}18.0$~mag to 0.06~mag at $i{\sim}25.5$~mag
in the Wide layer.
The 1\% error at the bright end is likely a systematic error in our measurement,
in agreement with our earlier tests.
For synthetic single-{S\'{e}rsic\ }
model galaxies, the typical uncertainties of HSC forced \texttt{cModel} photometry range
from 0.15~mag at $i{\sim} 20.0$~mag to 0.20~mag at $i{\sim} 25.2$~mag.  We will further
discuss this large error at bright magnitudes in Section \ref{sec:known_problems}.
Over the range of colors and magnitude that we have tested, we find that both forced PSF
and \texttt{cModel} photometry provide unbiased estimates of galaxy color.

One of the nice features of ingesting artificial sources is that we can evaluate
effects of blending.  
We find that the degree of galaxy blending (quantified by \texttt{blendedness\_abs\_flux},
or $b$ for short) has an important impact on photometry estimates. For stars with
$b > 0.1$, the forced PSF photometry on average overestimates the magnitudes of stars by
$0.1-0.2$~mag. For galaxies, high-blendedness typically adds an additional 0.05~mag
uncertainty in both magnitude and color estimates.  Further discussions can be found in
Murata et al. (in prep).

\subsubsection{Stellar sequence}
\label{sec:stellar_sequence}

\begin{figure*}
 \begin{center}
  \includegraphics[width=8cm]{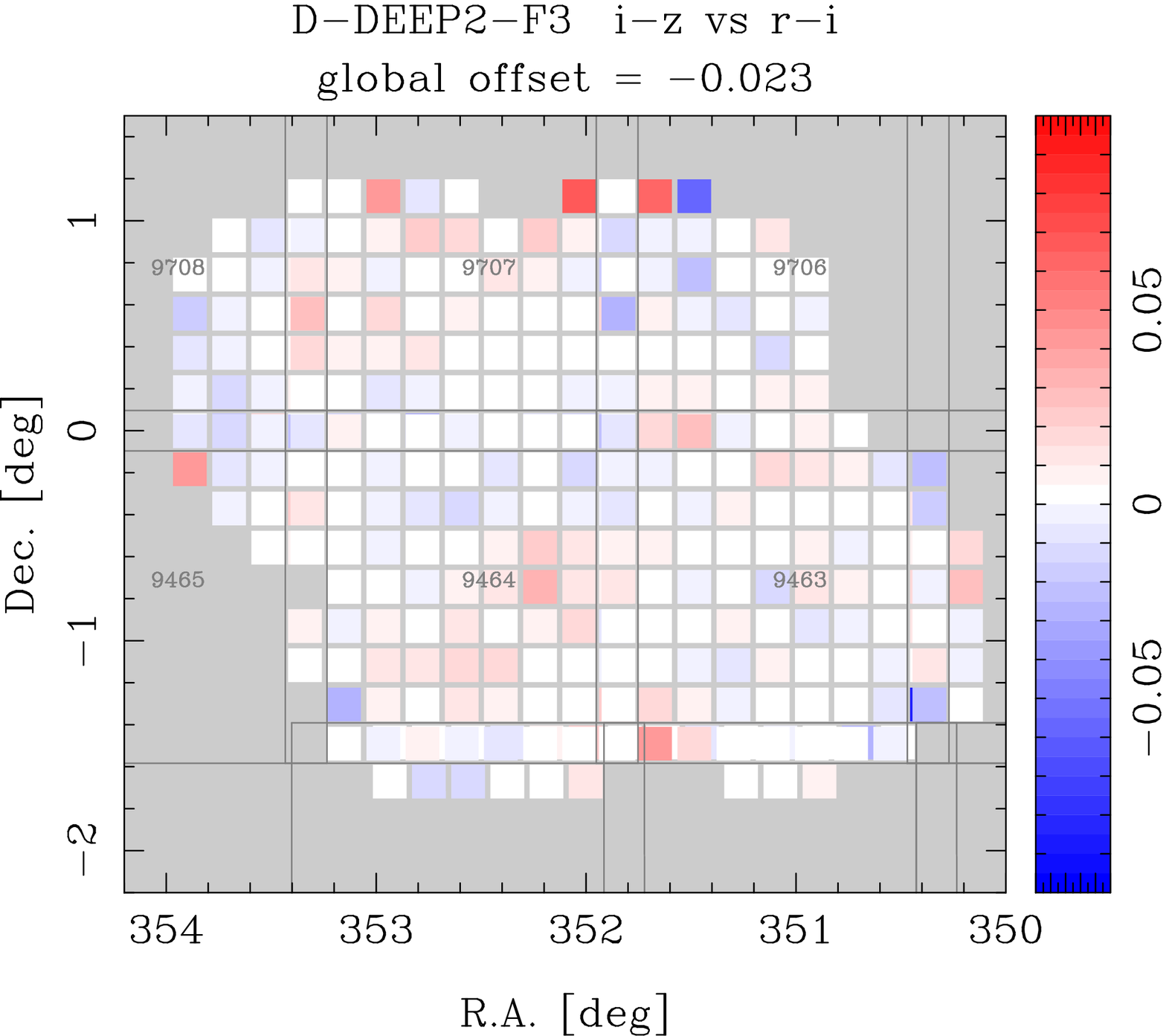}\hspace{0.5cm}
  \includegraphics[width=8cm]{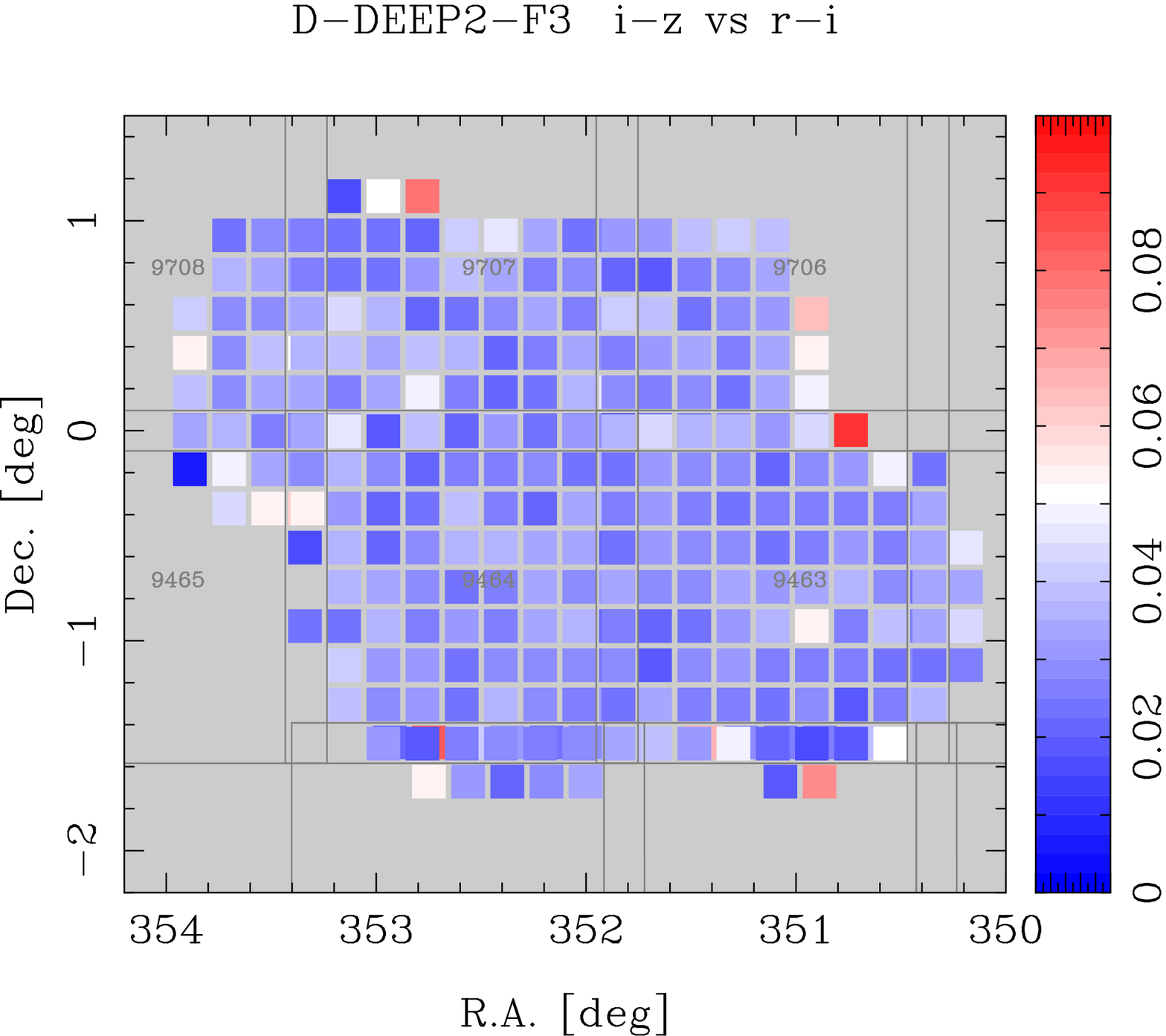} 
 \end{center}
 \caption{
   Color offset in the stellar sequence (left) and color scatter (right) in one of the fields on
   the $i-z$ vs. $r-i$ diagram.
   The median color offset across the field (0.023 mag) is subtracted
   to highlight the spatial inhomogeneity.
   The tract IDs and tract borders are shown in gray.
 }
 \label{fig:stellar_sequence}
\end{figure*}

As another test of photometry, we evaluate the uniformity of the photometric zero points across the survey area.
We estimate an offset between the location of the observed stellar sequence
and that of the synthetic \citet{1983ApJS...52..121G} stellar sequence on a color-color diagram
as a function of position on the sky.
We use only bright stars ($i_{PSF}<22$)
selected using \texttt{classification\_extendedness} with a set of flags applied to ensure
clean photometry (see Table \ref{tab:important_flags}).  At this magnitude range, the extendedness gives a fairly
clean sample of stars as shown in Section \ref{sec:star_galaxy_separation}, and the photometric
uncertainties are small enough for this task.  As the offsets are
degenerate between the two colors chosen, we assume that the offset is entirely in the vertical direction
on the color-color diagrams.  Galactic extinction is corrected for, but not all the stars
are behind the Galactic dust screen, which may introduce an additional offset and scatter in the stellar color.
The offset is evaluated for each patch and the sky distribution of
the stellar sequence offset in one of our fields is shown in
the left panel of Fig. \ref{fig:stellar_sequence}.
We have removed a global offset, which can be a systematic error in \citet{1983ApJS...52..121G} and/or
in our response functions and is generally at the level of 1-2 percent,
in order to enhance the spatial inhomogeneity of the photometric zero-point.
The figure shows that the zero-point is fairly uniform across the field at the level of a percent.
Some of the patches on the field edges have larger errors, but these are noisy regions and their
contribution to the overall area is fairly minor.

In the right panel, we estimate the color scatter around the stellar sequence.
The color scatter is also fairly small, 2-3 per cent.
Note that the color scatter shown in the figure is due to three filters, but 
the $\sqrt{3}$ reduction is not applied here.  Also, the intrinsic color scatter of stars
may contribute here.  Overall, we find that our photometry is accurate to 1-2 percent
in each band across the survey area.  There are a small number of patches with poorer quality
due to the PSF problem discussed in Sections \ref{sec:VVDS} and \ref{sec:poor_psf_modeling_in_good_seeing_areas}, but the photometric
quality in the vast majority of the area should be sufficient for science.

\subsection{Shapes}

A chief goal of the HSC survey is to measure galaxy shapes for weak lensing.
Our shape catalog is not included in this data release (section \ref{sec:the_release}), pending careful validation, but will be published separately.
Nevertheless, here we summarize in Table~\ref{tab:shapes} some basic measurements of the data quality as it impacts shape measurement.

Clearly the seeing over much of the survey area is exquisite, with a mean Gaussian FWHM ranging from 0.5~arcsec in the $i$-band in the HectoMap region to 0.9~arcsec in the $r$-band in GAMA09H and the $g$-band in HectoMap.
The $i$-band always has the best mean seeing for a given region; this is a result of our observing strategy, which prioritizes the $i$-band when the seeing is expected to be good (section \ref{sec:survey_progress}).


We compute the determinant radius of object as

\begin{equation}
  r_{det}=(I_{xx}\times I_{yy}-I_{xy}^2)^{1/4}
  \label{eq:detradius}
\end{equation}

\noindent
where $I_{xx}$, $I_{yy}$, and $I_{xy}$ are the second moments of the image.  
In practice, we measure the second moments with an adaptive window function using GalSim
\citep{2015A&C....10..121R}.  These ``adaptive moments'' are found by iteratively computing the moments
of the best-fitting elliptical Gaussian, using the fitted elliptical Gaussian as a weight function.
We use the difference in the determinant radii between the object and the PSF model for quantifying
the fidelity of the PSF model.

The mean of the determinant radius difference provides a rough measure of the fidelity of the PSF, while
the standard deviation (stdev) is a measure of how noisy our measurements are.
We find that our PSF models are slightly wider ($\sim0.2\%$) than the observations.
This behavior has been seen by other large surveys \citep{2016MNRAS.460.2245J,2015MNRAS.454.3500K} that use PSFEx
and thus is likely a feature of the software.
Its impact on the shear estimation is quantified in the shear catalog paper \citep{mandelbaum17} and
it turns out to be a subdominant component in our error budget.

\begin{figure*}
 \begin{center}
  \includegraphics[width=12cm]{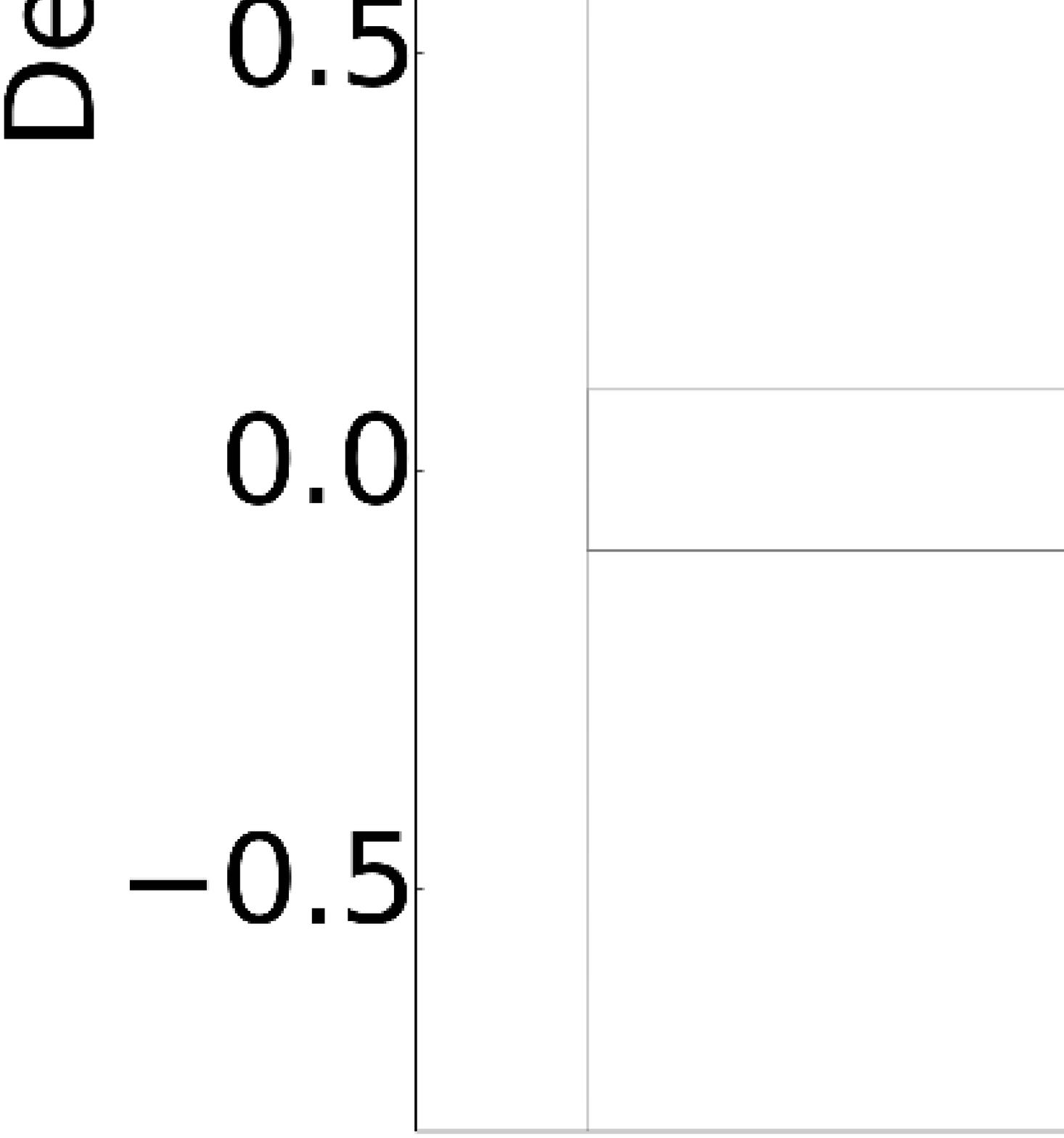}\\\vspace{0.5cm}
  \includegraphics[width=12cm]{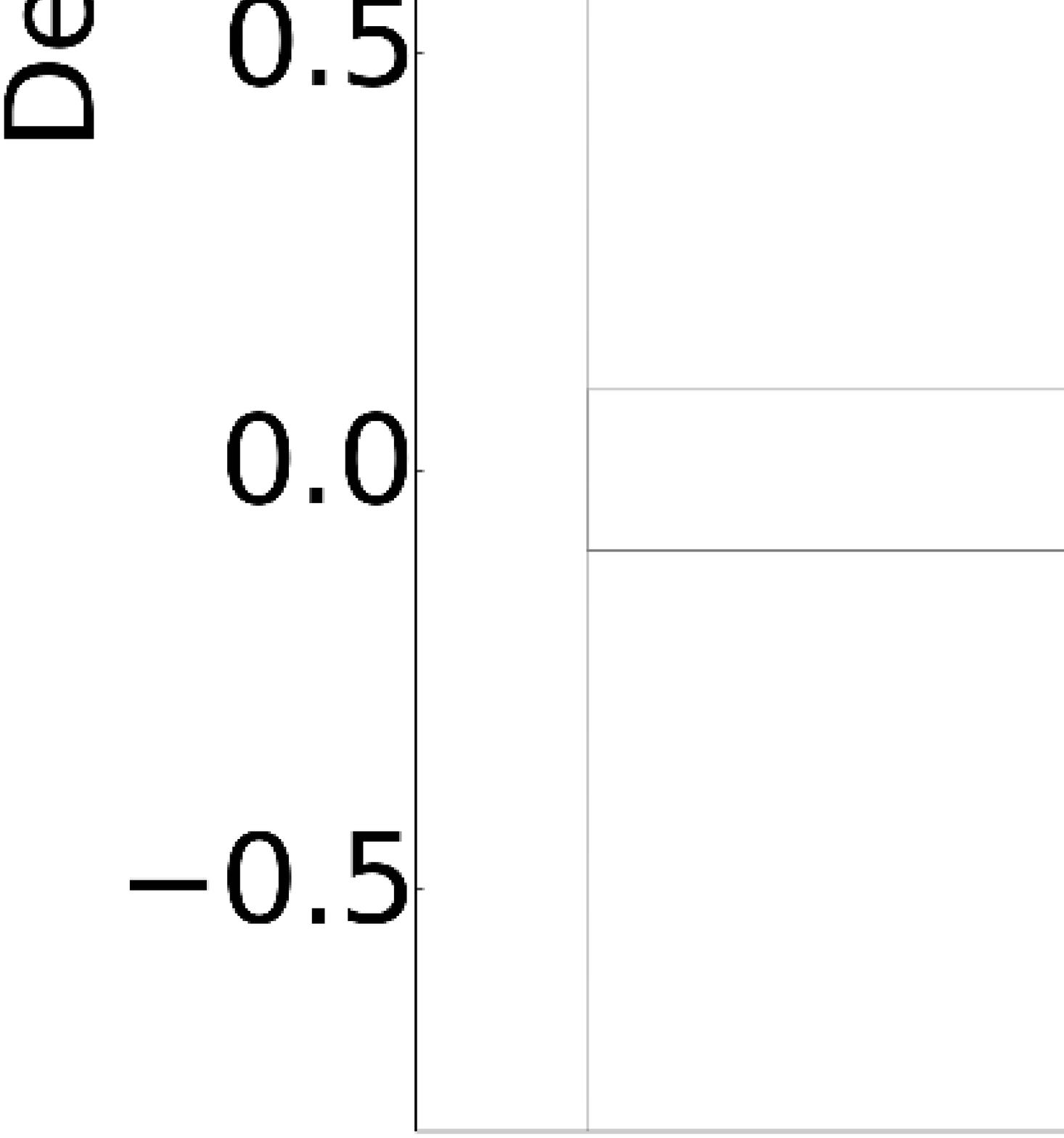}\\\vspace{0.5cm}
  \includegraphics[width=12cm]{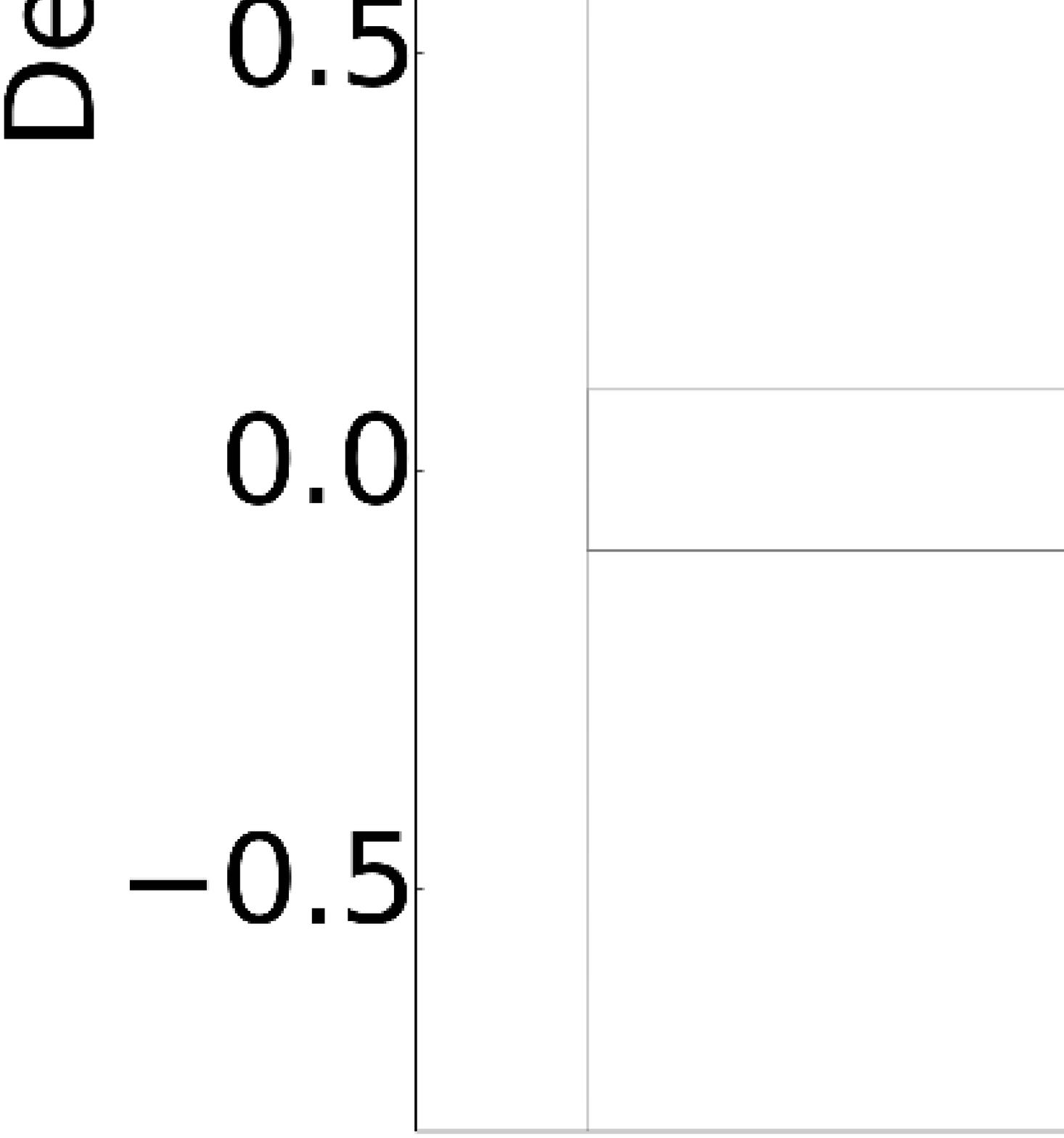}
 \end{center}
 \caption{
   PSF model shape quality measures plotted for an example Wide layer survey component, the VVDS region in the $r$-band.
   The first plot shows the seeing FWHM in arcsec assuming a Gaussian PSF (i.e., FWHM=$r_{det}\times2\sqrt{2\ln2}$).
   The second and third plots show the difference in the determinant radius between object and PSF, and its standard deviation.
   Each rectangle represents a patch.  
 }
 \label{fig:shapes1}
\end{figure*}

See Figure~\ref{fig:shapes1} for an example Wide layer field, the VVDS region in the $r$-band.
The seeing in the region can vary significantly (by a factor of 2) from patch to patch, because different
areas within the region have been observed under different conditions.
Generally, the standard deviation of the difference tracks the seeing because objects have higher S/N under better seeing.
It simply means that our shape measurements are noisier when the seeing is worse.

\begin{figure*}
 \begin{center}
  \includegraphics[width=8cm]{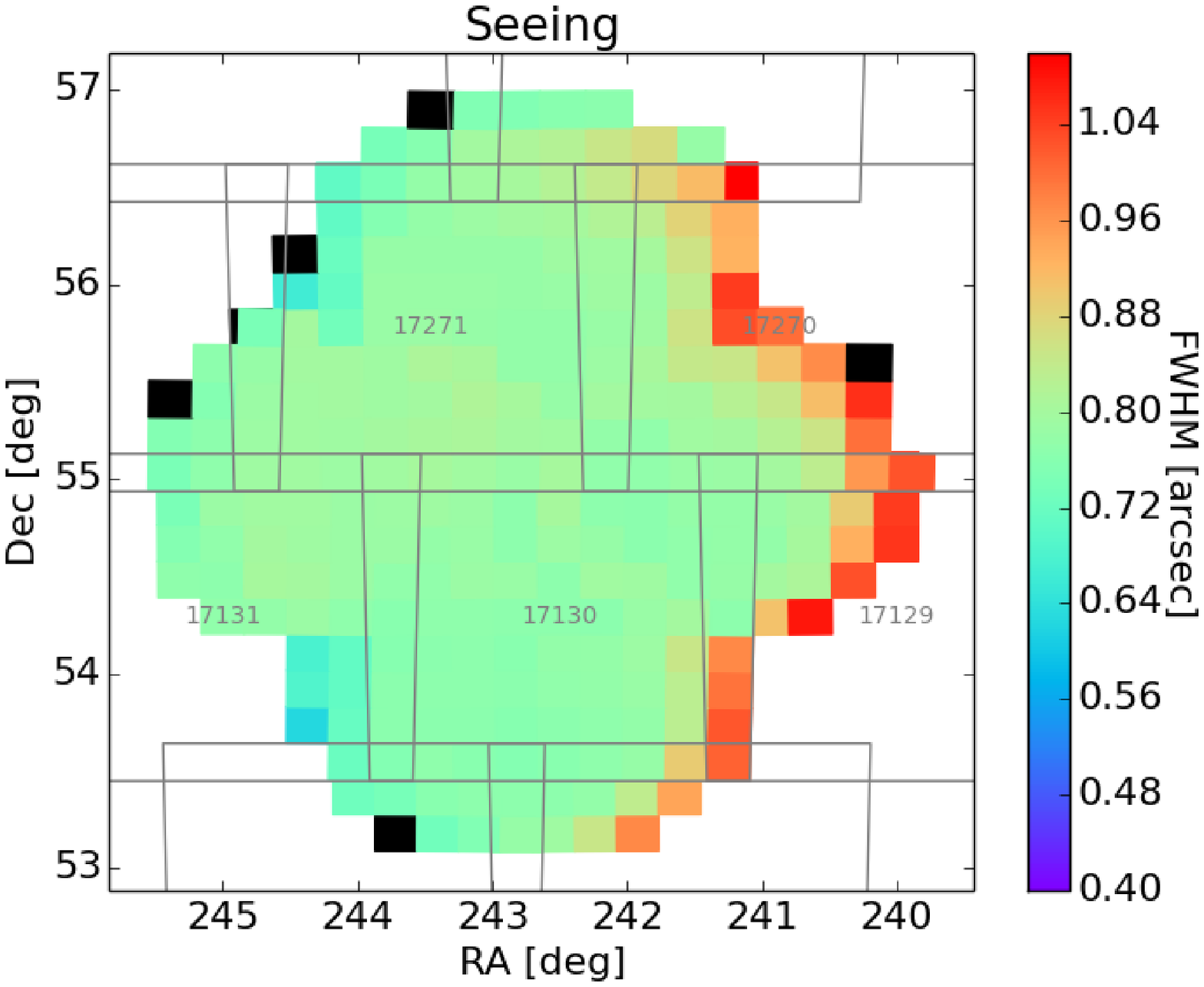}\hspace{0.5cm}
  \includegraphics[width=8cm]{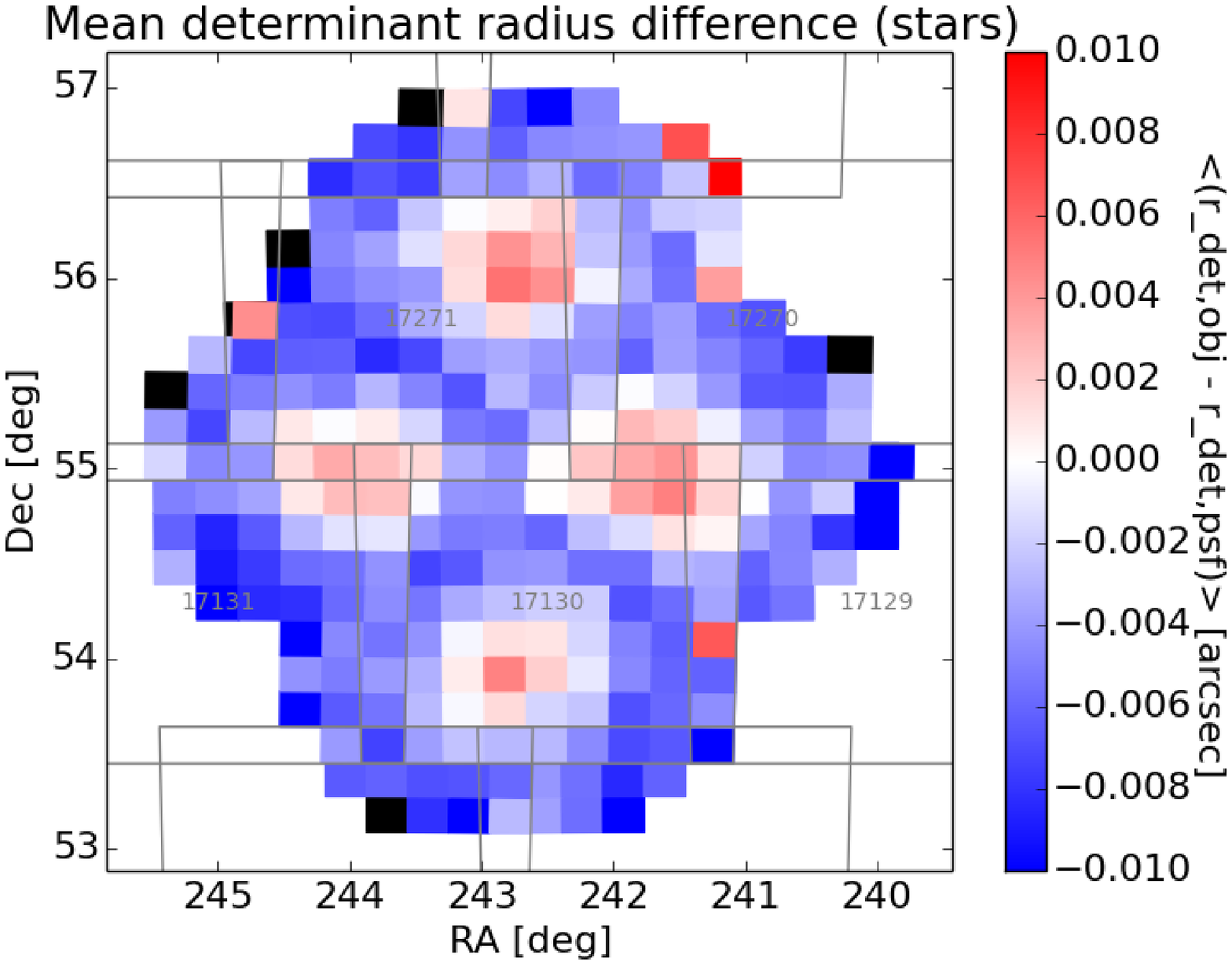}\\\vspace{0.5cm}
  \includegraphics[width=8cm]{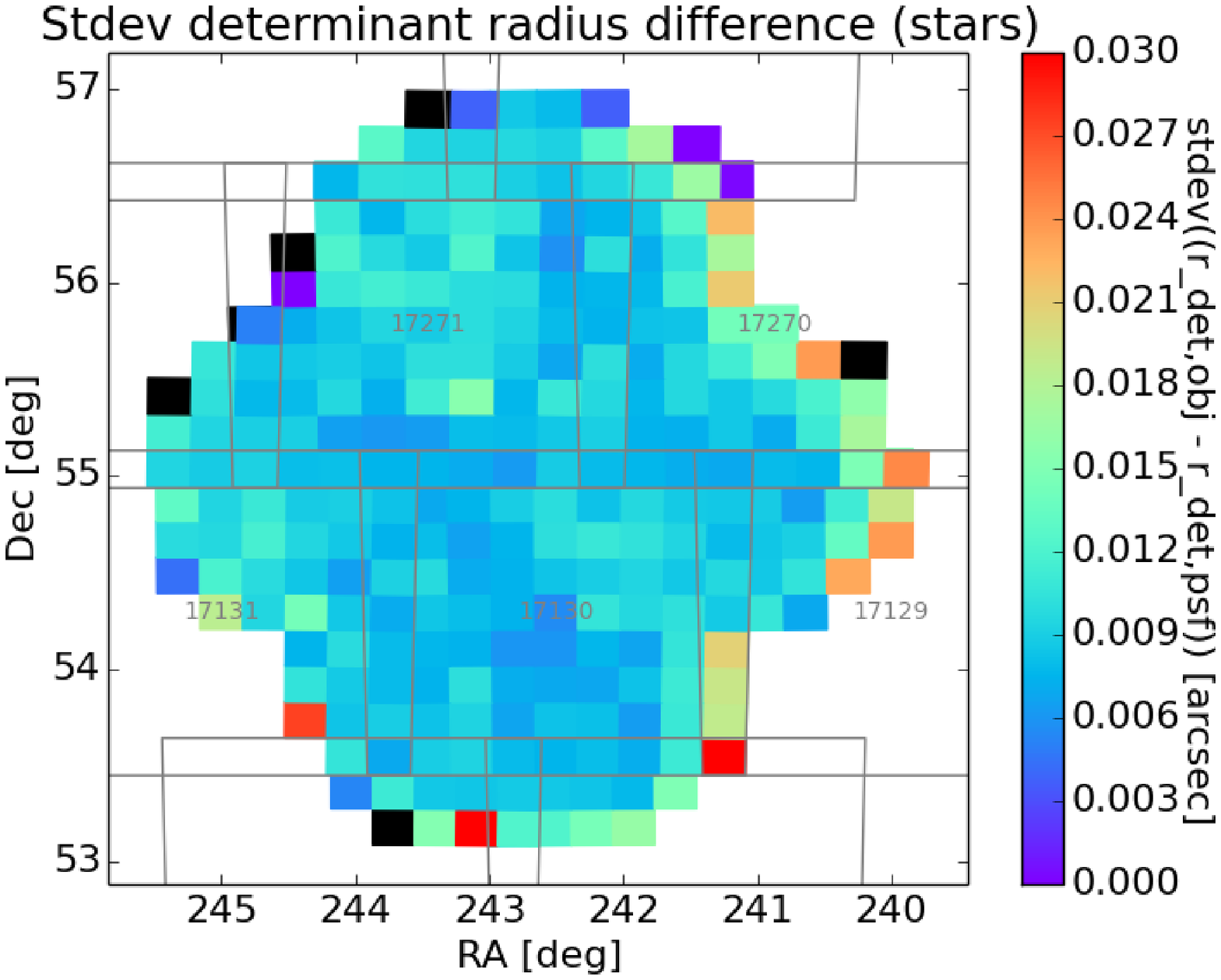}
 \end{center}
 \caption{
   Same as fig Fig.~\ref{fig:shapes1} but for the ELAIS-N1 field.
 }
 \label{fig:shapes2}
\end{figure*}

Figure~\ref{fig:shapes2} shows an example Deep layer field, the ELAIS-N1 region in the $r$-band, which consists of four pointings.
The mean difference varies over each pointing, while the standard deviation is constant.
This may indicate that the PSF is not being well fit in the center and extremities of each visit, even though the seeing is about 0.8~arcsec.
The same pattern is seen in Deep and Ultra-Deep layers and the AEGIS field, but generally not in the Wide layers.
This may be because of the large (1/3 of the field of view) dithers used in Wide, which balances out positive and negative errors, while the Deep, Ultra-Deep and AEGIS observations are done with small dithers (a few to several arcmin).  The problem is still being investigated, but we should emphasize that this is a very small effect (0.3\% variation in PSF size) and most science should be unaffected by this.

\begin{longtable}{llcc}
 \caption{
   Basic measurements of the data quality impacting shape measurement.
   The first statistical column is the seeing FWHM assuming a Gaussian PSF.
   The second statistical column is the determinant radius difference between object and the PSF written as mean $\pm$ RMS.
   In all cases, stars are identified as sources with \code{classification.extendedness = 0}.
   We use stars brighter than 21.5mag.
   In calculating statistics, we use all suitable sources in the stated region observed in the stated filter, clip at $3\sigma$ (where $\sigma$ is estimated from the inter-quartile range assuming a Gaussian distribution) and then calculate the mean or RMS as appropriate.
  }
  \label{tab:shapes}
  \hline
 Region & Filter & \centering seeing FWHM & determinant radius difference\\
 & & \multicolumn{1}{c}{(arcsec)} & \multicolumn{1}{c}{(arcsec)} \\
 \endfirsthead
 \endhead
  \hline
 \endfoot
  \hline
 \endlastfoot
 \hline
             & $g$ & $0.72$ & $-0.001\pm0.010$\\
            & $r$ & $0.71$ & $-0.003\pm0.008$\\
      AEGIS & $i$ & $0.52$ & $-0.004\pm0.007$\\
            & $z$ & $0.74$ & $-0.004\pm0.009$\\
            & $y$ & $0.65$ & $-0.001\pm0.012$\\
\hline
            & $g$ & $0.84$ & $-0.004\pm0.013$\\
            & $r$ & $0.84$ & $-0.003\pm0.014$\\
   W-XMMLSS & $i$ & $0.68$ & $-0.003\pm0.009$\\
            & $z$ & $0.70$ & $-0.002\pm0.009$\\
            & $y$ & $0.77$ & $-0.002\pm0.014$\\
\hline
            & $g$ & $0.80$ & $-0.004\pm0.011$\\
            & $r$ & $0.87$ & $-0.004\pm0.011$\\
  W-GAMA09H & $i$ & $0.63$ & $-0.003\pm0.008$\\
            & $z$ & $0.69$ & $-0.003\pm0.008$\\
            & $y$ & $0.69$ & $-0.002\pm0.010$\\
\hline
            & $g$ & $0.66$ & $-0.002\pm0.009$\\
            & $r$ & $0.55$ & $-0.003\pm0.006$\\
  W-WIDE12H & $i$ & $0.53$ & $-0.003\pm0.007$\\
            & $z$ & $0.66$ & $-0.003\pm0.007$\\
            & $y$ & $0.63$ & $-0.002\pm0.011$\\
\hline
            & $g$ & $0.70$ & $-0.003\pm0.010$\\
            & $r$ & $0.63$ & $-0.003\pm0.008$\\
  W-GAMA15H & $i$ & $0.56$ & $-0.002\pm0.006$\\
            & $z$ & $0.63$ & $-0.002\pm0.007$\\
            & $y$ & $0.67$ & $-0.002\pm0.011$\\
\hline
            & $g$ & $0.90$ & $-0.005\pm0.014$\\
            & $r$ & $0.73$ & $-0.003\pm0.007$\\
 W-HECTOMAP & $i$ & $0.48$ & $-0.003\pm0.006$\\
            & $z$ & $0.75$ & $-0.004\pm0.011$\\
            & $y$ & $0.59$ & $-0.002\pm0.011$\\
\hline
            & $g$ & $0.77$ & $-0.003\pm0.011$\\
            & $r$ & $0.66$ & $-0.003\pm0.009$\\
     W-VVDS & $i$ & $0.53$ & $-0.003\pm0.007$\\
            & $z$ & $0.57$ & $-0.003\pm0.009$\\
            & $y$ & $0.59$ & $-0.002\pm0.011$\\
\hline
\hline
            & $g$ & $0.74$ & $-0.003\pm0.013$\\
            & $r$ & $0.55$ & $-0.004\pm0.012$\\
            & $i$ & $0.85$ & $-0.003\pm0.014$\\
   D-XMMLSS & $z$ & $1.05$ & $-0.004\pm0.017$\\
            & $y$ & $0.79$ & $-0.002\pm0.017$\\
            & NB816 & ---   &  ---\\
            & NB921 & ---   &  ---\\
\hline
            & $g$ & $0.96$ & $-0.005\pm0.016$\\
            & $r$ & $0.59$ & $-0.003\pm0.008$\\
            & $i$ & $0.55$ & $-0.001\pm0.009$\\
   D-COSMOS & $z$ & $0.57$ & $-0.004\pm0.007$\\
            & $y$ & $0.65$ & $-0.003\pm0.012$\\
            & NB816 & ---   &  ---\\
            & NB921 & $0.67$ & $-0.002\pm0.009$\\
\hline
            & $g$ & $0.61$ & $-0.002\pm0.011$\\
            & $r$ & $0.78$ & $-0.003\pm0.010$\\
            & $i$ & $0.55$ & $-0.004\pm0.008$\\
 D-ELAIS-N1 & $z$ & $0.80$ & $-0.003\pm0.013$\\
            & $y$ & $0.55$ & $-0.002\pm0.009$\\
            & NB816 & ---   &  ---\\
            & NB921 & $0.65$ & $-0.003\pm0.009$\\
\hline
            & $g$ & $0.93$ & $-0.004\pm0.015$\\
            & $r$ & $0.69$ & $-0.004\pm0.009$\\
            & $i$ & $0.54$ & $-0.004\pm0.009$\\
  D-DEEP2-3 & $z$ & $0.68$ & $-0.002\pm0.010$\\
            & $y$ & $0.50$ & $-0.002\pm0.007$\\
            & NB816 & $0.51$ & $-0.004\pm0.008$\\
            & NB921 & $0.57$ & $-0.003\pm0.008$\\
\hline
\hline
            & $g$ & $0.84$ & $-0.004\pm0.014$\\
            & $r$ & $0.58$ & $-0.003\pm0.009$\\
            & $i$ & $0.65$ & $-0.003\pm0.011$\\
  UD-COSMOS & $z$ & $0.59$ & $-0.004\pm0.010$\\
            & $y$ & $0.74$ & $-0.003\pm0.015$\\
            & NB816 & ---   &  ---\\
            & NB921 & $0.76$ & $-0.003\pm0.012$\\
\hline
            & $g$ & $0.74$ & $-0.004\pm0.014$\\
            & $r$ & $0.67$ & $-0.003\pm0.015$\\
            & $i$ & $0.68$ & $-0.005\pm0.010$\\
    UD-SXDS & $z$ & $0.57$ & $-0.009\pm0.025$\\
            & $y$ & $0.65$ & $-0.001\pm0.011$\\
            & NB816 & $0.64$ & $-0.003\pm0.047$\\
            & NB921 & $1.05$ & $-0.006\pm0.018$\\
    \hline
 \hline
\end{longtable}

\begin{figure}
  \begin{center}
    \includegraphics[width=8cm]{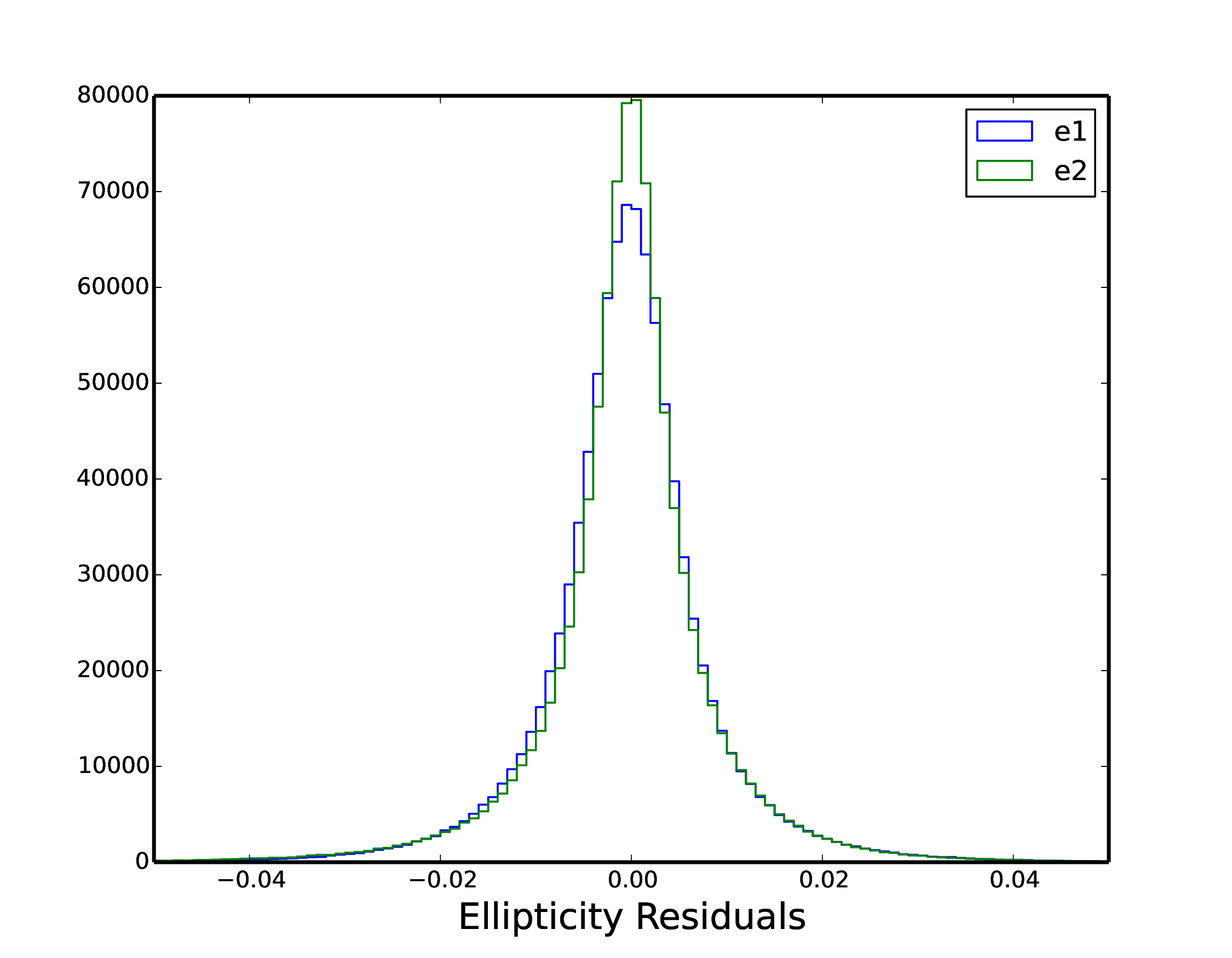}
  \end{center}
  \caption{
    The distribution of ellipticity residuals for the stars across the whole survey that were selected in
    the PSF modeling in the i-band.  The plot shows the two components of the ellipticity where $e1$
    corresponds to changes along the coordinate axes and $e2$ corresponds to elongation at 45$\deg$ from the axes.
  }
  \label{fig:psf_plots}
\end{figure}

To further evaluate the performance of the PSF modeling, we compare the ellipticity for
individual stars, measured by fitting Gaussian moments, to their corresponding PSF images.
The former are measured on the coadded image, while the PSF modeling is done on the individual visits.
The PSF model on the coadds is evaluated by warping and stacking the models from individual visits \citep{bosch17}.
Fig. \ref{fig:psf_plots} shows the ellipticity residuals for a selection of stars used
in the PSF modeling (typically $18-22.5$~mag stars) across the whole survey. 
The plot shows that we model the PSF at the percent level; the scatter in the ellipticity residuals is
$\sim 1\%$.  
More in-depth analysis can be found in the shear catalog paper \citep{mandelbaum17}.

\subsection{Star galaxy separation}
\label{sec:star_galaxy_separation}


We have used the \texttt{classification\_extendedness} parameter to separate stars from
galaxies in the previous sections.  Here, we test how well the parameter works as a function of magnitude.
The parameter is based on the magnitude difference between PSF and CModel as mentioned above
and is currently a binary classifier with extendedness 0 being point-like and 1 being extended in each band.
We use the HST/ACS catalog
in COSMOS \citep{2007ApJS..172..219L} as the truth table, which is a reasonable assumption given
the higher angular resolution of HST.  The star/galaxy classification in the catalog is reliable
down to $i\sim25$.  As the performance of the star-galaxy separation
depends on the image depth and seeing, we cross-matched the ACS catalog with the COSMOS
UltraDeep as well as with the best, median, and worst seeing Wide-depth stacks.

\begin{figure*}
 \begin{center}
  \includegraphics[width=8cm]{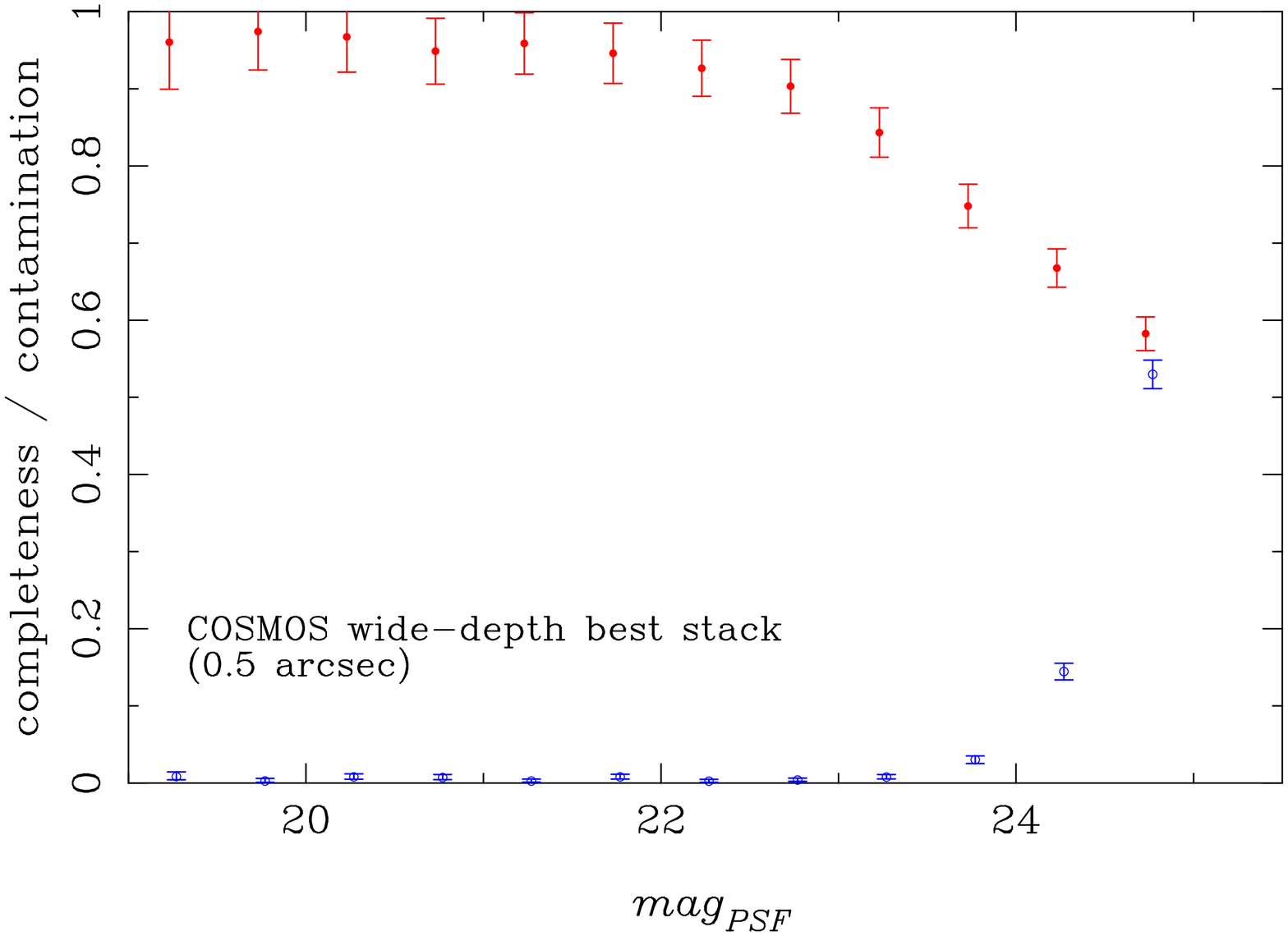}
  \includegraphics[width=8cm]{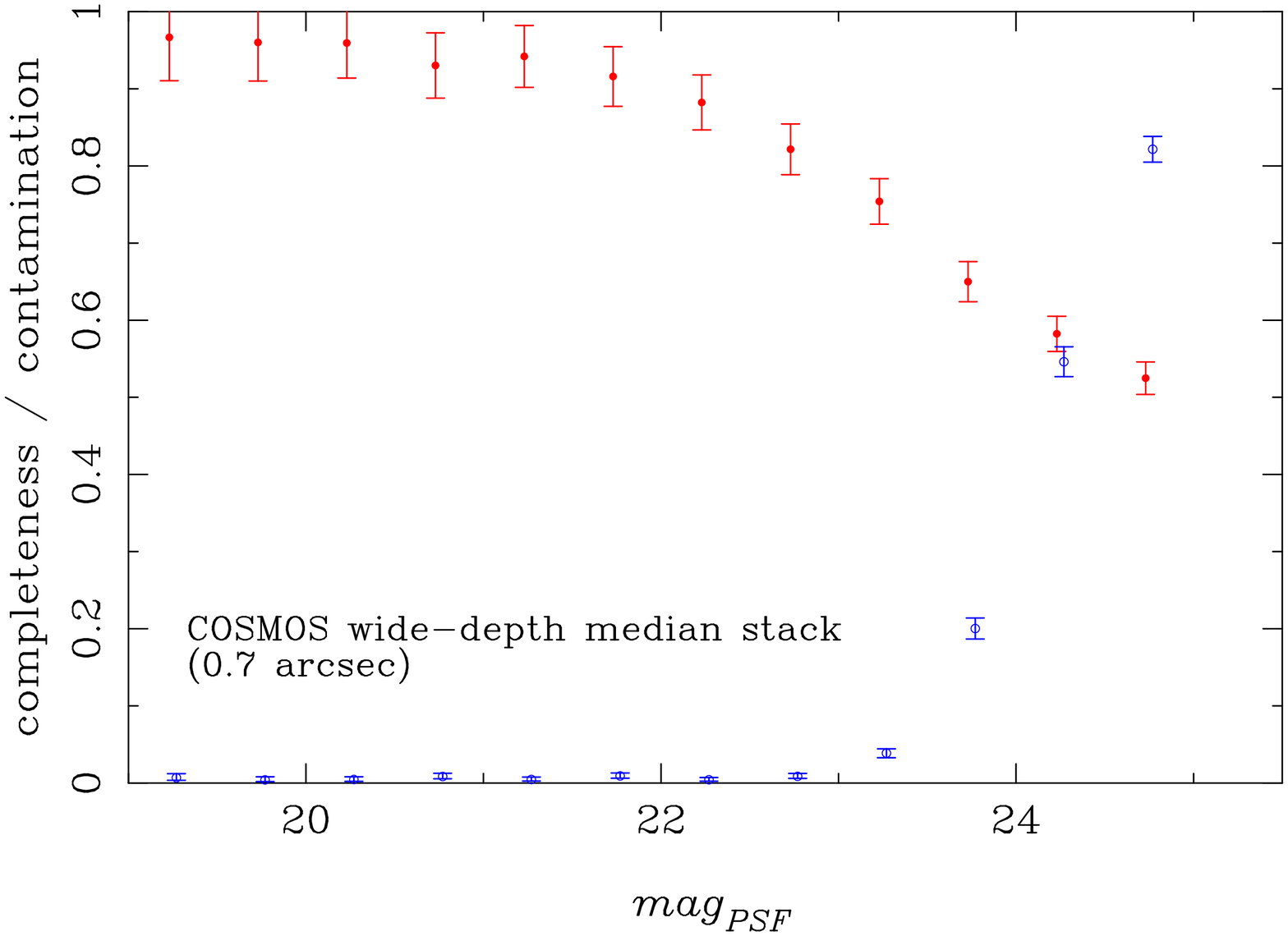}\\
  \includegraphics[width=8cm]{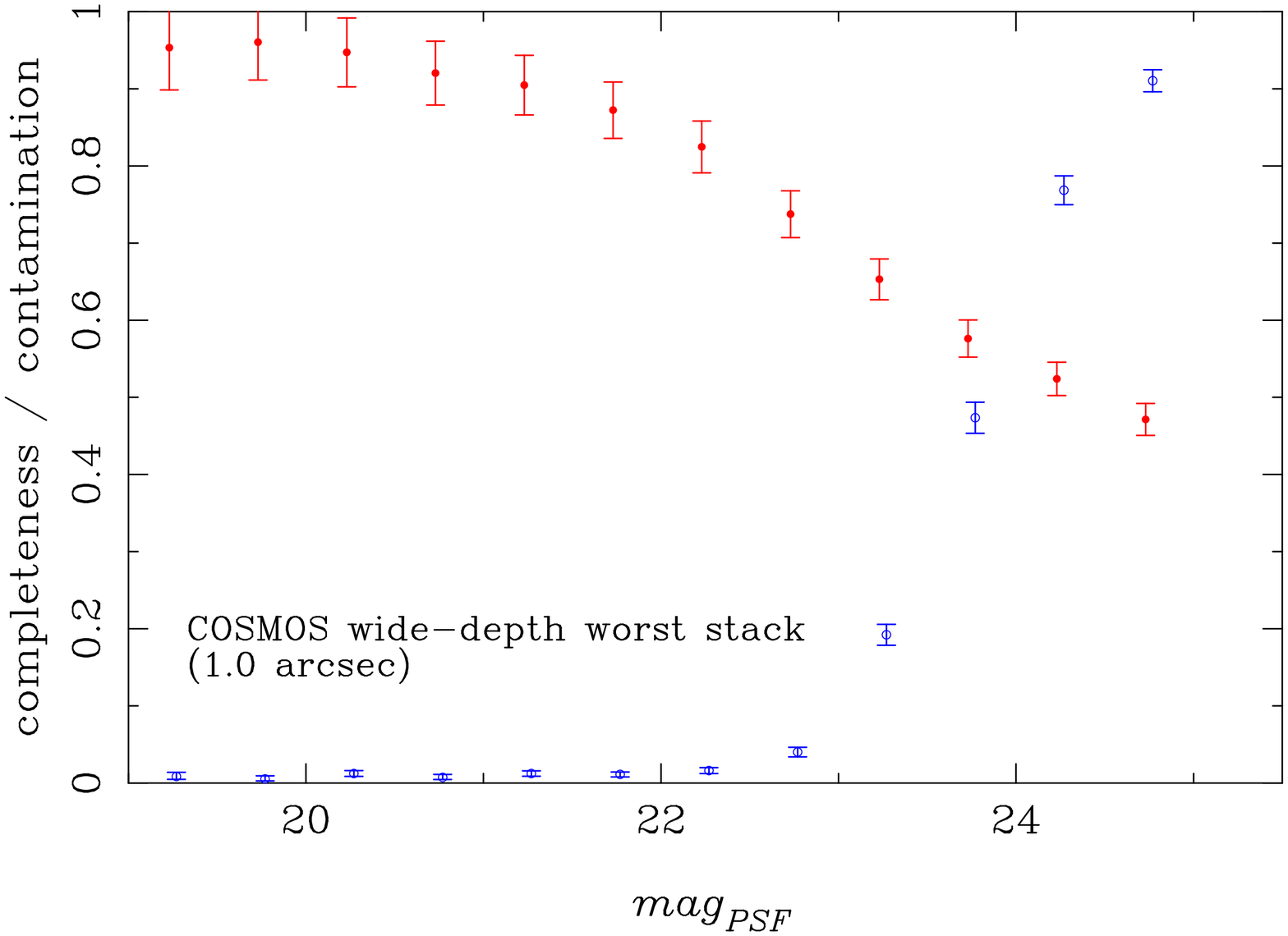}
  \includegraphics[width=8cm]{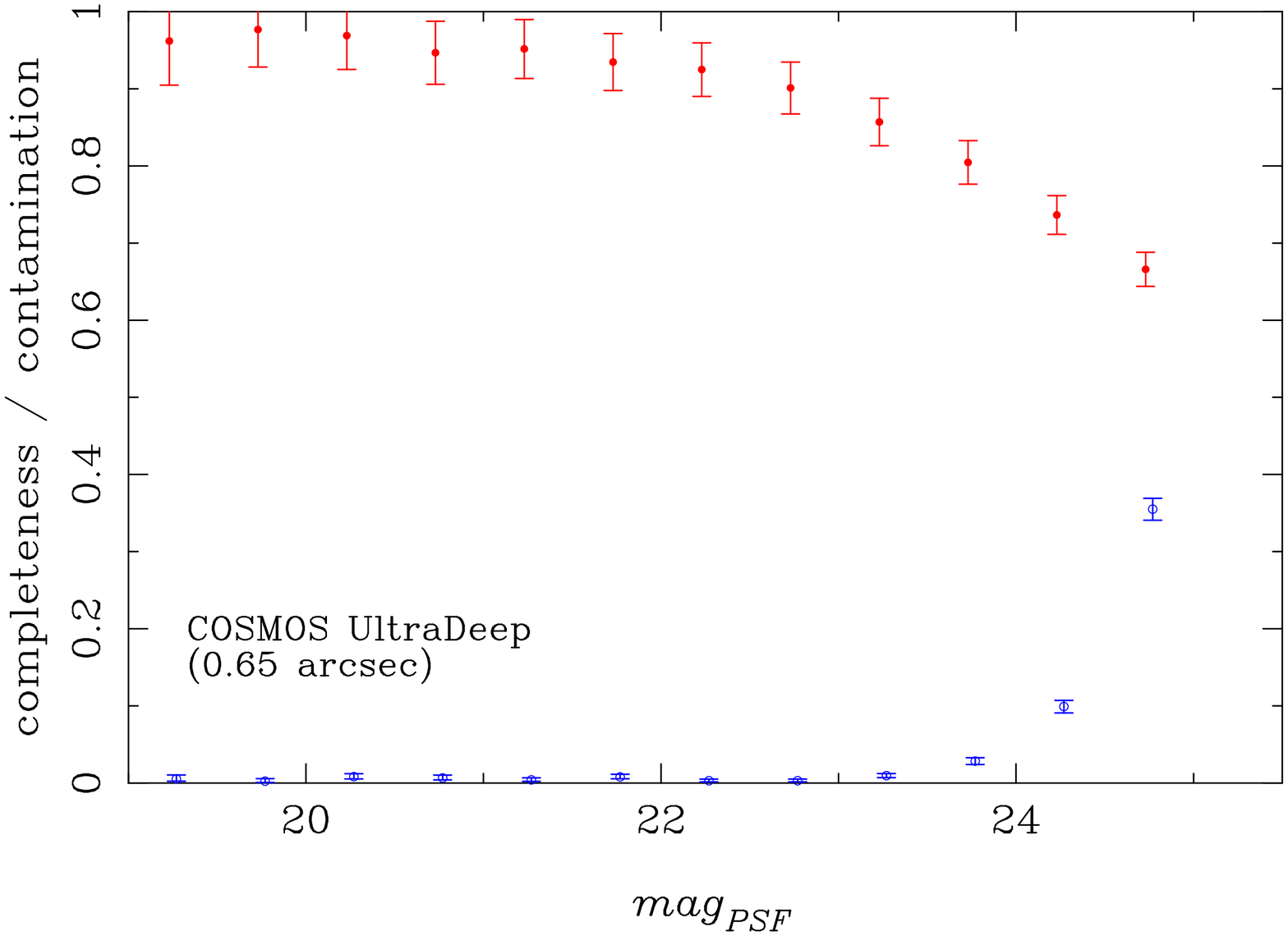}\\
 \end{center}
 \caption{
   Completeness (red) and contamination (blue) as a function of the $i$-band
   PSF magnitude for the COSMOS Wide-depth best (top-left), median (top-right),
   and worst (bottom-left) seeing stacks.  The bottom-right figure is for the UltraDeep depth.
   The error bars are Poisson errors.  This is star/galaxy separation compared to HST observations,
   where star-galaxy separation is taken as truth.
 }
 \label{fig:sg_sep}
\end{figure*}

Fig. \ref{fig:sg_sep} shows completeness and contamination of our classifications
for the COSMOS Wide-depth stacks with three different seeing sets and also for
the UltraDeep depth.  The seeing is shown in each panel.
The completeness is defined as the fraction of ACS stars properly classified as stars in HSC.
The contamination is the fraction of ACS galaxies among objects classified as stars in HSC.
Under the typical seeing conditions of 0.7 arcsec, the star-galaxy separation is reasonable down to
$i\sim24$, although the completeness is somewhat low (60\%).
At fainter magnitudes, the classification is rather difficult.
The classification accuracy is a strong function of seeing and depth as expected;
e.g., the completeness is still 60\% at $i\sim25$ when the seeing is $0.5$ arcsec,
but the same level of completeness can be achieved only at $i\sim23.5$ under 1 arcsec seeing.
In deeper imaging, the classification is still reasonable even in the faintest magnitude bin
with completeness above 60\%.

We plan to include another star/galaxy classifier using the size and color information
in a future incremental release.
The new classifier gives a continuous probability between 0 and 1 and is known to outperform
the extendedness parameter \citep{bosch17}.

\subsection{Survey Depth}
\label{sec:survey_depth}

\begin{figure}
 \begin{center}
  \includegraphics[bb= 0 0 530 438,clip,width=8cm]{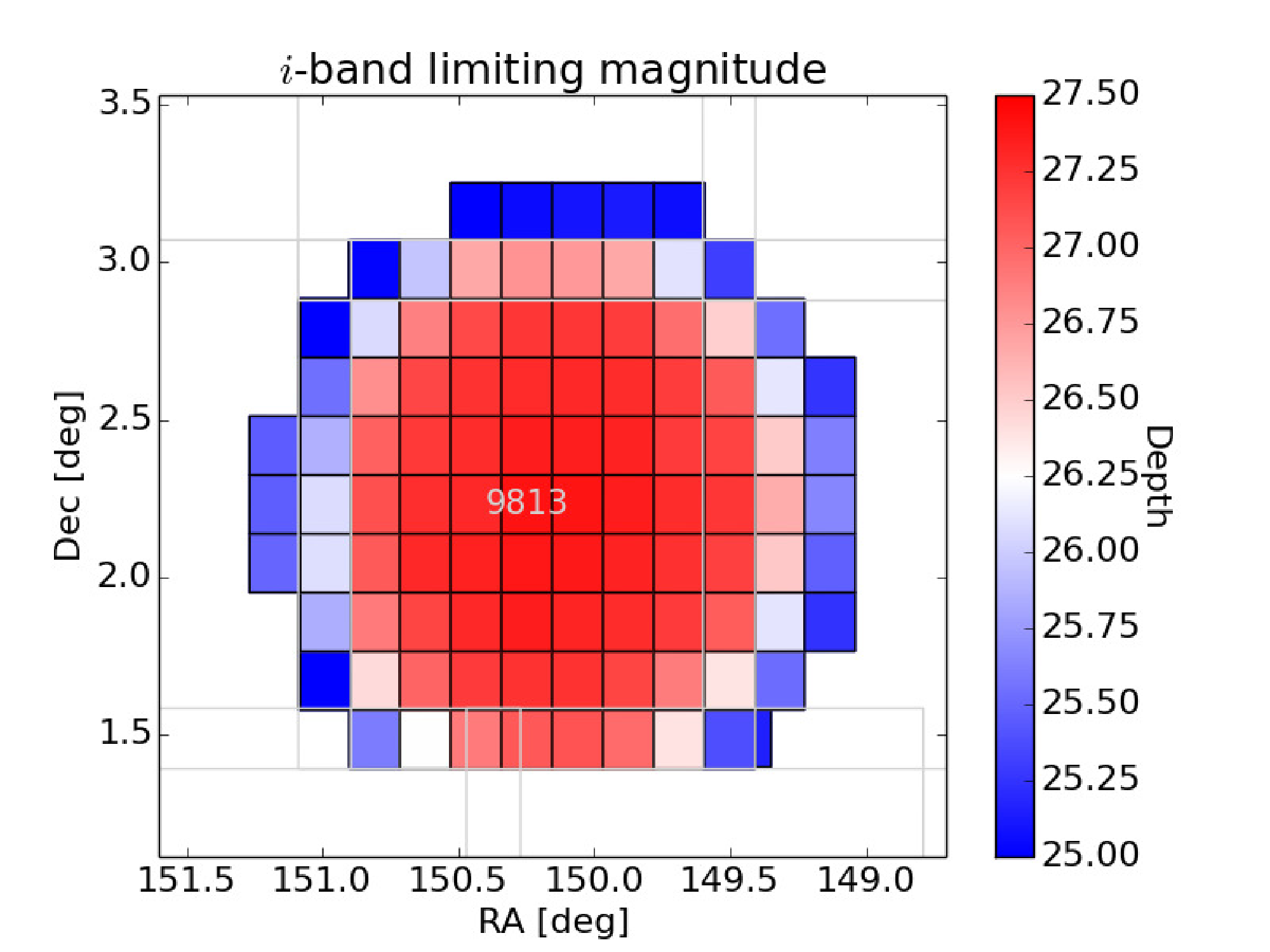} 
 \end{center}
 \caption{
   Depth map of the $i$-band limiting magnitude for 5$\sigma$ point-source detection in the COSMOS UltraDeep field.
   Each square represents a patch.
 }
 \label{fig:fig_depth_i_obj}
\end{figure}

We estimate $5\sigma$ limiting magnitudes in our survey fields.  There are a number of ways
to estimate the depth, but here we take a simple approach --- we estimate magnitudes at which
the PSF photometry has $S/N\sim5\sigma$, where the flux uncertainties are as quoted by the pipeline.
Because we use the PSF photometry on coadds,
we tend to underestimate the flux uncertainty due to covariances between the pixels introduced
in the warping, leading us to somewhat optimistic estimates of the depth.  Also, systematic uncertainties
such as PSF modeling error are not accounted for.
Despite these caveats, this is still a useful way to evaluate the depth over
the entire survey region.  We first apply a set of pixel flags (\texttt{flags\_pixel\_saturated\_center},
\texttt{flags\_pixel\_interpolated\_center}, \texttt{detect\_is\_primary}) and select
objects that have PSF magnitudes $S/N = 4-6 \sigma$ in each patch.  We then take their mean
magnitude to represent the $5\sigma$ depth for point sources, assuming that the source distribution
is flat within this range.
As an example, Fig. \ref{fig:fig_depth_i_obj} shows the $i$-band limiting magnitude map of
the UltraDeep COSMOS field.  We reach an impressive depth of $i\sim27.5$ in the central
$\sim1.5$ square degrees of the COSMOS field.  This is surely one of the deepest images of
the COSMOS field  (cf. \cite{2007ApJS..172...99C}).
Once again, this is the depth using only the data gathered in DR1 and we will go deeper in the future.
The $5\sigma$ depths for each filter and for each patch measured in this way over the entire survey fields are available in the database.


\subsection{Detection Completeness}

Another approach to characterize the survey depth is to evaluate detection completeness
by inserting artificial point sources in the coadds and repeating the detection.
One could add objects in individual visits instead of coadds for better estimates (and that is exactly
what SynPipe does; \cite{huang17}), but we work with the coadds in order to save computing time.
The detection completeness is dependent on size and shape of objects,
but we focus on point sources for simplicity.  We put artificial point sources at random
positions in the coadds using the PSF model (\texttt{coaddPsf}) at each position.
We make a series of magnitude bins and generate and detect point sources.  When matching
the input and output catalogs, we use a matching radius of 0.5 arcsec.
As we put artificial point sources at random positions, some of them may be located close to real
objects and matched with them just by chance even when an input object is too faint to be detected.
We find that the probability of this random matching is about 10\% (the exact number depends on
the filter).  We assume that the completeness at 30th magnitude, where we should find no matches,
represents the random matching probability and we correct for it in the following discussion.
To be specific, we apply

\begin{equation}
  P_{corr}=(P-P_{random})/(1-P_{random}),
\end{equation}

\noindent
where $P$ is the measured matching probability, $P_{random}$ is the random matching probability,
and $P_{corr}$ is the corrected probability.

\begin{figure}
 \begin{center}
   \includegraphics[width=8cm]{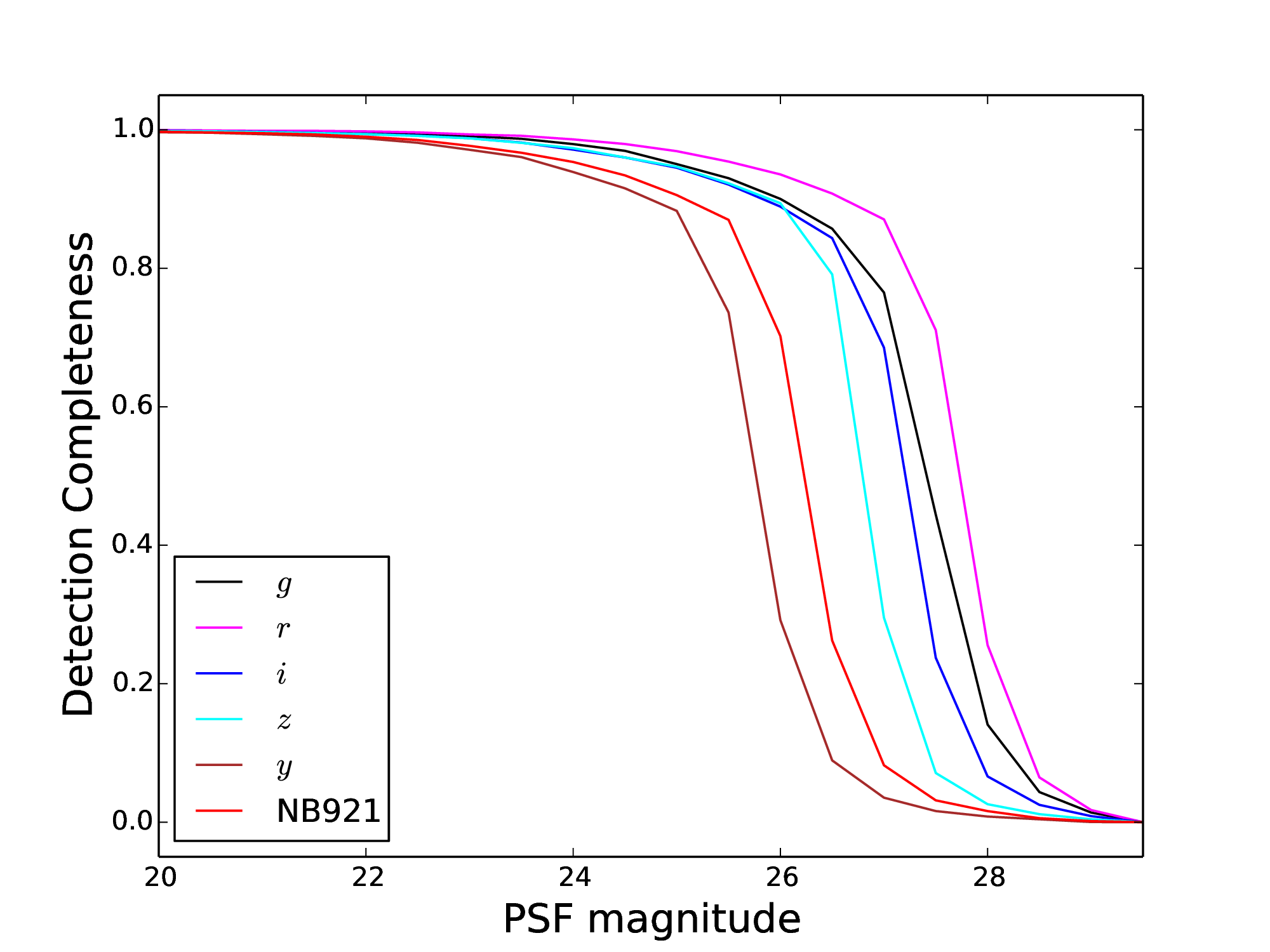} 
 \end{center}
 \caption{
   Detection completeness as a function of magnitude in the central region of the COSMOS UltraDeep field (tract=9813, patch=4,5).
   The different colors show different filters as indicated in the figure.  Effects of random matching have been corrected for in this plot.
 }
 \label{fig:fig_detect_all}
\end{figure}

Fig. \ref{fig:fig_detect_all} shows the detection completeness in the central region of the COSMOS field.
We are 80\% complete for point sources at $g\sim26.8$, $r\sim27.2$, $i\sim26.6$, $z\sim26.5$,
$y\sim25.3$, and $\rm NB921\sim25.7$. 
The $r$-band is the deepest band in COSMOS due to the superb seeing ($\sim0.5$ arcsec).  Comparisons with
the $5\sigma$ magnitude limits quoted earlier suggests that the $5\sigma$ limits correspond
roughly to 50\% completeness limits.
For reference, we find that 3$\sigma$ and 10$\sigma$ limits estimated in the same way
correspond to 15\% and 85\% detection completeness.

\begin{figure}
  \begin{center}
    \includegraphics[width=8cm]{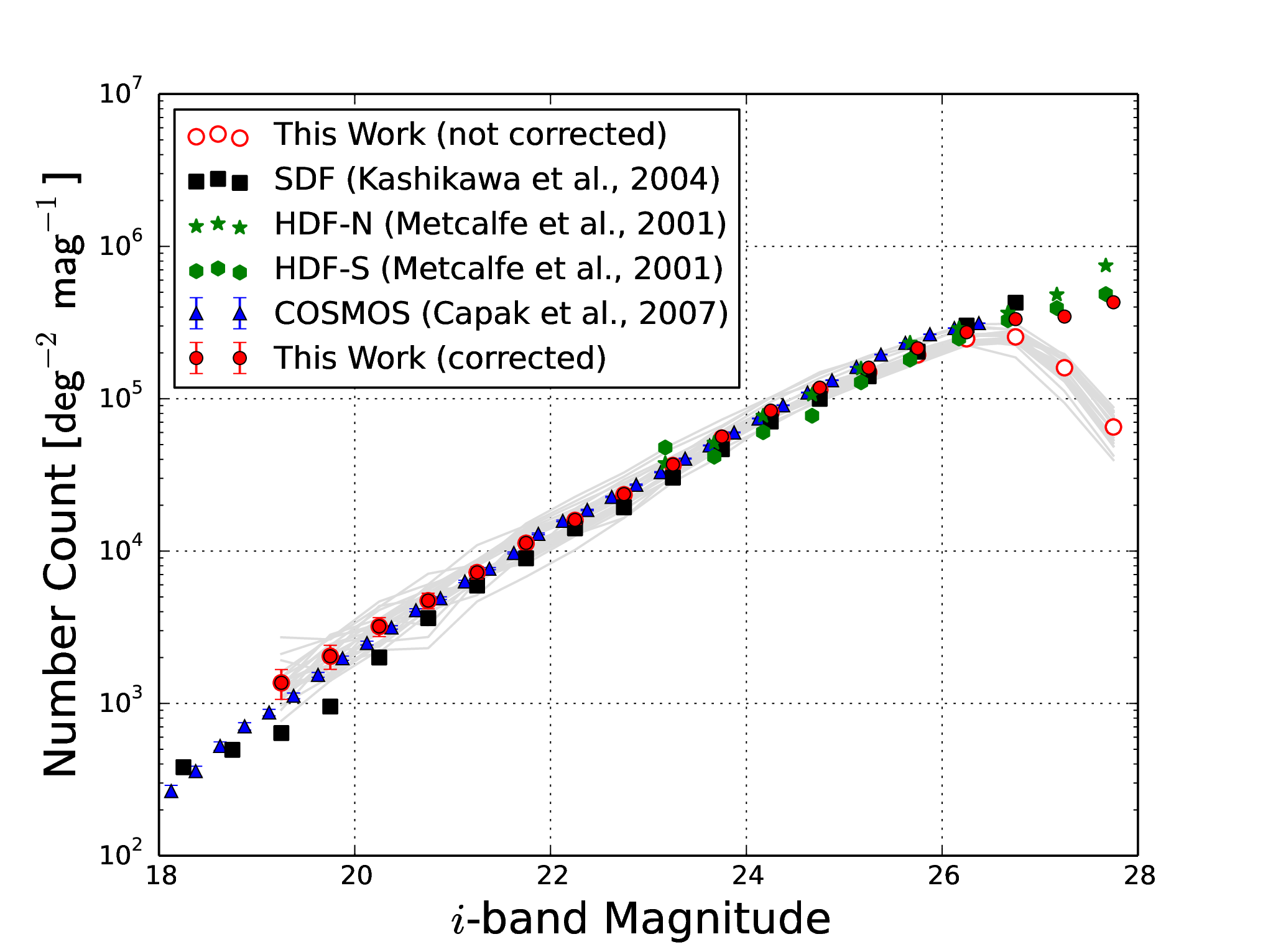} 
 \end{center}
 \caption{
   Galaxy number counts in the $i$-band.
   Open and solid circles show the raw and corrected galaxy number counts from the COSMOS UltraDeep catalog.
    Black, blue, and green points show the galaxy number counts from the literature
    \citep{2001MNRAS.323..795M,2004PASJ...56.1011K,2007ApJS..172...99C}.
    The gray lines show the number counts in each patch.
 }
 \label{fig:fig_number_count}
\end{figure}

As a further test of the detection completeness, we compare the galaxy number counts as
a function of magnitude with literature results.  In addition to the pixel flags used in
the previous section, we impose \texttt{classification\_extendedness}=1 at $i<24.5$ to eliminate
point sources.  At fainter magnitudes, we assume all the sources are extended because galaxies
significantly outnumber stars at such faint magnitudes at high latitudes.  As we now
focus on galaxies, we use the CModel photometry.  The open circles in Fig. \ref{fig:fig_number_count}
show the observed counts of galaxies in the $i$-band and they agree with the literature results
\citep{2001MNRAS.323..795M,2004PASJ...56.1011K,2007ApJS..172...99C} 
down to $i\sim26.5$.  At fainter magnitudes, the completeness drops rapidly, which is consistent
with Fig. \ref{fig:fig_detect_all}.
Using the completeness estimates, we can apply a correction
to the observed galaxy counts to reconstruct the real counts.
This is only a rough correction because we apply the completeness correction for point sources
to galaxies.  The filled circles in Fig. \ref{fig:fig_number_count} show the corrected counts.   
The corrected galaxy counts agree reasonably well with HDF-S down to $i\sim28$, suggesting that
our completeness estimates are reasonable.



\subsection{Known problems}
\label{sec:known_problems}

As demonstrated in the previous section, our data are of high quality, but
they are not without problems.  In this section, we summarize known issues in our data.
We will keep the list of known problems up-to-date at the data release site.
We will continue to improve the pipeline to mitigate these problems for future data releases.

\subsubsection{Disabled junk suppression}
\label{sec:disabled_junk_suppression}
  
We often detect a large number of spurious sources in the outskirts of bright
stars and galaxies because some pixels go above the detection threshold just
by chance due to noise fluctuations in the presence of an elevated background.
In order to suppress these spurious detections,
we subtract the very local ``sky'' in the detection step.  However, this junk suppression
procedure was mistakenly left disabled in the main processing.  It was turned
on in the afterburner processing (Section \ref{sec:afterburner}) and users are encouraged to use the afterburner
table in the database to reduce spurious sources.
About 10\% of the sources have been flagged as junk.
They are mostly faint noise peaks, but photometry of parent
objects may also be affected by the spurious sources as the afterburner only flags
them and does not re-perform photometry.  Photometry of bright objects or objects with
extended outskirts should thus be handled with caution (see also galaxy shredding
in Section \ref{sec:shredded_bright_galaxies}).

\subsubsection{Missing patches}

Some of the patches are missing due to processing failures, which are in part
caused by the disabled junk suppression and also by bright stars contaminating the patches.
This results in holes in the survey footprint.
To be specific, there are three missing patches in the Hectomap region in all the bands,
and nearly a whole tract is missing in VVDS in the $y$-band (tract 9936).
These missing patches are summarized at the data release site.

\subsubsection{Shredded bright galaxies}
\label{sec:shredded_bright_galaxies}

Large galaxies that have significant sub-structure are often overly deblended into many
smaller objects.  The fact that we did not enable junk suppression makes this even worse.
This ``shredding'' of objects results in poor photometry because a significant fraction of
light is assigned to the child objects.  The effect is more severe for late-type galaxies
than for early-type galaxies due to spiral arms and knots therein.
Comparisons with the SDSS photometry shows that, for bright ($i<19$) blue galaxies,
about 15\% of them suffer from shredding, half of which have their photometry underestimated
by $>0.25$~mag.  Shredding is a larger problem for brighter sources with $i\lesssim18$.
In the future, we plan to use techniques similar to those used in the SDSS pipeline
\citep{SdssPhoto} to identify such objects and remove the appropriate child objects from the blend.

\subsubsection{Poor PSF modeling in good seeing areas}
\label{sec:poor_psf_modeling_in_good_seeing_areas}

We are unable to model the PSF accurately for the visits with extremely good seeing
as already mentioned in Section \ref{sec:VVDS}.  The problem severely affected the $i$-band
in the VVDS field and has been mitigated by reprocessing the data with these
visits removed as a temporary solution.  However, there are about 20
affected patches in the $z$-band in VVDS ($\sim0.035$ square degrees, which is about
0.035 per cent of the Wide data in this release).  The other fields are also affected
(but less severely).  These bad patches should not be used for science analysis as
the photometry is poor.  They can be easily identified as having a large scatter
and offset of the stellar sequence in color-color diagrams performed as part of
the validation test in section  \ref{sec:stellar_sequence}, and
these color scatter and offset values for each patch can be found in the \texttt{patch\_qa} database table.

\subsubsection{Over-subtracted sky around large objects}

\begin{figure}
 \begin{center}
  \includegraphics[width=8.5cm]{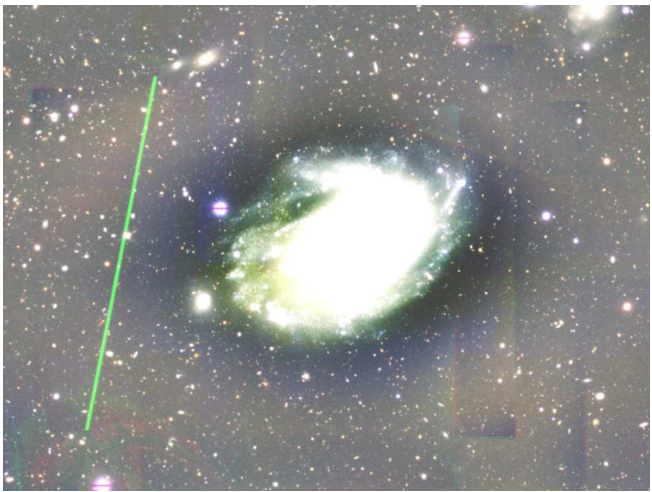} 
 \end{center}
 \caption{
   Remaining satellite trail (the slanted green line on the left) and the over-subtracted
   sky background around the large galaxy at the center.  The level is stretched to enhance
   the background.  The image is approximately 9 arcmin $\times$ 7 arcmin.
 }
 \label{fig:known_issues1}
\end{figure}

The sky around large objects with size $\gtrsim1$ arcmin is often over-subtracted (Fig. \ref{fig:known_issues1}).
As described in the pipeline paper, we apply the background subtraction on a CCD by CCD basis
using 128 pixel grids.  The grid size is a trade-off between how well we subtract the sky
on small scales and how well we keep the large-scale light profile of objects unaffected.
The current choice is tuned for the former, and the outskirts of large objects are often
misinterpreted as part of the sky, resulting in the over-subtraction.  A new algorithm to
subtract the sky using the entire field of view has been developed and it will improve
the sky subtraction in our future releases.

\subsubsection{Poor CModel photometry for large galaxies}

Despite the galaxy shredding and over-subtraction of the sky, CModel tends to
overestimate fluxes of large galaxies.  Compared to the SDSS photometry in the $i$-band,
about 50\% of bright ($i<19$) blue galaxies have overestimated CModel fluxes
by $-0.1$ to $-0.7$~mags.  In rare cases (2\%),
magnitude differences can be $-0.7$ to $-1.0$~mag.
On the other hand, only $\sim20\%$ of galaxies with $i<19$
have consistent CModel photometry with SDSS within $0.05$~mag.
Although CModel photometry of red galaxies appears less biased, we still observe a large
scatter and obtain similar numbers to the blue galaxies (e.g., only $\sim20$\% have consistent
photometry within $0.05$~mag).
The exact cause of this somewhat discrepant CModel photometry is being investigated.
More extensive tests of the CModel photometry can be found in \citet{huang17}.

\subsubsection{Satellite trails}

We detect and mask satellite trails by identifying outlying pixels in individual visits
used in the coadd, but a fraction of satellite trails still remain unmasked (Fig. \ref{fig:known_issues1}).
This is more severe in the narrow bands, in which individual exposures are longer and thus
we have fewer visits.  This results in detected ``objects'' with nonsense colors and very high ellipticity,
which can be used to reject them at a catalog level.
But, users searching for objects detected in a small number of filters
(e.g., Lyman $\alpha$ emitters) should be careful and are advised to visually
check the images.  A satellite trail finder on single exposure using Hough
transform is being developed.  Also, difference imaging will be implemented as part of
the processing in our future releases, allowing us to detect and reject satellite trails
as they can be identified as residuals in difference images.

\subsubsection{Ghosts and scattered light due to bright stars}
  
\begin{figure}
  \begin{center}
  \includegraphics[width=4.cm]{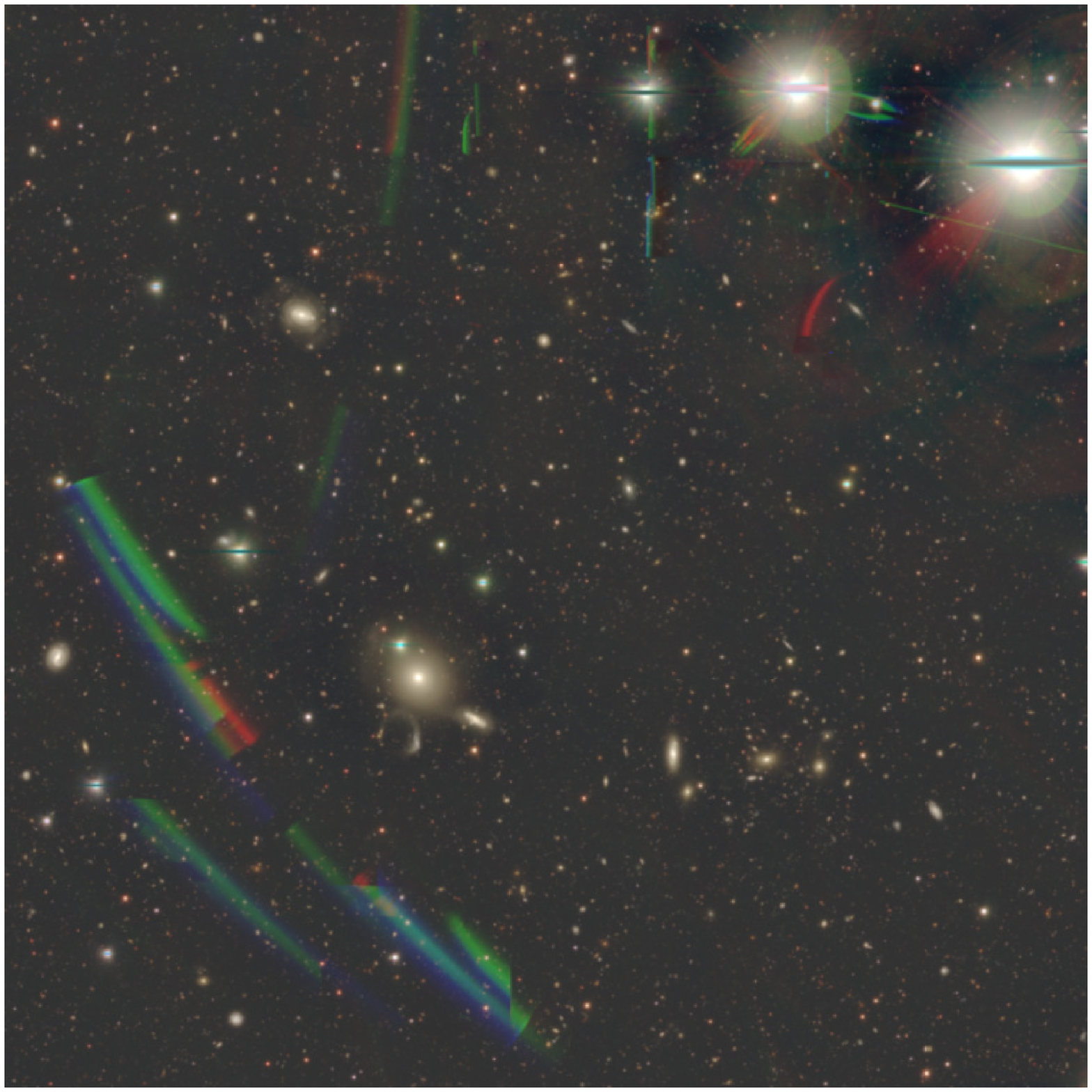} 
  \includegraphics[width=4.cm]{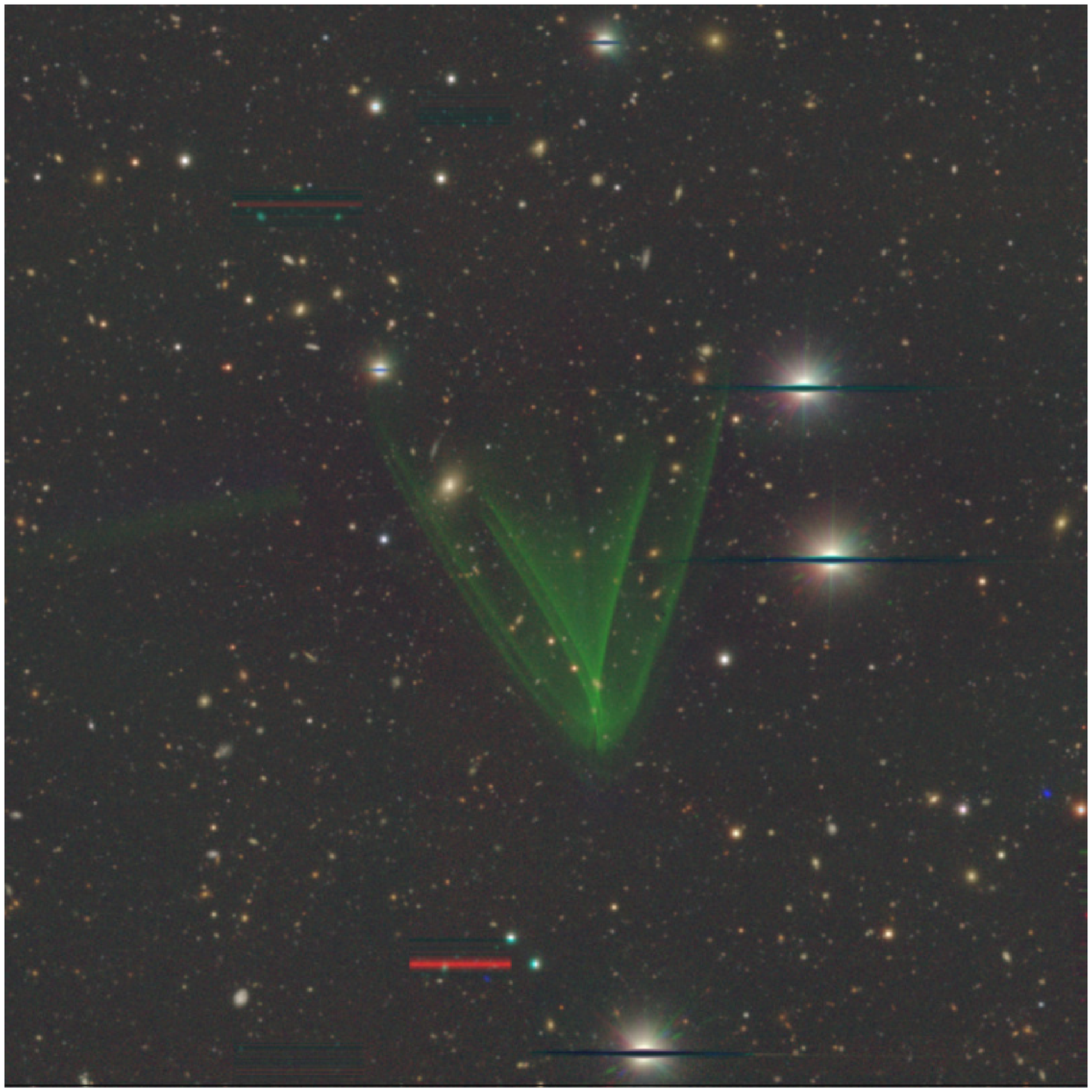} 
  \end{center}
 \caption{
   Scattered light from nearby bright stars.
 }
 \label{fig:known_issues23}
\end{figure}

Ghosts and scattered light due to bright stars are often left unmasked
(Fig.~\ref{fig:known_issues23}).  The frequency of these optical artifacts
depends on the density of bright stars, but for reference, about $1-2\%$ of the area in UD-COSMOS
is affected by ghosts and scattered light.  The ghosts and scattered light as well as the satellite
trails mentioned above are worse in the UltraDeep and Deep data than in Wide because of the small dithers.
Cataloged objects that are located coherently on the sky over
$\gtrsim1$ arcmin should be taken with caution and should be visually checked.
An algorithm to predict the location of ghosts from a list of bright stars is being developed.
The difference imaging mentioned above will also reduce the ghosts in our future processing.

\subsubsection{Overly conservative bright object masks}

Objects close to bright stars are flagged (\texttt{flags\_pixel\_bright\_object\_\{center,any\}}) because they are likely to have bad photometry.
We use a catalog of bright stars from Tycho-2 \citep{2000A&A...355L..27H} and the bright object catalog from
Naval Observatory Merged Astrometric Dataset (NOMAD; \cite{2005yCat.1297....0Z})
in the current version.
We mask objects brighter than 17.5 mag in any of the $BVR$ filters in these catalogs,
which is approximately the saturation limit of the HSC data (see Table \ref{tab:exptime}).
The current bright object masks may be overly conservative, e.g., a whole tract can be masked where
there is a very bright (e.g., mag$<5$) star, although many objects far from the stars are actually
unaffected.  
Another known feature is that nearby bright galaxies are often misinterpreted as stars in the NOMAD catalog.
About 8\% of the masked objects are actually galaxies.
It is advised not to use the bright object
masks for studies of nearby galaxies.  Improvements will be made in a future version of the pipeline.

\subsubsection{Deblending failure in crowded areas}
\label{sec:deblending_failure_in_crowded_areas}

\begin{figure}
 \begin{center}
  \includegraphics[width=8.cm]{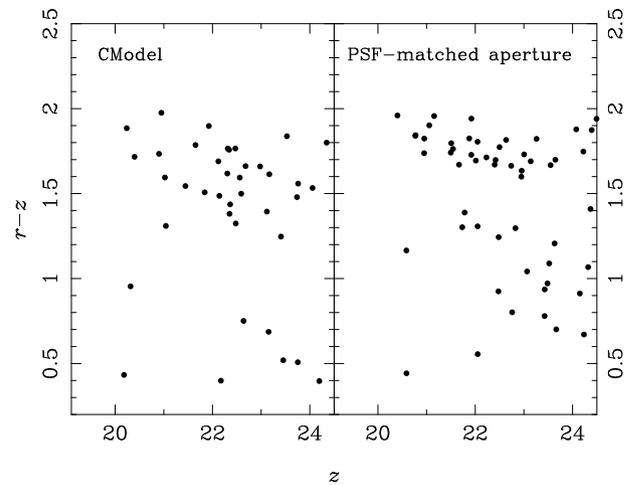} 
 \end{center}
 \caption{
   $r-z$ plotted against $z$ for galaxies in the core of a redshift $\sim0.7$ cluster.
   The left and right panels are for CModel and afterburner photometry, respectively.
   Note the tighter red sequence in the right panel.
 }
 \label{fig:cmd_cluster_core}
\end{figure}

The deblender tends to fail in very crowded areas such as the cores of galaxy clusters,
resulting in poor photometry.  This is a major problem for cluster science especially
at high redshifts, where clusters appear more compact.
As described in section \ref{sec:afterburner}, PSF-matched aperture photometry is performed as part of
the afterburner processing to mitigate the problem.  A color-magnitude diagram of
a $z\sim0.7$ cluster shown in Fig. \ref{fig:cmd_cluster_core} illustrates the improvement.
The cluster red sequence has a large scatter in CModel, while it is tighter when
the afterburner photometry is used.  Users working on high density
environments should check if their objects are affected by this problem and
use the afterburner photometry where appropriate.

\subsubsection{Underestimated flux uncertainties in the afterburner photometry}

Flux uncertainties in the afterburner photometry are underestimated because
significant covariances are introduced in the Gaussian smoothing process to match
PSFs and they are not accounted for.  The amount of underestimation depends on
the difference between the target seeing and native seeing, but it can be
a factor of several or larger.  As a rough proxy, one could use flux uncertainties
from the aperture photometry on the native PSF with the same aperture size.

\subsubsection{Incorrect prior weighting in CModel}

The CModel galaxy fitting algorithm utilizes a Bayesian prior on radius and
ellipticity, largely as a way to regularize fits to low S/N and/or poorly
resolved galaxies.  When combining this with the likelihood to form the
posterior probability (which is then maximized by the fitter), the relative
weighting of these terms is incorrect, giving the prior much greater
influence over the result than intended. This is essentially equivalent to
utilizing a prior that decreases much more rapidly than it should at large radius or large
ellipticity.  As a result, CModel sizes and ellipticities are biased low,
which almost certainly biases CModel fluxes low as well.
The prior is only
used when fitting the size and ellipticity, however, and this measurement is
done in only one band (albeit a different one for each object) before
performing forced photometry in all bands (see
Section~\ref{sec:multi_band_measurement}).  As a result, colors are much less
affected by this bug.
In fact, imposing such a strong penalty for large
radii -- even a physically unreasonable one -- seems to decrease the number
of catastrophic outliers in CModel colors.  However, any galaxy photometry
algorithm that operates on images with different PSFs in different bands can
yield inconsistent colors if the model assumed for the galaxies is incorrect
(as is always the case to some degree), and using the wrong prior can
exacerbate this.  We have not seen any evidence that incorrect prior
weighting is degrading the colors significantly in this respect,
but because
we do not know the true distribution of colors, these tests are generally
limited to comparisons with other flux measures and experiments on
simulations \citep{huang17}.  A more complete description of this problem can
be found in \citet{bosch17}.

\subsubsection{Poor astrometry in the corner of the UltraDeep COSMOS field}

The South-East corner of the UltraDeep COSMOS field has an astrometric error
in the $z$-band, likely introduced by a bad astrometric fit in the mosaic process.
Only a few patches suffer from the poor astrometry, but these patches should
not be used for science.
See the online document for a list of the patches that are affected.


\subsubsection{Residual background in the $y$-band}

\begin{figure}
 \begin{center}
  \includegraphics[width=8.cm]{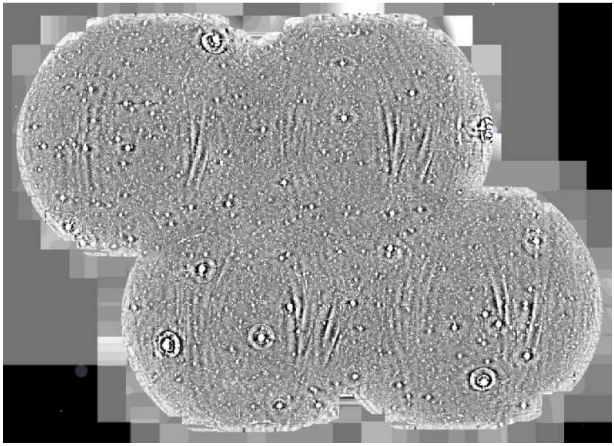}
 \end{center}
 \caption{
   $y$-band coadd image of DEEP2-3 field.  The level is stretched to enhance
   the background features.  The image is approximately 4 degree by 3 degree.
 }
 \label{fig:deep2f3_y}
\end{figure}

The y-band suffers from scattered light and it was not removed very well
in the sky subtraction, leaving arc/linear features with both positive and negative fluxes
in the coadds.  Fig. \ref{fig:deep2f3_y} shows the y-band image of the DEEP2-3 field.
The level is strongly stretched to enhance the features. These features are most prominent
in the Deep and UltraDeep fields where we apply small dithers, but can also be seen in the Wide layer.
The amplitude of the feature varies, but roughly $\pm$0.1 DN per pixel
(note that the zero-point is 27~mag/DN).
Sources close to these features may have poor photometry due to the improper background subtraction.
We have identified the source of the scattered light and are working on improved removal of the feature.

\subsubsection{Shallow $i$-band depth in the COSMOS Wide-depth median stack}

The COSMOS Wide-depth stacks can be used for various tests, but we discovered that one of
the visits used for the median seeing stack in the i-band had a guiding error and the visit
was actually not included in the coadd.  As a result, we have a shallower i-band data than
the Wide depth by 0.16 mag (the integration is 15min as opposed to 20min).  For many tests,
a depth change at a level of $0.1-0.2$ mag does not significantly matter, but it can be a major
problem for tests around the detection limits.  The problem exists
only the median seeing stack, and the other stacks are unaffected.

\subsubsection{BAD and CR flags do not propagate to the coadds}

When making coadds we ignore pixels with BAD or CR set. However, we neglected to set the corresponding
mask bits on the coadd to indicate that we have done this. We also did not set any mask bit on the coadd
for regions that are at the boundary of CCDs.
This problem makes our coadded PSF model inconsistent with images in these areas, since the coadded
PSF model does not account for the fact that these pixels were excluded in the coadd. The same is
true for pixels removed with safe clipping algorithms, and the CLIPPED
flag can be helpful to filter these pixels.
We have already found some discrepancies between the coadd PSFs and the per-visit PSFs. 
The effects of this problem are still under investigation and we will report results at the data release site.

\section{Catalog and Data Archives}
\label{sec:catalog_and_data_archives}


The processed images and catalogs are both made available in this data release,
and this section briefly describes the functionality of our dedicated database
and user interfaces.  Details of the database can be found
in \citet{takata17}. The current design of our data distribution scheme is similar to that of SDSS --
catalog data can be retrieved from postgreSQL database servers, while
custom-designed user interfaces allow users to retrieve binary data such as images.
We discuss each of the catalog and data archive servers in what follows.
Once again, the data release site is at \url{https://hsc-release.mtk.nao.ac.jp/}.

\subsection{Catalogs}

The catalog data are stored in postgreSQL database tables and can be
retrieved with SQL scripts.  Each of the Wide, Deep, and UltraDeep layers has its own
schema and \texttt{meas} (unforced), \texttt{forced}, and \texttt{afterburner}
tables, and a number of meta tables are available for each. As we have mentioned
earlier, we have some issues with the data such as poor photometry in
a very small number of patches.  The \texttt{patch\_qa} table can be used to identify
these problematic patches.  It also gives approximate depths ($5\sigma$ limiting magnitudes
for point sources) as well as the seeing sizes for each filter and for each patch.
The schema browser should be referred to for details of the table columns.
The online SQL editor provides an easy environment to write, check, and submit SQL queries.
In addition, queries can be sent from a local client using a Python-based script,
which will be useful for sequential data retrieval.

\subsection{Binary data}

The image files described in Section \ref{sec:data_products}, both individual CCDs and coadds,
are available for direct download.  As mentioned in Section \ref{sec:the_release}, some of the catalog
fits files will be released in a future incremental release.  There is an online search tool to
find files by constraining, e.g., filters and coordinates.  As patch images are large, an image
cutout interface is also available to generate postage stamps of objects by uploading a coordinate list.











In addition to the binary data, we offer a browser-based image viewer, hscMap.
A user can pan and zoom in and out of the HSC images,
change the filter combination for color composites, and tweak
flux levels.  Both the standard RGB color scheme and the SDSS color scheme \citep{2004PASP..116..133L}
are available.  hscMap accepts a user catalog to mark objects in the browser.
Also, it talks to the database and a catalog can be retrieved from the database
and loaded into browser.  More useful functions in hscMap are described in the online manual.


\subsection{Acknowledging the HSC data}

For any scientific publications based on the HSC-SSP data, please quote the first four
paragraphs in the acknowledgment section of this paper.  In addition, the following
publications should be referred to where appropriate:
the survey design paper \citep{aihara17},
\citet{miyazaki17} for the camera system, \citet{komiyama17}
for the camera dewar, \citet{kawanomoto17} for the filter response functions,
\citet{bosch17} for the processing pipeline,
\citet{takata17} for the database, \citet{furusawa17} for the on-site system,
\citet{tanaka17} for photometric redshifts, \citet{mandelbaum17} for
the lensing shear catalog,  \citet{huang17} for SynPipe, and this paper for the public data.
The pipeline is developed as part of LSST and therefore LSST should be referenced, too:
\citet{2008arXiv0805.2366I}, \citet{2010SPIE.7740E..15A}, and \citet{2015arXiv151207914J}.
We have calibrated our data against an early version of the Pan-STARRS data and 
this release would not have been possible without it.  We would like to encourage
users to reference Pan-STARRS as well:
\citet{2012ApJ...756..158S}, \citet{2012ApJ...750...99T}, and \citet{2013ApJS..205...20M}.


\section{Future Releases}
\label{sec:future_releaes}

Our current plan is to make major data releases every two years: DR2 in 2019 and DR3 in 2021.
Each of these future releases will include data from more than 100 additional nights
and we expect to make major improvements in the data quality as well as in the data retrieval
tools.

In addition to these major data releases, we will make incremental data releases,
likely once or twice a year.  Incremental releases are intended to deliver data
products to add value to the current major data release, not to increase the area.
The first incremental release happened in June 2017 and it included joint COSMOS data
by HSC-SSP and the University of Hawaii \citep{2017arXiv170600566T} and photo-$z$ products for the Wide layer
\citep{tanaka17}.
Another incremental release is planned and will include fully-validated
shape measurements for weak-lensing.
There are two surveys that are collaborating with us by obtaining deep
observations in the HSC-SSP fields.  The CFHT Large Area U-band Deep Survey (CLAUDS; Sawicki et al.,
in prep) has recently obtained very deep u-band imaging over 20 square degrees of the Deep and UltraDeep
layers to HSC-matched depths ($\sim$27.0 mag, 5$\sigma$ in 2 arcsec apertures);  these observations are
complete and the data are being processed.  Additionally, Steward Observatory is leading a near-IR
$JHK$ imaging campaign with UKIRT.   In the future, we plan to release u-band and near-IR enhanced products in
collaboration with our CLAUDS and Steward partners.

Updates of the user interfaces and data retrieval tools are also within the scope of an incremental release.
Currently, the catalog archive (i.e., database) and the data archive (i.e., flat files) are somewhat
separate, but
a python environment that will allow users to retrieve catalog products and image products
in the same fashion is being developed.  A major upgrade of hscMap is in progress and users
will be able to control hscMap from the console, which is a very powerful feature when combined
with the python environment.  Also, we plan to allow users to make their own tables on our database,
so that they can join their tables with the main database tables.

We note that an incremental release may happen without any publications (e.g., in case only the data
retrieval tools are updated) and users are encouraged to check our website regularly.
Registered users to the data release site will be notified.













\section{Summary}

We have presented the first data release of HSC-SSP.  The release includes data from the first 61.5 nights
of the survey and covers over 100 square degrees of the Wide area and $\sim30$ square degrees
of the Deep and UltraDeep area.  We have processed the data with a version of LSST stack, hscPipe,
and demonstrated the quality of our data; we achieve 1-2\% PSF photometry and $\sim10/40$ mas
internal/external astrometry, and we reach $i\sim26.4$, $\sim26.5$, and $\sim27.0$, in the Wide,
Deep, and UltraDeep layers, respectively.  These are the depths thus far and we will go even deeper
in the Deep and UltraDeep layers.  There are a number of known issues in the data, but we have plans to
fix them in our future releases.  The processed images and catalogs are served to the community
through dedicated databases and user interfaces, allowing users to retrieve the data easily.
Only a brief outline of the data products is given in this paper, but more detailed
information can be found at the data release site as well as in companion papers.

We plan to make incremental data releases to enhance the scientific value of this data release.
The first incremental release has happened already as mentioned above and we plan to make another
one to release detailed shape measurements.
On a longer term, we will make two more
major data releases as the survey progresses, each of which will include
additional $>100$ nights of data.  We hope to make significant improvements in the data quality
as well as in the database and user interfaces for the community to fully exploit even larger
sets of HSC data.

\section*{Acknowledgments}
The Hyper Suprime-Cam (HSC) collaboration includes the astronomical communities of Japan and Taiwan,
and Princeton University.  The HSC instrumentation and software were developed by the National
Astronomical Observatory of Japan (NAOJ), the Kavli Institute for the Physics and Mathematics of
the Universe (Kavli IPMU), the University of Tokyo, the High Energy Accelerator Research Organization (KEK),
the Academia Sinica Institute for Astronomy and Astrophysics in Taiwan (ASIAA), and Princeton University.
Funding was contributed by the FIRST program from Japanese Cabinet Office, the Ministry of Education,
Culture, Sports, Science and Technology (MEXT), the Japan Society for the Promotion of Science (JSPS),
Japan Science and Technology Agency  (JST),  the Toray Science  Foundation, NAOJ, Kavli IPMU, KEK, ASIAA,
and Princeton University.

This paper makes use of software developed for the Large Synoptic Survey Telescope. We thank the LSST
Project for making their code available as free software at http://dm.lsst.org.

The Pan-STARRS1 Surveys (PS1) have been made possible through contributions of the Institute for Astronomy, the University of Hawaii, the Pan-STARRS Project Office, the Max-Planck Society and its participating institutes, the Max Planck Institute for Astronomy, Heidelberg and the Max Planck Institute for Extraterrestrial Physics, Garching, The Johns Hopkins University, Durham University, the University of Edinburgh, Queen's University Belfast, the Harvard-Smithsonian Center for Astrophysics, the Las Cumbres Observatory Global Telescope Network Incorporated, the National Central University of Taiwan, the Space Telescope Science Institute, the National Aeronautics and Space Administration under Grant No. NNX08AR22G issued through the Planetary Science Division of the NASA Science Mission Directorate, the National Science Foundation under Grant No. AST-1238877, the University of Maryland, and Eotvos Lorand University (ELTE) and the Los Alamos National Laboratory.

This paper is based on data collected at the Subaru Telescope and retrieved from the HSC data archive system, which is operated by Subaru Telescope and Astronomy Data Center at National Astronomical Observatory of Japan.

We thank the anonymous referee for a useful report, which helped improved the paper.

This work is also based on zCOSMOS observations carried out using the Very Large Telescope at the ESO Paranal Observatory under Programme ID: LP175.A-0839, on observations taken by the 3D-HST Treasury Program (GO 12177 and 12328) with the NASA/ESA HST, which is operated by the Association of Universities for Research in Astronomy, Inc., under NASA contract NAS5-26555, on data from the VIMOS VLT Deep Survey, obtained from the VVDS database operated by Cesam, Laboratoire d'Astrophysique de Marseille, France, on data from the VIMOS Public Extragalactic Redshift Survey (VIPERS). VIPERS has been performed using the ESO Very Large Telescope, under the "Large Programme" 182.A-0886. The participating institutions and funding agencies are listed at http://vipers.inaf.it.  Funding for SDSS-III has been provided by the Alfred P. Sloan Foundation, the Participating Institutions, the National Science Foundation, and the U.S. Department of Energy Office of Science. The SDSS-III web site is http://www.sdss3.org/.  SDSS-III is managed by the Astrophysical Research Consortium for the Participating Institutions of the SDSS-III Collaboration including the University of Arizona, the Brazilian Participation Group, Brookhaven National Laboratory, Carnegie Mellon University, University of Florida, the French Participation Group, the German Participation Group, Harvard University, the Instituto de Astrofisica de Canarias, the Michigan State/Notre Dame/JINA Participation Group, Johns Hopkins University, Lawrence Berkeley National Laboratory, Max Planck Institute for Astrophysics, Max Planck Institute for Extraterrestrial Physics, New Mexico State University, New York University, Ohio State University, Pennsylvania State University, University of Portsmouth, Princeton University, the Spanish Participation Group, University of Tokyo, University of Utah, Vanderbilt University, University of Virginia, University of Washington, and Yale University.  GAMA is a joint European-Australasian project based around a spectroscopic campaign using the Anglo-Australian Telescope. The GAMA input catalogue is based on data taken from the Sloan Digital Sky Survey and the UKIRT Infrared Deep Sky Survey. Complementary imaging of the GAMA regions is being obtained by a number of independent survey programmes including GALEX MIS, VST KiDS, VISTA VIKING, WISE, Herschel-ATLAS, GMRT and ASKAP providing UV to radio coverage. GAMA is funded by the STFC (UK), the ARC (Australia), the AAO, and the participating institutions. The GAMA website is http://www.gama-survey.org/.  Funding for the DEEP2 Galaxy Redshift Survey has been provided by NSF grants AST-95-09298, AST-0071048, AST-0507428, and AST-0507483 as well as NASA LTSA grant NNG04GC89G.  Funding for PRIMUS is provided by NSF (AST-0607701, AST-0908246, AST-0908442, AST-0908354) and NASA (Spitzer-1356708, 08-ADP08-0019, NNX09AC95G). HM is supported by the Jet Propulsion Laboratory, California Institute of Technology, under a contract with the National Aeronautics and Space Administration.
This work is in part supported by MEXT Grant-in-Aid for Scientific Research on Innovative 
Areas (No.~15H05887, 15H05892, 15H05893).

\bibliographystyle{apj}
\bibliography{references}

\end{document}